\documentclass{psp-book9x6}

\usepackage{psp-book-van} 
\usepackage{cite,graphicx,color,amsmath,booktabs,microtype,afterpage,subfigure,bm}
\usepackage{mathpazo} 
\usepackage[scaled]{helvet} 
\usepackage[colorlinks,plainpages=false,linkcolor=black,urlcolor=black,citecolor=black,pdfpagemode=UseNone,pdfstartview=FitBH]{hyperref}
\usepackage[dvipsnames]{xcolor}

\let\Gamma\varGamma
\let\Delta\varDelta
\let\Theta\varTheta
\let\Lambda\varLambda
\let\Xi\varXi
\let\Pi\varPi
\let\Sigma\varSigma
\let\Upsilon\varUpsilon
\let\Phi\varPhi
\let\Psi\varPsi

\copyline{Rare-Earth Borides}{Dmytro S. Inosov (ed.)}{2020}
\title{Rare-Earth Borides} 

\makeindex

\graphicspath{{Figures/}} 

\begin{document}






\tableofcontents



\setcounter{page}{1}
\setcounter{chapter}{0}

\newcommand{\be}{\begin{equation}}
\newcommand{\ee}{\end{equation}}
\newcommand{\bea}{\begin{eqnarray}}
\newcommand{\eea}{\end{eqnarray}}
\newcommand{\non}{\nonumber}\def\ve{\varepsilon} \def\ep{\epsilon}

\def\lam{\lambda}
\def\boldsigma{{\bm \sigma}}
\def\boldtau{{\bm \tau}}
\def\boldrho{{\bm \rho}}
\def\boldeta{{\bm \eta}}
\def\boldepsilon{{\bm \epsilon}}
\def\boldLambda{{\bm \Lambda}}
\def\boldchi{{\bm \chi}}
\def\boldkappa{{\bm \kappa}}
\def\om{i\omega_n}

\def\br{{\bf r}} \def\bR{{\bf R}}
\def\bk{{\bf k}} \def\bK{{\bf K}} \def\bq{{\bf q}}
\def\bp{{\bf p}} \def\bQ{{\bf Q}}
\def\bI{{\bf I}}
\def\bJ{{\bf J}}
\def\bh{{\bf h}}
\def\bG{{\bf G}}
\def\bH{{\bf H}}
\def\bM{{\bf M}}
\def\bD{{\bf D}}
\def\bO{{\bf O}}
\def\bT{{\bf T}}
\def\bun{{\bf 1}}
\def\bx{{\bf x}}
\def\bX{{\bf X}}
\def\bz{{\bf z}}

\def\tg{\tilde{g}}

\def\ua{\uparrow} \def\da{\downarrow}
\def\la{\leftarrow} \def\ra{\rightarrow}

\def\dg{\dagger}
\def\bra{\langle} \def\ket{\rangle}

\def\fs{\frac{1}{2}}
\def\fq{\frac{1}{4}}

\def\ftb{\frac{30}{\sqrt{5}}}
\def\ftbj{\frac{\sqrt{15}}{6}}

\def\LB{LaB$_6$}
\def\CB{CeB$_6$}
\def\CBL{Ce$_{1-x}$La$_x$B$_6$}
\def\YB{YbB$_{12}$}
\def\YBB{YbB$_{6}$}
\def\SB{SmB$_6$}

\def\UR{URu$_2$Si$_2$}
\def\CC{CeCoIn$_5$}

\newcommand{\bl}{\begin{aligned}}
\newcommand{\el}{\end{aligned}}

\chapter{Multipolar order and excitations in rare-earth boride Kondo systems \label{Chapter:Thalmeier}}

\chapauth{Peter Thalmeier$^{\text{a},\ast}$, Alireza Akbari$^{\text{b},\text{c},\text{d}}$ and Ryousuke Shiina$^{\text{e}}$
\chapaff{\noindent
$^\text{a}$Max Planck Institute for Chemical Physics of Solids, D-01187 Dresden, Germany\\
$^\text{b}$Asia Pacific Center for Theoretical Physics, Pohang, Gyeongbuk 790-784, Korea\\
$^\text{c}$Department of Physics, POSTECH, Pohang, Gyeongbuk 790-784, Korea\\
$^\text{d}$Max Planck POSTECH Center for Complex Phase Materials, POSTECH, Pohang 790-784, Korea\\
$^\text{e}$Dept. of Physics and Earth Sciences, University  of Ryukyus, Nishihara, Okinawa 903-0213, Japan\\
$^\ast$E-mail address: \href{mailto:Peter.Thalmeier@cpfs.mpg.de}{Peter.Thalmeier@cpfs.mpg.de}}}

\section*{Abstract}

The cubic rare-earth boride series displays diverse electronic states like localized 4$f$ electron multiplets split by the crystal electric field (CEF), itinerant heavy-fermion quasiparticle bands of the Kondo lattice as well as gapped Kondo insulator or mixed-valent semiconductor states. Furthermore, at low temperatures fairly exotic ordered states may appear due to the ''hidden'' order of multipoles carried by degenerate CEF multiplets, in addition to common (dipolar) magnetic order present in many $R$B$_6$ ($R$ = rare earth) systems. Most prominent are \CB\ and its La-diluted alloys which exhibit quadrupolar and octupolar ordering enabled by the cubic $\Gamma_8$ quartet state. The associated collective excitations are multipolar waves with a dispersion characteristic for the underlying order and accessible by inelastic neutron scattering.

This localized multipolar-moment picture of $R$B$_6$ has to be complemented by the itinerant Kondo-lattice approach. Due to the presence of hybridization and collective ordering gaps, a singular magnetic response can lead to the appearance of collective spin exciton modes inside the gap around symmetry points of the Brillouin zone (BZ). This has been observed in heavy-fermion metal \CB\ and in particular in the Kondo insulators \YB\ and \SB. The latter, which has no Landau-type local symmetry breaking is also the prime candidate for a strongly correlated insulator with topological order, caused by odd number of band crossings of 4$f$ and 5$d$ bands in the BZ. The signature of topological order is the existence of massless Dirac surface states with helical spin polarization, a topic of intense investigation in \SB.

\section{Introduction}
\label{sect:intro}

\begin{figure}[b]
\begin{center}
\includegraphics[width=0.6\textwidth]{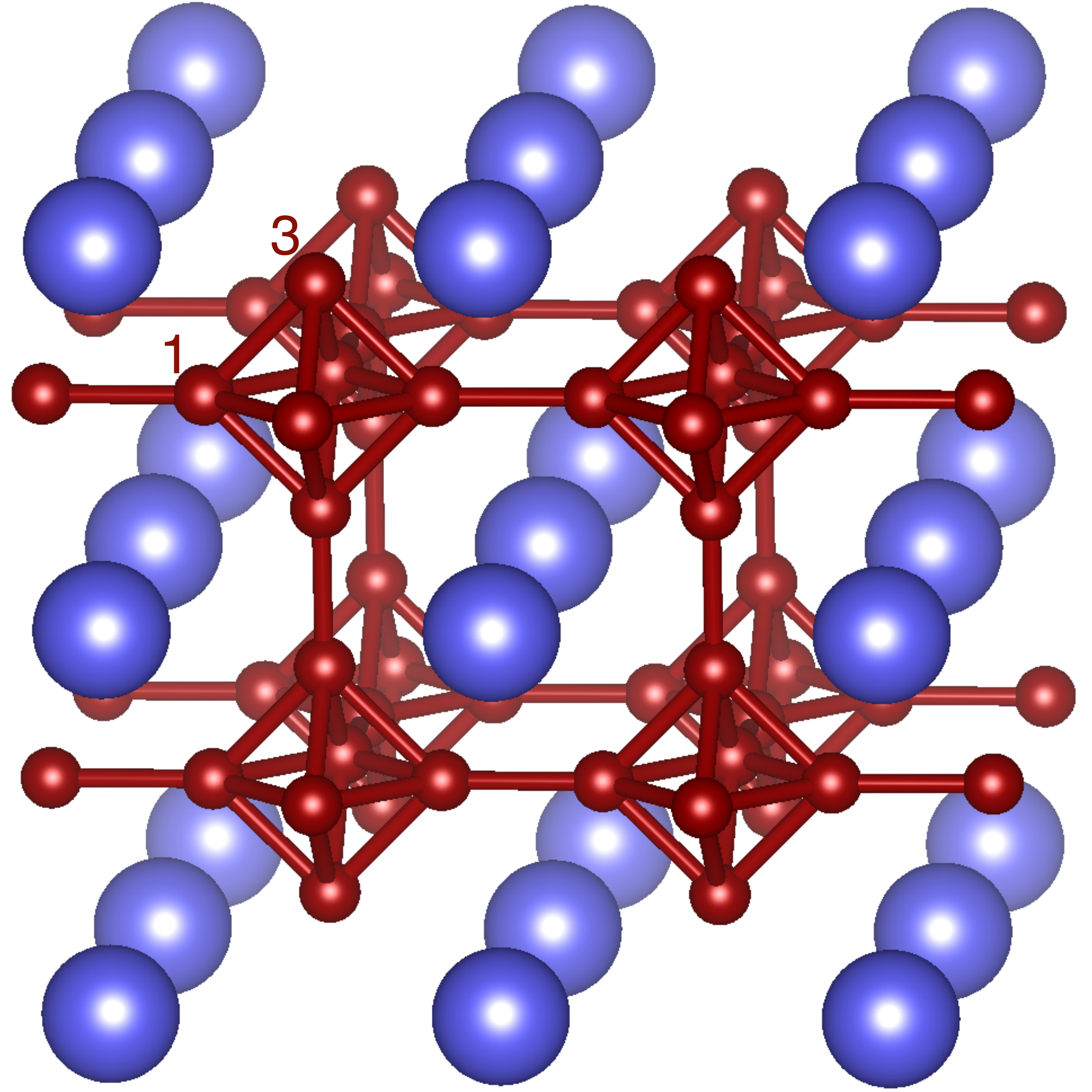}\hfill
\end{center}
\caption{Cubic CaB$_6$-type structure of $R$B$_6$ borides. $R^{3+}$ ions (blue) form a simple cubic sublattice. The B$_6$ octahedra (red) play the role of the anions at body-centered positions forming a cage for RE. The lattice constant for $R=\text{Ce}$ is $a\approx4.14$~\AA. Boron sites 1 and 3 are inequivalent in NMR (Sec.~\ref{sect:HOexp}).}
\label{fig:CeB6struc}
\end{figure}

The $R$B$_6$ ($R$ = rare earth) compounds are a versatile model series for strongly correlated 4$f$ electron materials. These compounds have the cubic CaB$_6$ structure (space group $Pm\bar{3}m$) where the B$_6$ octahedra play the role of anions (Fig.~\ref{fig:CeB6struc}). The rare-earth (RE) ions are mostly in the $3+$ configuration, but $2+$ and mixed valence \index{mixed valence} also occur (see Table~\ref{tbl:REB6}). The continuous interest in the series in the last fifty years stems largely from the fact that they show a great variety of low temperature ordered phases. They are triggered by the lifting of the degeneracies by the 4$f$-electron CEF \index{crystal electric field} ground state multiplets and the interplay with RE 4$f$-5$d$ conduction electron {\it inter-site} hybridization. In most cases the ordered phases show magnetic order of (non-)collinear type depending on temperature and external field. The degenerate CEF ground states, however, not only support ordering of dipolar (rank~1) magnetic moments but also of higher quadrupole (rank 2), octupole (rank 3) and generally rank $p\leq 2l$ $(l=3)$ multipoles of 4$f$ electrons.  Here even and odd rank $p$ correspond to the preservation and breaking of time reversal symmetry by the order parameter \cite{kusunose:08,santini:09}. The hybridization with conduction states leads (via a generalized Schrieffer-Wolff mechanism) to effective inter-site interaction between those multipoles which drive their ordering. Multipole order with rank $p\geq 2$ is generally termed `'hidden order'' (HO) because it cannot easily be detected by the conventional (dipolar) neutron and x-ray scattering which yield no diffraction peaks except when it induces a considerable secondary lattice distortion. More involved methods like neutron diffraction (ND) in external field \cite{erkelens:87}, high momentum transfer neutron scattering \cite{kuwahara:07}, resonant x-ray scattering \cite{nagao:01,matsumura:09,nagao:06} or ultrasonic investigations \cite{nakamura:96,yanagisawa:18} and NMR method \cite{takigawa:83} have to be applied. This has given new impetus to investigate the multipolar ordering in $f$-electron compounds. The most well-known examples of higher rank HO are cubic hexaborides \CB\ and \CBL\ (rank 2 quadrupole and rank 3 octupole) \cite{shiina:97}\index{CeB$_6$!La doped, Ce$_{1-x}$La$_x$B$_6$!octupolar order}, NpO$_2$ (rank 3 octupole) \index{NpO$_2$}\cite{shiina:07} and tetragonal \UR\ (proposed rank 5 dotriakontapole) \index{URu$_2$Si$_2$} \cite{ikeda:12,shibauchi:14,thalmeier:14,akbari:15}. Further examples are found in the cubic 4$f$ skutterudites \cite{kuramoto:09,takimoto:06,shiina:13} and 1-2-20 cage compounds \cite{onimaru:16}.

\index{rare-earth hexaborides!crystal structure}

In fact, the hexaborides may also be considered as cage compounds where rare-earth ions on the simple cubic sublattice are surrounded by a cage of eight $B_6$ octahedra (Fig.~\ref{fig:CeB6struc}). In both series this leads to the interesting possibility of 'rattling'  or strongly anharmonic motion of RE ions in the cages. It is also known from the RE clathrate cage compounds \cite{zerec:04} where it strongly influences transport properties, in particular thermoelectric power. The rattling motion in the boride series leads to flat phonon branches that can be interpreted as low energy ($\hbar\omega_\text{E} \simeq 10$~meV) Einstein modes of RE ions in oversized $B_6$ cages. It is most prominent for some heavier RE (Gd, Dy, Tb) hexaborides due to the lanthanide contraction of RE ionic radii \cite{iwasa:11,iwasa:14,serebrennikov:16}. The anharmonic low energy rattling phonons are in contrast to the extremely stiff motion of the boron cage as witnessed by the very large longitudinal elastic constants \cite{luethi:84}.\index{elastic constants}

\subsection{Conduction bands and Fermi surface}\label{sect:Fermisurf}

\begin{figure}[t]
\begin{center}
\includegraphics[width=1.02\textwidth]{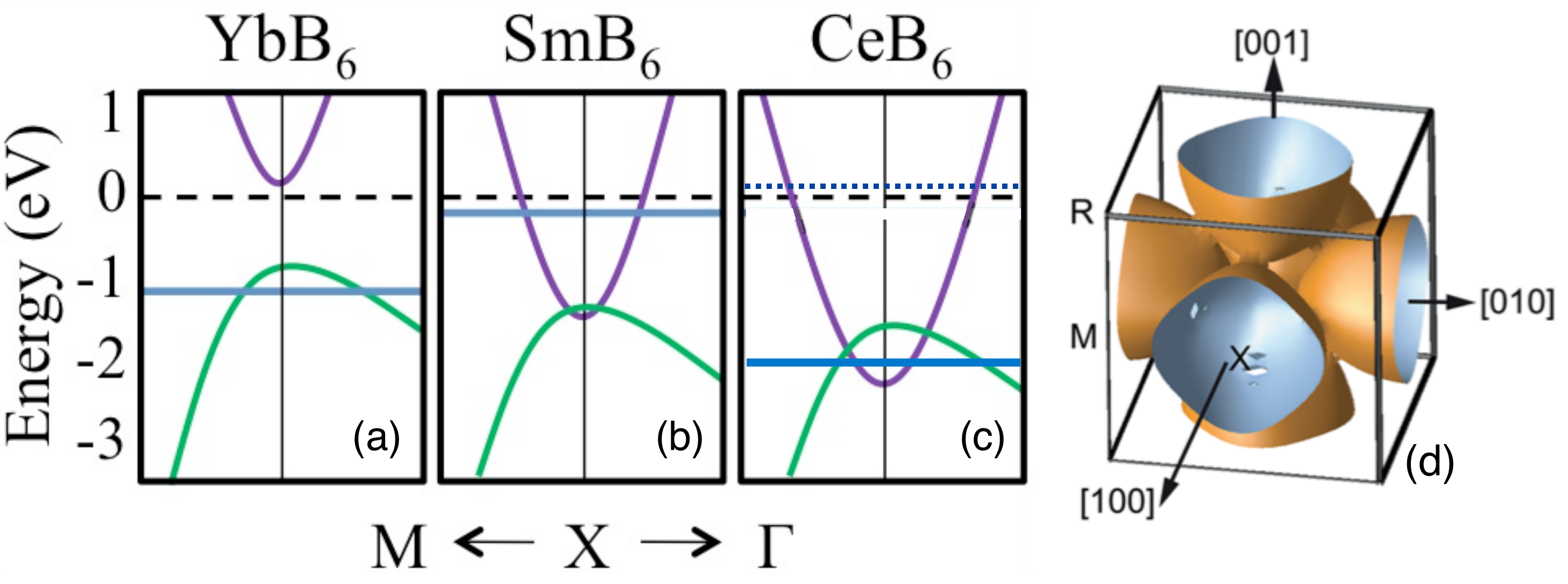}\vspace{-2pt}
\end{center}
\caption{(a-c)~Schematic bulk band structure (without hybridization, Fermi level $\epsilon_{\rm F}=0$) consisting of RE 5$d$ (purple) and 4$f$ (blue) as well as B 2$p$ (green) type bands for  exemplary compounds with integer as well as mixed valences; Yb$^{2+}$, Sm$^{2.5+}$ and Ce$^{3+}$, respectively. Here \YBB\ is a $p$-$d$ band semiconductor due to the filled 4$f$ shell ($4f_{7/2}$ binding energy $|\epsilon_f|\simeq 1~\text{eV}$. Due to (inter-site) $f$-$d$ hybridization and on-site $f$-$f$ correlation \SB\ becomes a mixed valence semiconductor and \CB\ ($|\epsilon_f|\simeq 2.1~\text{eV}$) a Kondo heavy-fermion metal (adapted from Ramankutty \textit{et al.}~\cite{ramankutty:16}). (d)~Main large Fermi surface sheet ($\alpha_3$-orbit on cubic faces) of hexaborides. The dimensions of connected $X$-point ellipsoids is almost identical for \LB, \CB, and \SB. Adapted from Tan \textit{et al.}~\cite{tan:15}.}
\label{fig:REB6_bands}
\end{figure}

In this review we focus exclusively on the correlated electronic properties of the series, in particular hidden 4$f$ multipole order and its excitations. We also discuss the consequences of Kondo effect and associated 4$f$-5$d$ hybridization, i.e. Kondo insulator state \index{Kondo insulator}, spin resonance formation, band crossing and topological properties like protected helical surface states. Therefore it is useful to get first a schematic picture of the electronic degrees of freedom, itinerant 5$d$ and localized 4$f$ of the rare earth as well as 3$p$ valence state of the B$_6$ cages. A sketch of the position and dispersion of these electronic states (excluding hybridization effects) is given in Fig.~\ref{fig:REB6_bands} for typical cases of the RE valences.
\footnote[1]{In all figures, panels a,\,b,\,... are labeled from left to right and top to bottom.}

\index{CeB$_6$!Fermi surface}\index{LaB$_6$!Fermi surface}\index{SmB$_6$!Fermi surface}\index{rare-earth hexaborides!Fermi surface}\index{Fermi surface}

The 5$d$-like bands show large dispersions and, except for the two $2+$ valence cases of semimetal EuB$_6$ \index{EuB$_6$} and semiconductor YbB$_6$, \index{YbB$_6$} lead to the large 5$d$-type electron pockets in the Fermi surface (FS) (see Fig.~\ref{fig:REB6_bands}) around the $X$ point $(0\fs 0)$ (in r.l.u., 1\,r.l.u.~=~$\frac{2\pi}{a}$) and equivalent ones. The FS ellipsoids are touching and form small necks between them. In the cases \LB, \CB, PrB$_6$ \index{PrB$_6$} and NdB$_6$ \index{NdB$_6$} where the dimensions of the $X$-point pockets have been determined by dHvA experiments (Table~\ref{tbl:masses}) \cite{onuki:89,kubo:93} \index{de~Haas\,--\,van~Alphen} the orbital cross sections are very close for all compounds and their field-angular dependence identifies an almost spherical shape. Although the large electron FS (corresponding to an $\alpha_3$-orbit of dHvA results shown in Fig.~\ref{fig:REB6_bands}) are similar in \LB\ and \CB, their effective masses are vastly different \cite{joss:87,onuki:89} while that of PrB$_6$ is in between (Table~\ref{tbl:masses}). In \LB~there are no $f$-electrons  and the 5$d$ band mass is observed. In \CB\ the valence is close to $3+$ with a $4f^1$ $\Gamma_8$ CEF ground state \index{crystal electric field}\index{CeB$_6$!crystal electric field}\index{$\Gamma_8$ quartet|(} but 4$f$-5$d$ hybridization and 4$f$-4$f$ Coulomb repulsion lead to very narrow 4$f$-quasiparticle bands due the Kondo lattice formation (Sec.~\ref{sect:PAM}). They may be interpreted as lattice-periodic coherent bands formed by the sharp single-site Kondo resonance states (dotted blue line in Fig~\ref{fig:REB6_bands}). In PrB$_6$ the integer $3+$ valent  4$f^2$ CEF $\Gamma_5$ ground state has a much smaller hybridization that leads only to a small perturbative renormalization to an effective 5$d$ mass. While for NdB$_6$ there is no mass enhancement as compared to \LB~due to negligible hybridization. The boron 2$p$ states do not cross the Fermi level in the series. Nevertheless they have an important indirect influence in \SB\ where the $f$-band obtains an {\it upward} dispersion in $\varGamma X$ direction (thin blue line in Fig.~\ref{fig:STI_disp}) due to the (on-site) hybridization with lower but close 2$p$ bands at $X$. This effect is essential for being able to form the topological insulator state as discussed in Sec.~\ref{sect:SmB6TI}.

\begin{table}[t]
\tbl{Area and the effective cyclotron mass $(\bH\parallel [001])$ of $\alpha_3$ orbits (intersection of $X$-point pockets in Fig.~\ref{fig:REB6_bands} with cubic $[001]$ faces). An average $\alpha_3$ area of 8000~T corresponds to about $1/3$ of the cubic BZ face area $(2\pi/a)^2$ in Fig.~\ref{fig:REB6_bands}.\index{heavy-fermion metal!effective mass}\vspace{3pt}}
{\begin{tabular}{@{}cccccc@{}}
\toprule
$\alpha_3$ orbit &\;\;\;\;\;\; \LB \;\;\;\;\;\;\;\; & \CB \;\;\;\;\;\;\;\; &  PrB$_6$ \;\;\;\;\;\;\;\; & NdB$_6$ \;\;\;\;\;\;\;\;  & \SB \;\;\;\;\;\;    \\
\midrule
$F_{\alpha_3}$ [T] &  7890    &  8670   & 8190 & 7980 &  7800  \\
$m^*/m_e$ & 0.64 &   14-21  & 1.95    & 0.60 &  *  \\
Ref. &  \cite{onuki:89}  & \cite{onuki:89}  &  \cite {onuki:89} & \cite{kubo:93} & \cite{tan:15}  \\
\botrule
\end{tabular}
}
\begin{tabnote}
$^*$ reliable mass assignment was not possible
\end{tabnote}
\protect\label{tbl:masses}
\end{table}

\subsection{Localized 4$f$ shells, their CEF states, multipoles and RKKY interactions}
\label{sect:CEF}\index{crystal electric field|(}

\begin{figure}[t]
\begin{center}
\includegraphics[width=\textwidth]{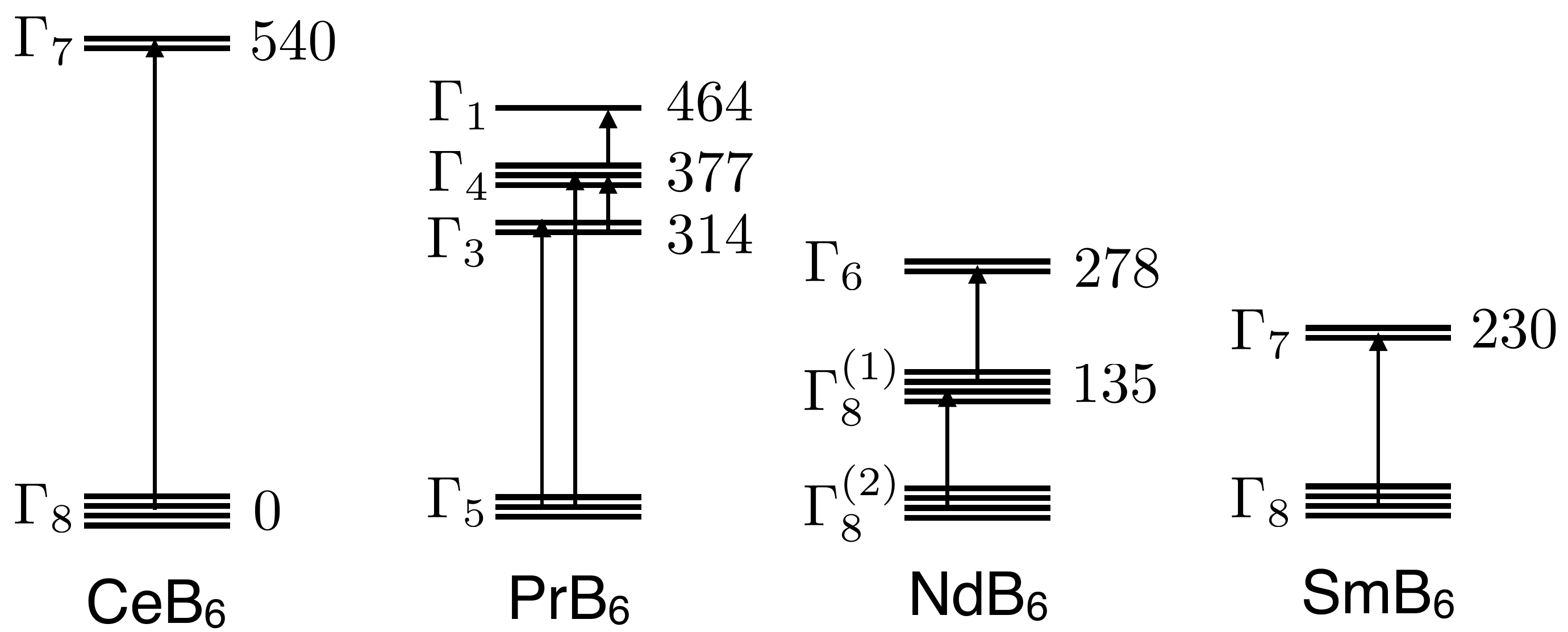}\hfill
\end{center}
\caption{Spectroscopically determined CEF splittings in [K] of the RE-hexaborides. Results from INS (Ce-Nd) \cite{loewenhaupt:86} and from RIXS (Sm) \cite{amorese:19}.}
\label{fig:REB6_CEF}
\end{figure}

In the cubic structure of $R$B$_6$ the RE ion point group is $O_h$. This leads to a cubic CEF potential for the shell of localized spherically symmetric $4f^n$ states. For finite temperature and low energies one may restrict to the ground state characterized by shell angular, spin and total angular momentum $(LSJ)$, the latter being determined by the large spin-orbit coupling [$\zeta_{\rm s.o.}=0.045~\text{eV (Ce)} - 0.364~\text{eV (Yb)}$] in the $R^{3+}$ ions. In the common Stevens representation \index{Stevens representation} the cubic CEF potential is written as an operator in terms of symmetrized polynomials of $(J_x,J_y,J_z)$ in the $(2J+1)$-dimensional Hilbert space of the total angular momentum $J$ ground state multiplet. This is the well known expression
\bea
{\cal H}_{CEF} = B_4^0\bigl[O_4^0+5O_4^4\bigr] + B_6^0\bigl[O_6^0-21O_6^4\bigr]
\eea
where $O_4^m$ and $O_6^m$ represent, respectively, fourth and sixth order symmetrized polynomials of $J_z$ $(m=0)$ and $J_\pm,J_z$ $(m=4)$  \cite{hutchings:64} corresponding to the real space tesseral harmonics. Furthermore $B_4^0, B_6^0$ are CEF parameters that may be formally given within a point-charge model \cite{hutchings:64}. The latter determine the splitting into generally degenerate CEF multiplets that belong to the $O_h$ representations
$\Gamma_\alpha(d_\alpha)$ where $\alpha=1,2,3,4,5$ for non-Kramers ions (integer $J$) with corresponding degeneracy  $d_\alpha=1,1,2,3,3$ and $\alpha=6,7,8$ for Kramers ions (non-integer $J$) having corresponding degeneracy  $d_\alpha=2,2,4$. The degenerate states of each CEF multiplet are designated by $|\Gamma_\alpha^i\ket$ with $1\leq i \leq d_\alpha$. They are tabulated in \cite{lea:62} for all J as function of CEF parameters $B_4$,$B_6$ or alternatively related parameters $W,x$. In practice the splittings and wave functions, i.e. the CEF parameters have to be determined experimentally, mainly by two methods: i) analysis of high-temperature (single-ion) susceptibility and b) fitting to peak positions and intensities of inelastic neutron scattering (INS) spectra that determines directly splitting and dipolar magnetic matrix elements between the CEF states. In CEF schemes with
high $J$ (or inequivalent RE sites) this is, however not a unique procedure to determine the CEF parameters. Recently x-ray techniques like NIXS and RIXS have also contributed to unravel the CEF states and energies \cite{sundermann:18,hamamoto:17,amorese:19}. The CEF level schemes, and in particular, CEF ground states for the $R$B$_6$ series known so far are summarized in Fig.~\ref{fig:REB6_CEF}, they concern mostly the light RE, there is surprisingly little information of the heavier $R$B$_6$ in the literature.\index{rare-earth hexaborides!CEF level schemes}

\begin{table}[b]
\protect\label{tbl:REB6}\index{rare-earth hexaborides!magnetic properties}\index{rare-earth hexaborides!CEF level schemes}\index{rare-earth hexaborides!ordering vectors}
\tbl{Compilation of essential data for $R$B$_6$ compounds with hidden (e.g. quadrupolar) and/or magnetic (dipolar) order characteristics. \CB\ is a Kondo-lattice heavy-fermion metal while semiconducting \SB\ is the only compound with strongly mixed valence.\vspace{3pt}}
{\begin{tabular}{@{}lccccccc@{}} \toprule
compound & J & valence & O$_h$ CEF g.s. & T$_\text{HO}$ &  \bq$_\text{HO}$  &  T$_\text{N}$  & \bq$_\text{m}$  \\
$R$B$_6$$^\text{a}$  &  & & (degeneracy) & [K] & r.l.u. & [K] & [r.l.u.]  \\ \midrule
CeB$_6$ (HFM)  & $\frac{5}{2}$ & 3+& $\Gamma_8(4)$ & 3.3  & $ (\fs\fs\fs)$  & 2.3    & $ (\fq\fq\fs)$  \\
\hline
PrB$_6$ (m)  & $4$ &3+& $\Gamma_5(3)$ &    & $ (\fs\fs0)$  & 7 (IC)  & $ (\fq\!-\!\delta\fq\fs)$ \\
  &  & & & 4.2  & $ (\fs\fs 0)$  & 4.2 (C)   &   $(\fq\fq\fs)$ \\
NdB$_6$ (m)  & $\frac{9}{2}$ &3+& $\Gamma^{(2)}_8(4)$ & -  & - & 8  &  $(00\fs)$ \\
EuB$_6$ (sm)  & $\frac{7}{2}$ &2+& $(L=0)$& -  & - & 12.5  &  $(000)$ \\
GdB$_6$ (m) &  $\frac{7}{2}$  &3+& $(L=0)$& -  & - & 16 &  $(\fq\fq\fs)$ \\
TbB$_6$ (m)  & $6$ &3+&  $\Gamma_2(1)$ or $\Gamma_3(1)$ & - & - &20 &  $(\fq\fq\fs)$ \\
DyB$_6$ (m)  & $ \frac{15}{2}$ &3+& $\Gamma_8^{(1)}(4)$ & 31 & (000) & 26 &  $(\fq\fq\fs)$ \\
HoB$_6$ (m) & $ 8$ &3+&$\Gamma_5(3)$ & 6.1 & (000) & - &  $-$ \\
YbB$_6$ (sc) & $0$&2+ &-&-&-&-&-\\
\hline
SmB$_6$ (MV,TI) & $0\sim\frac{5}{2}$&2.55+ &$\Gamma_8(4)$ &-&-&-&-\\
\botrule
\end{tabular}
}
\begin{tabnote}
$^\text{a}$ HFM = heavy-fermion metal; m = metal; sm = semimetal; sc = semiconductor; MV = mixed valent; TI = topological insulator\\
\end{tabnote}
\protect\label{tbl:REB6}
\end{table}

A CEF level scheme with \mbox{$(2J+1)$} states can carry $1\leq  n \leq (2J+1)^2-1$ multipole operators $X_n$ (the identity has been subtracted) which is simply equal to the number of standard basis operators (minus the identity) defined by $L_{\alpha\beta}^{ij}=|\Gamma_\alpha^i\ket\bra\Gamma_\beta^j|$. The multipole operators are linear combinations of the standard basis operators that belong to specific cubic representations. They may be expressed as rank $p$ polynomials $P_p(J_x,J_y,J_z)$ and their explicit form (for $p\leq 4$) is tabulated in \cite{nagao:06}. Their treatment is discussed in more detail in Sec.~\ref{sect:CeB6_loc} for a special case. These multipoles are the physical 4$f$-shell degree of freedoms at every site. When the hybridization with 5$d$ conduction electrons is taken into account the localized 4$f$ multipoles may be effectively coupled at adjacent sites by a generalized RKKY mechanism \index{RKKY interactions} well known for the rank $p=1$ multipoles (magnetic dipoles).  For a degenerate CEF ground state then at low temperatures the multipole with the maximum effective inter-site coupling at a particular wave vector \bq\ will be the primary order parameter. Most frequently this is either a multipole of rank 1 ( magnetic order) or rank 2 (quadrupolar HO) but more general HO, in particular rank 3 (octupolar order), can occur (see Sec.~\ref{sect:CeLaB6}). General expressions for the effective RKKY-type multipole interactions may be derived \cite{teitelbaum:76,schmitt:84,schlottmann:00,kuramoto:02} and quantitative first principle results for \CB\ were presented recently \cite{yamada:19}. They demonstrate that quadrupolar and octupolar nearest-neighbor (n.n.) interactions are maximally enhanced supporting the parameterized form \cite{shiba:99} used in the following sections. When the \bk,\bq-dependent multipole matrix elements between conduction band states are replaced by a constant their generalized RKKY interaction is proportional (for every multipole) to the Lindhard function \index{Lindhard function} $\chi_L(\bq)=\sum_\bq(f_{\bk +\bq}-f_\bk)/(\epsilon_\bk-\epsilon_{\bk +\bq})$ where $\epsilon_\bk$ and $f_\bk$ are conduction band and Fermi function. For the FS topology \index{Fermi surface} of Fig.~\ref{fig:REB6_bands} it has (sub-)maxima at the wave vectors $\bQ_\parallel=(\fq\fq 0)$ and $\bQ'=(\fs\fs\fs)$, respectively due to various nesting \index{nesting}\index{Fermi surface!nesting properties} properties of the FS \cite{koitzsch:16}. Indeed $\bQ_\parallel$ is the in-plane component of most common AFM structure (Fig.~\ref{fig:PrB6_morder}) in $R$B$_6$ and $\bQ'$ is the HO wave vector in~\CB.
\index{crystal electric field|)}

\section{Overview of RE-boride compounds}
\label{sect:overview}

The RE borides, in particular hexaborides, have the advantageous property of existing for almost the whole series of rare-earth atoms within the same crystal structure, thus allowing for the study of systematic variations in physical properties. These change greatly due to the varying degree of localization or itineracy of 4$f$ electrons, all the while the basic CEF multiplet states are the same due to the universal cubic $O_h$ point symmetry of this class. One can roughly distinguish two cases:

First, the stable moment compounds where the hybridization of well localized 4$f$ electrons and conduction electrons can be treated perturbatively, leading to an on-site interaction with 4$f$ multipoles in second order and to their inter-site effective RKKY coupling in fourth order of the hybridization. These compounds are listed in the central part of Table \ref{tbl:REB6} with their salient ordering characteristics.

In the second case the hybridization is strong (as in \CB, \YB\ and \SB\ compounds) and may destabilize the moment by screening and formation of a local singlet state that forms coherent heavy electron bands at low temperature, or, in the large hybridization case, may lead to a mixed valent state with pronounced non-integer 4$f$-electron occupation. In both instances a hybridization (pseudo-) gap in the renormalized electron bands appears. Most frequently a metallic ground state with moderately (MV) \index{mixed valence} or strongly (HF) enhanced effective $m^*$ electron mass is realized as, e.g. in \CB. The heavy electron liquid may become instable at the lowest temperatures leading to HF superconductivity as frequently in $122$ and $115$ Ce-compounds \cite{thalmeier:91} or to multipolar order as in \CB. If the number of conduction electrons is suitable the Fermi level may fall into the hybridization gap \index{hybridization gap}\index{heavy fermion metal!hybridization gap} producing a rare ``MV semiconductor'' or ``Kondo insulator'' (Fig.~\ref{fig:Quasiband}).\index{Kondo insulator} In the borides two famous hybridization gap insulators are known: \YB\ and \SB. In particular the latter has raised enormous attention due to its  nontrivial topological state. On the other hand \YBB\ is a conventional $p$-$d$ semiconductor with nonmagnetic Yb$^{2+}$ state.  The strong hybridization compounds are the most investigated ones and will be the main focus of this review.

Here, as an overview we first briefly discuss some salient features of the stable moment compounds in Table~\ref{tbl:REB6}. Among the light rare earth, PrB$_6$ and NdB$_6$ are the most studied. Also their CEF level schemes shown in Fig.~\ref{fig:REB6_CEF} are well known \cite{loewenhaupt:86, pofahl:87}.

\begin{figure}[t]
\begin{center}
\includegraphics[width=0.85\textwidth]{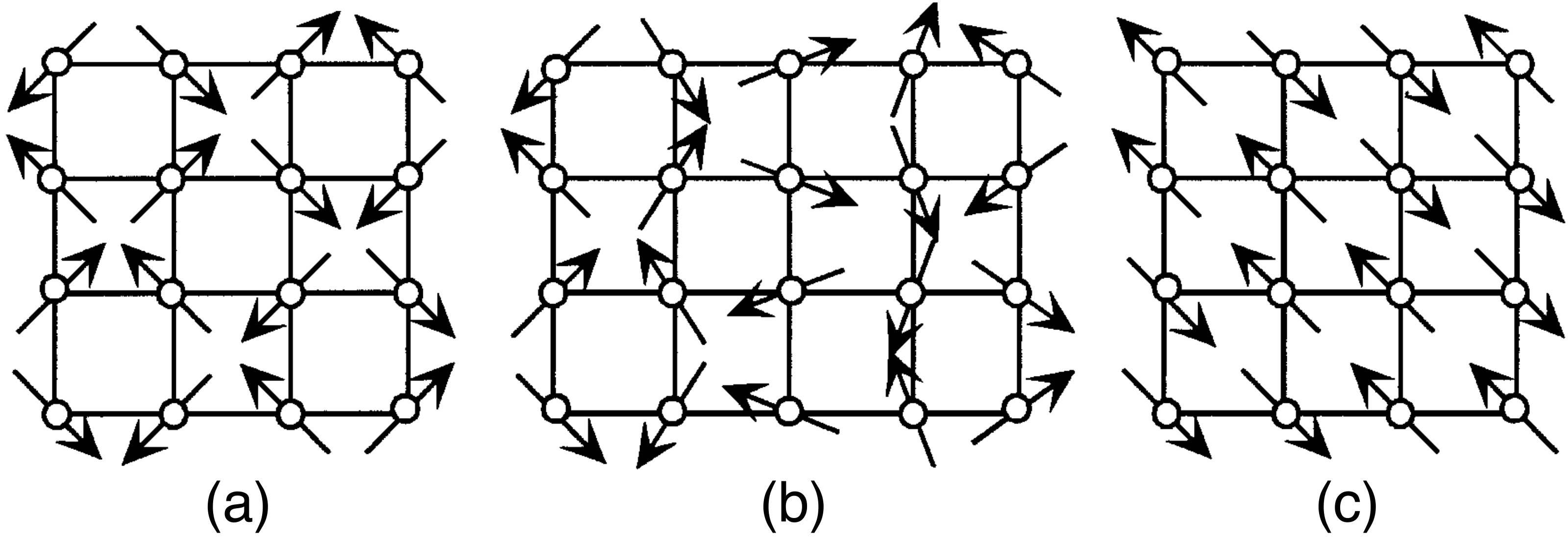}\\[0.25cm]
\includegraphics[width=0.50\textwidth]{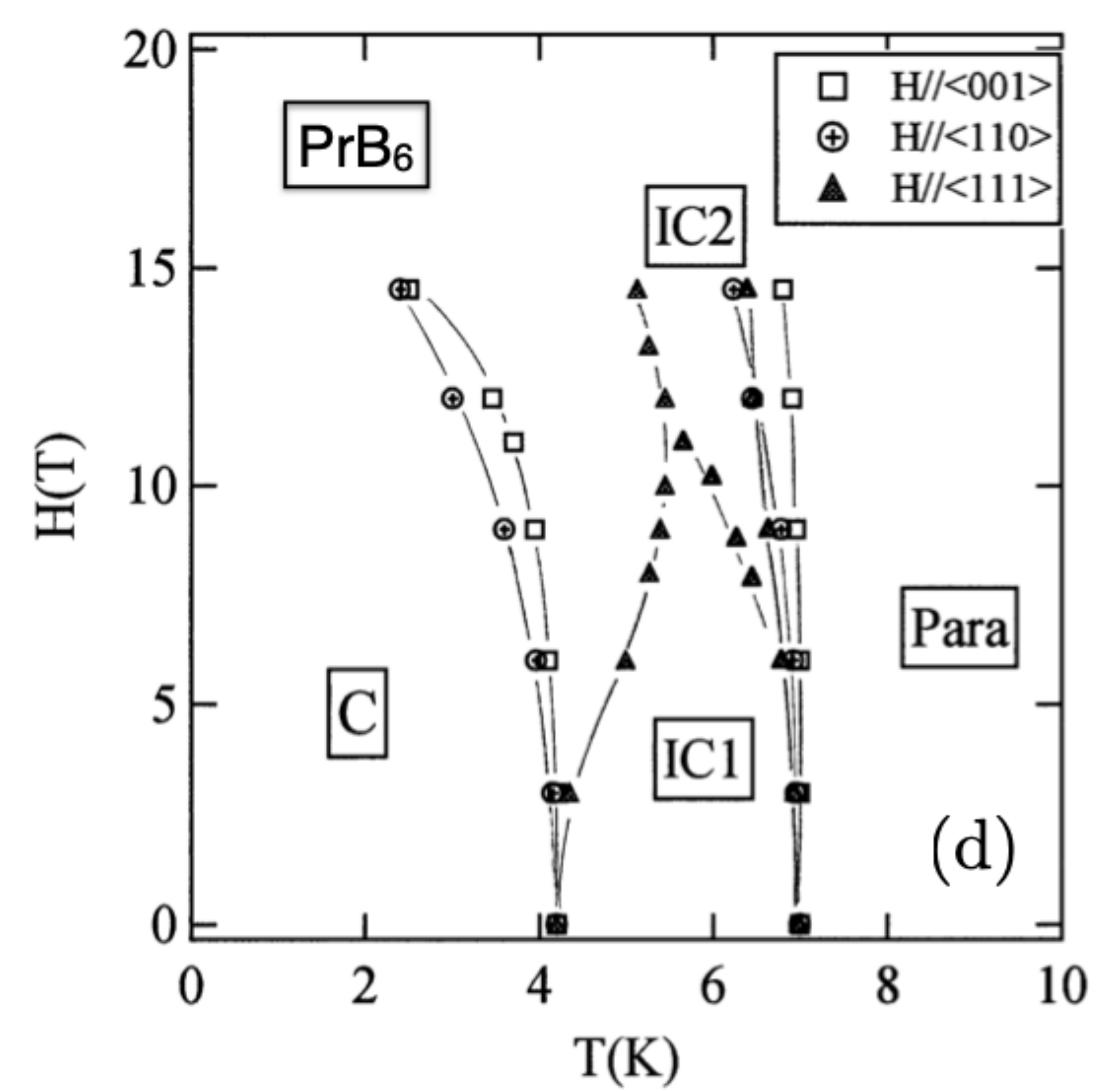}
\end{center}
\caption{Magnetic structures of PrB$_6$ and the $H$-$T$ phase diagram. (a)~Double-$\mathbf{q}$ structure with $\bQ_1=(\fq\fq\fs)$, $\bQ_2=(\fq\bar{\fq}\fs)$ (b) IC structure with $\bQ_\text{IC}= (\fq\!-\!\delta~\fq~\fs)$ $(\delta=0.05)$ (c) single-$\mathbf{q}$ structure with $\bQ=(\fq\fq\fs)$ in external field $\bH\parallel [110]$. (d) $H$-$T$ phase diagram for \bH\ along symmetry directions. Separate low and high field IC phases IC$_1$ and IC$_2$ are observed. The C-phase has coexisting easy-plane magnetic and $O_{xy}$ quadrupolar order. Reproduced from Kobayashi~\textit{et~al.}~\cite{kobayashi:01}.}
\label{fig:PrB6_morder}
\end{figure}

\index{PrB$_6$}
{\bf PrB}$_{\bf 6}$ with a $\Gamma_5$ triplet ground state exhibits two consecutive first order transitions to an incommensurate (IC) phase at $T^\text{IC}_\text{N} = 7$~K and a further lock-in transition to a magnetic C-phase with wave vector $\bQ=(\fq\fq\fs)$. The latter coexists with an induced AFQ order presumably with  $\bQ=(\fs\fs 0)$ below $T^\text{C}_\text{N} =4.2$~K. These wave vectors together with $\bQ'=(\fs\fs\fs)$ for HO characterize most of the ordered structures in the hexaboride compounds and they are related to the nesting structure \index{nesting}\index{Fermi surface!nesting properties} of the Fermi surface (Fig.~\ref{fig:REB6_bands}) that translates into the preferred magnetic wave vectors  via the RKKY mechanism (Sec.~\ref{sect:CEF}). \index{RKKY interactions} The ordered arrangements of moments are illustrated in Fig.~\ref{fig:PrB6_morder}\,(a--c). In zero field the magnetic structure is of the noncollinear double-$\mathbf{q}$ type (a) which switches to single-$\mathbf{q}$ type in applied field $\bH \parallel [110]$ (c). The incommensurate structure with wave vector $(\fq\!-\!\delta~\fq~\fs)$ is shown in (b). A theoretical investigation for PrB$_6$ has been presented in \cite{kuromaru:02} based on the $\Gamma_5$ CEF ground state which carries 3 $\Gamma^-_4$ dipoles and 5 $(\Gamma_3^+,\Gamma_5^+)$ quadrupoles (here $\pm$ denotes even/odd behaviour under time reversal). In distinction to \CB\  (Sec.~\ref{sect:CeB6_loc}) the non-Kramers triplet carries no octupoles. The presence of n.n. isotropic and next-nearest neighbor (n.n.n.) pseudo-dipolar exchange interactions was proposed to obtain the stability of the IC phase (although with a slightly different wave vector  $\bQ_\text{IC}= (\fq\!-\!\delta~\fq\!-\!\delta~\fs)$ with $\delta=0.16$. The transition to the primary IC phase is of second order whereas the lower lock-in transition to the AFM C-phase is accompanied by a secondary $O_{xy}$ quadrupole HO at wave vector $(\fs\fs 0)$  and therefore is of first order. The corresponding $H$\,--\,$T$ phase diagram of PrB$_6$ is shown in [Fig.~\ref{fig:PrB6_morder}\,(d)]. The transverse elastic constants $c_{44}$ show a pronounced softening due to the Curie-Weiss type $\Gamma_5^+$ quadrupolar susceptibility resulting from the orbitally degenerate $\Gamma_5^+$ CEF ground state \cite{nakamura:94}. The softening is arrested, however, at the magnetic phase transition where the CEF ground state splits due to the molecular field.

\index{NdB$_6$}
{\bf NdB}$_{\bf 6}$ has a quartet $\Gamma_8^{(2)}$ ground state similar to \CB. The observed high-field magnetization anisotropy $M^{[100]} < M^{[110]} < M^{[111]}$  with [111] easy axis supports this CEF model \cite{awaji:99}. But its ordered magnetic structure is much simpler than the non-collinear double-$\mathbf{q}$ structure of \CB\ (Sec.~\ref{sect:CeLaB6}), corresponding to a {\it collinear} single-$\mathbf{q}$ type-I AFM below $T_\text{N}=8$~K with $\bQ=(00\fs)$ which implies three possible domains. The ordered moment is, however, oriented  along the fourfold [001] axis instead of the CEF [111] easy axis. This has been attributed to a small ferro-type interaction of $O_2^0$ quadrupoles \cite{yonemura:09}. In high fields (\mbox{$\sim20$~T}) along $[111]$ an unexpected metamagnetic transition to a non-collinear triple-$\mathbf{q}$ magnetic structure with $\bQ_1=(\fs 00)$, $\bQ_2=(0\fs 0)$, $\bQ_3=(\fs 00)$ occurs which is stabilized by an interplay with induced triple-$\mathbf{q}$ AFQ order $\bQ'_1=(0\fs\fs)$,  $\bQ'_2=(\fs  0\fs)$,  $\bQ'_3=(\fs\fs 0)$ for $O_{yz}$, $O_{zx}$, $O_{xy}$ quadrupoles, respectively \cite{awaji:99}. Therefore, although the $\Gamma_8^{(2)}$ ground state of NdB$_6$ does not lead to primary AFQ order as in phase II of \CB\ (Sec.~\ref{sect:CeB6_loc}), the presence of $\Gamma_8^{(2)}$-sustained quadrupoles and their interactions shows a subtle influence on the low- and high-field magnetic order of this compound. Similar to PrB$_6$ the orbitally degenerate $\Gamma_8^{(2)}$ causes a softening of $c_{44}$ elastic constants which is again arrested by the magnetic phase transition \cite{nakamura:94}.

\index{EuB$_6$}
{\bf EuB}$_{\bf 6}$ is an outlier of the series in two aspects: Firstly it contains europium in the half-filled 4$f^7$ Eu$^{2+}$ S-state ionic configuration with $J=\frac{7}{2}$ and hence has no CEF splitting. Secondly the main interest in EuB$_6$ does not stem from the 4$f$ electrons which show simple bulk ferromagnetism below $T_\text{C}=12.5$~K due to their S-state but rather from the peculiar transport properties of conduction electrons. Since this is not the central focus here we only comment briefly on it. The compound is a ferromagnetic, partly spin-polarized semimetal with a small valence/conduction band overlap for the majority band of order $0.5~\text{eV}$ at zero temperature while the spin minority band is gapped by a similar amount. This leads to small majority spin electron pockets at the $X$ point \cite{denlinger:02}. When approaching $T_\text{C}$ from below the overlap of majority bands and concomitantly the carrier density and plasma frequency decrease \cite{kim:08}. This behaviour can be described by a two-band Kondo lattice type model with FM/AFM coupling \cite{kreissl:05}. Above $T_\text{C}$ ferromagnetic features remain due to a phase separation into paramagnetic regime and percolating magnetic polarons \cite{pohlit:18} that become isolated at the percolation temperature $T_\text{M} = 15.5$~K. At this temperature a cusp in the resistivity and giant magnetoresistance are observed \cite{wigger:04}.

\index{GdB$_6$}
{\bf GdB}$_{\bf 6}$ and {\bf TbB}$_{\bf 6}$: In the center of the $R$B$_6$ series we have again a half-filled S-state ion Gd$^{3+}$ with \mbox{$J=\frac{7}{2}$} as for Eu$^{2+}$ and therefore no CEF splitting. The Tb$^{3+}$ \mbox{$(J=6)$} CEF ground state should be a singlet (Table~\ref{tbl:REB6}) as concluded from elastic constants measurements \cite{nakamura:94}. Both compounds show a first order transition to an AFM state with $\bQ=(\fq\fq\fs)$. For GdB$_6$ \mbox{($T_\text{N} = 15$~K)} the moments are parallel to the $\fs$ component of the ordering vector, while they are perpendicular for TbB$_6$ ($T_\text{N} = 21$~K). The moments are large and correspond to the expected value of $3+$ ions (Table~\ref{tbl:REB6}). Therefore magnetoelastic effects are noticeable \cite{amara:10,iwasa:18} and lead to various lattice distortions and concomitant superlattice reflections. Recently is was observed that TbB$_6$ shows an additional ordering vector  $\bQ'_\parallel = (\fq\fq 0)$ \cite{iwasa:18}.

\begin{figure}[t]
\begin{center}
\includegraphics[width=\textwidth]{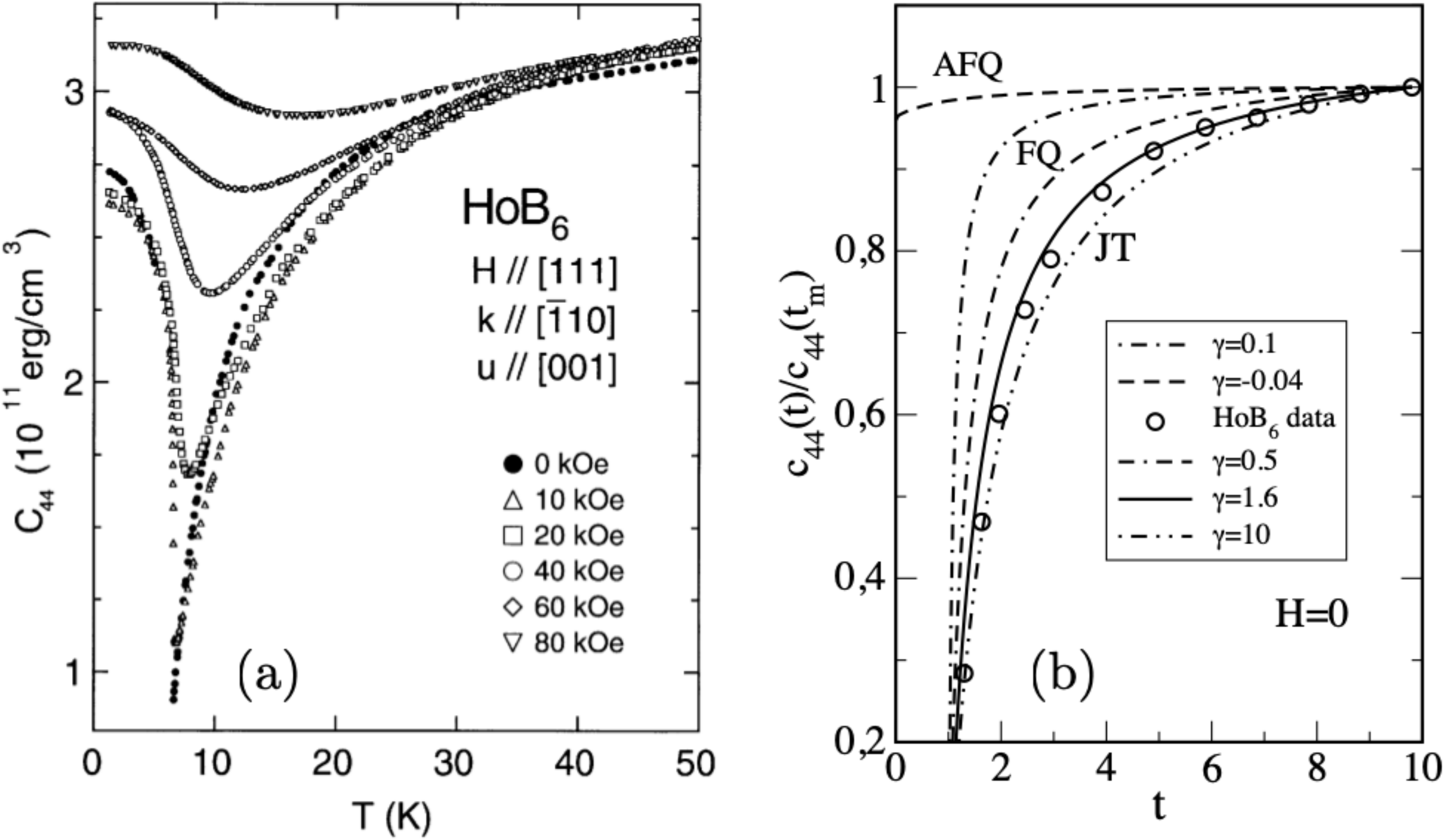}\vspace{-4pt}
\end{center}
\caption{(a)~Transverse $c_{44}$ elastic constant softening in HoB$_6$ above the JT transition at $T_\text{Q}(H)$. Reproduced from Goto~\textit{et al.}~\cite{goto:00}. (b) Comparison for $H=0$ data with results from Eq.~(\ref{eqn:elastic}). Best fit for $\gamma >1$ therefore HoB$_6$ is a cooperative Jahn-Teller compound. The $c_{44}(t)$ for FQ case $(\gamma <1)$ behaves qualitatively different, AFQ case with $\gamma < 0$ (similar to \CB) is shown for comparison. Here $t=T/T^*_\text{Q}$ with $T^*_\text{Q}=5.1$~K. $c_{44}$ is normalized at $t_m=9.8$.\index{HoB$_6$!elastic constants}\index{elastic constants}}
\label{fig:HoB6_c44}
\end{figure}

\index{DyB$_6$}\index{HoB$_6$}
\mbox{{\bf DyB}$_{\bf 6}$ and {\bf HoB}$_{\bf 6}$}: These heavy RE hexaborides are least investigated. Therefore in both cases the CEF ground states of Dy$^{3+}$ \mbox{$(J=\frac{15}{2})$} and Ho$^{3+}$ \mbox{$(J=8)$} can only be conjectured (Table \ref{tbl:REB6}) from specific heat, magnetization and elastic constant measurements \cite{goto:00,sera:19}. In the latter a huge softening of transverse $c_{44}$ elastic constants is observed for both compounds. This means the CEF ground state has orbital degeneracy leading to a Curie-Weiss-type quadrupolar $(\bq=0)$ susceptibility for $\Gamma=\Gamma^+_5$-type quadrupoles $(O_{yz},O_{zx},O_{xy})$\cite{thalmeier:91}. At the ordering temperature (Table~\ref{tbl:REB6}) these quadrupoles acquire an expectation value as signature of a ferroquadrupolar HO which splits the orbital degenerate ground state and distorts the lattice . This is the well-known cooperative Jahn-Teller (JT) effect. In this case we can denote $T_\text{HO}=T_\text{Q}$ in Table~\ref{tbl:REB6} also as $T_{\rm JT}$ because the driving mechanism is not primarily the intersite-coupling $(g'_\Gamma)$ of quadrupoles but rather the linear magnetoelastic coupling \index{magnetoelastic coupling} $(g_\Gamma)$ of $\Gamma=\Gamma^+_5$ quadrupoles to the trigonal  $\Gamma^+_5$-type homogeneous ($\bq=0$) lattice strains $\epsilon_{\Gamma_5}=(\epsilon_{yz},\epsilon_{zx},\epsilon_{xy})$. The softening of the symmetry elastic constants $c_\Gamma=c_{44}$  in HoB$_6$ as a precursor to the JT transition is shown in Fig.~\ref{fig:HoB6_c44}. Unlike in PrB$_6$ and NdB$_6$ it is not arrested by a preceding magnetic transition. It is determined by the $T$-dependence of the ($\bq=0$) quadrupolar susceptibility $\chi_\Gamma^\text{Q}$ according to \cite{luethi:73,mullen:74,thalmeier:91}
\bea
\frac{c_\Gamma(T)}{c_\Gamma^0}&=&\frac{1-(\tg^2_\Gamma+g'_\Gamma)\chi^\text{Q}_\Gamma(T)}{1-g'_\Gamma\chi^\text{Q}_\Gamma(T)}\non\\
&\simeq&\frac{T-T^*_\text{Q}}{T-\Theta_\Gamma^\text{Q}}=\frac{t-1}{t-\alpha_\Gamma}
\label{eqn:elastic}
\eea
\begin{table}[t]
\tbl{Magnetoelastic JT ($g_{\Gamma_5}$) and quadrupolar ($g'_{\Gamma_5}$) coupling constants from $c_{44}$ ($\Gamma_5$ symmetry) elastic constant measurements \cite{nakamura:94,goto:00}. Here $\gamma<0$ corresponds to AFQ coupling,  $0<\gamma<1$ to FQ case and  $\gamma >1$ to dominant JT coupling.\vspace{3pt}}
{\begin{tabular}{@{}cccccc@{}}
\toprule
compound &\;\;\;\;\;\; $g_{\Gamma_5}$[K] \;\;\;\;\;\;\;\; & $\tg^2_{\Gamma_5}=\frac{g^2_{\Gamma_5}}{c^0_{44}\Omega_c}$[K] \;\;\;\;\;\;\;\; &    $g'_{\Gamma_5}$[K] $^{(a)}$\;\;\;\;\;\;\;\; & $\gamma=\frac{\tg^2_{\Gamma_5}}{g'_{\Gamma_5}}$ \;\;\;\;\;\;\;\; & type \;\;\;  \\
\midrule
\CB &  190    &  0.078   & -2.1 & -0.037 & AFQ \\
PrB$_6$ & 200 &  0.093  & -0.16    & -0.58& AFQ \\
NdB$_6$ &  83 & 0.016  &  0.032 & 0.50 & FQ\\
DyB$_6$ &  -    &     -      &     -     &   0.70 & FQ \\
HoB$_6$ &  -    &     -      &     -     &   1.43 & JT \\
\botrule
\end{tabular}
}
\begin{tabnote}
$^{(a)}$  In the convention of experimental literature $g'_\Gamma <0$ and  $g'_\Gamma >0$ correspond to AFQ and FQ intersite coupling, respectively. This is opposite to the conventions for $zD=g'_{\Gamma_5}$ in Secs.~\ref{sect:CeB6_loc}--\ref{sect:CeB6excloc}.
\end{tabnote}
\protect\label{tbl:softening}
\end{table}
where $\tg^2_\Gamma=g^2_\Gamma/(c^0_\Gamma v_{\rm c})$ with $c^0_\Gamma$ and $v_{\rm c}$ denoting the background elastic constant and volume per RE ion, respectively. The monotonic behaviour in Fig.~\ref{fig:HoB6_c44} is dominated by the Curie contribution $\sim (m^\text{Q}_\Gamma)^2/T$  to the quadrupolar $\chi^\text{Q}_\Gamma(T)$ which can only come from a degenerate $\Gamma_5$ ground state. Then the approximation in Eq.~(\ref{eqn:elastic}) holds where $T^*_\text{Q}=(m^\text{Q}_\Gamma)^2(\tg^2_\Gamma+g'_\Gamma)$ is the quadrupolar transition temperature, neglecting the effect of excited CEF states ($m^\text{Q}_\Gamma$ is the quadrupolar ground state matrix element). It is treated as a fitting parameter for elastic constants and may differ somewhat from the real $T_\text{Q}$ where the specific heat jump occurs. Furthermore $\Theta_\Gamma^\text{Q}=\alpha_\Gamma T^*_\text{Q}$  with $\alpha_\Gamma=(1+\gamma_\Gamma)^{-1}$, $\gamma_\Gamma=\tg^2_\Gamma/g'_\Gamma$ and $t=T/T^*_\text{Q}$ denoting the reduced temperature. Here $\gamma_\Gamma$ is the ratio of magnetoelastic to intersite quadrupolar coupling which determines the qualitative temperature dependence of $c_\Gamma(T)$ in Fig.~\ref{fig:HoB6_c44}\,(b). For $\gamma <1$ when quadrupole interactions dominate the $T$-dependence is mostly flat and then a sudden softening occurs (FQ). When $\gamma >1$  and magnetoelastic JT interaction dominates the softening occurs over a large temperature range above $T^*_\text{Q}$. Fig.~\ref{fig:HoB6_c44}\,(b) demonstrates that HoB$_6$ is in the JT driven regime of softening whereas all other hexaborides (Table~\ref{tbl:softening}) are in the quadrupolar interaction dominated regime, in particular \CB. Thus elastic constant measurements can identify the driving mechanism of quadrupolar order. In the AFQ case ($\tg^2_\Gamma+g'_\Gamma<0$) no softening occurs (in the approximate Eq.~(\ref{eqn:elastic}) minus signs will be replaced by plus signs). Under applied field the softening around $T_\text{Q}$ turns into a minimum that shifts to higher temperature in Fig.~\ref{fig:HoB6_c44}\,(a). Therefore $T_\text{Q}(H)$ increases with applied field. For DyB$_6$ the zero field behaviour is similar (although still $\gamma_\Gamma < 1$), but no field-dependence of $c_{44}$ is observed, indicating field-independent  $T_\text{Q}$ up to 8~T. Furthermore an additional AFM phase transition appears at $T_\text{N} = 23~\text K <T_\text{Q}$, again with the canonical $\bQ=(\fq\fq\fs)$ ordering wave vector. From the interpretation of thermodynamic measurements \cite{sera:19} it was concluded that the order parameters are carried by a $\Gamma_8^{(1)}(0)$ quartet ground state and a closeby \mbox{$\Gamma_7$ (9~K)} doublet. A total splitting of the five multiplets of $\sim$160~K was proposed although no spectroscopic confirmation of the level scheme exists to date.

\index{ErB$_6$}\index{TmB$_6$}
{\bf ErB$_6$} and {\bf TmB$_6$} heavy rare earth-hexaborides have not been successfully synthesized \cite{iwasa:14} and may not be stable, presumably due to the small radius of their heavy $3+$ ions.

\index{YbB$_6$}
{\bf YbB$_6$} is the last in the hexaboride series and has been one of the most controversial. For a while it was thought it might be a topologically nontrivial material similar to \SB\ but recent ARPES experiments \cite{kang:16} for the non-polar [110] surface have established a different picture: The binding energy of the 4f$_{7/2}$ state is quite large, about 1~eV (Fig.\ref{fig:REB6_bands}). Therefore the stable purely divalent Yb$^{2+}$ 4f ground state is realized like Eu$^{2+}$ in EuB$_6$. Hence there is no band crossing with 5$d$ states. The latter exhibit a semiconducting gap $\sim 0.3~\text{eV}$ with respect to the lower B 2$p$ states. Therefore the electronic structure (schematically shown in Fig.~\ref{fig:REB6_bands}) is reminiscent of EuB$_6$ except that there is no magnetic order in YbB$_6$ and hence no overlap of spin-split bands, consequently it stays semiconducting. Under pressure, however, the 2$p$ and 5$d$ bands overlap transforming YbB$_6$ into a slightly mixed valent semimetal \cite{kang:16}. The ambient-pressure semiconductor  may exhibit band bending effects and therefore 2D confined surface states can exist which has led to previous misguided conclusions on the electronic structure.

After this brief survey of $R$B$_6$ materials with stable magnetic moments we turn now to  species with larger 5$d$-4$f$ hybridization which show either Kondo-lattice heavy-fermion behaviour with hidden multipolar order like \CB\ and its La diluted alloys or are Kondo insulators like \YB\ or strongly non-integer mixed valent semiconductors with topological order like \SB.

\section{Multipolar hidden order in CeB$_6$ in the localized 4$f$ scenario}
\label{sect:CeB6_loc}

\begin{figure}[b]
\begin{center}
\includegraphics[width=0.72\textwidth]{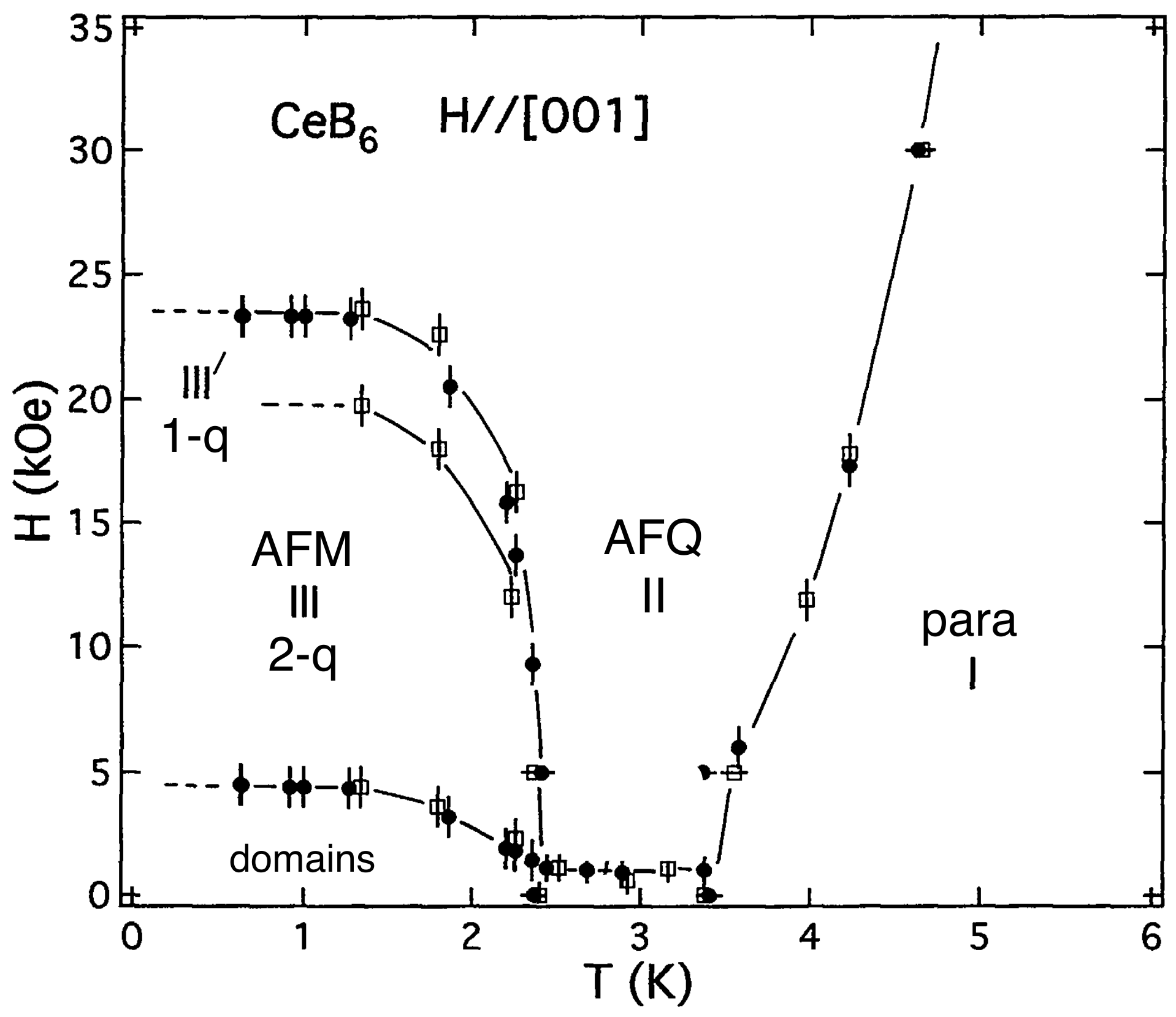}
\end{center}
\caption{Low-field phase diagram of \CB\ as obtained by tracking elastic constant anomalies. Ordering wave vectors are AFQ (II): $\bQ'=(\fs\fs\fs)$; AFM double-$\mathbf{q}$~(III) $\bQ_1=(\fq\fq\fs)$; $\bQ_2= (\fq\frac{\smash{\overline{1}}}{4}\fs)$ [structure of Fig.~\ref{fig:PrB6_morder}\,(a)]. Below the lowest line $H<5$~kOe, domain formation occurs (10~kOe = 1~T). Adapted from Nakamura~\textit{et~al.}~\cite{nakamura:95}.\index{CeB$_6$!$H$-$T$ phase diagram}}
\label{fig:CeB6_PDmod}
\end{figure}

The high degeneracy and strong Coulomb repulsion of $f$-electrons in lanthanide and actinide compounds can lead to exotic quantum matter states at low temperatures \cite{onuki:04, thalmeier:05, kusunose:08, kuramoto:09}. The hybridization of valence (conduction) electrons with strongly correlated f electrons may result in the formation of heavy-fermion metals with quasiparticles that have large effective masses and opening of hybridization gaps \index{hybridization gap}\index{heavy fermion metal!hybridization gap} that can lead to Kondo insulator state.\index{Kondo insulator} This is most frequently observed in intermetallic Ce or Yb compounds which commonly have one 4$f$ electron or hole state with orbital energy $\epsilon_f$ not too far below the Fermi level and having considerable hybridization with conduction bands. Furthermore, at even lower temperatures, broken symmetry phases (usually magnetic or superconducting) appear, driven by residual quasiparticle interactions \cite{thalmeier:05, thalmeier:05a}. Of particular interest are the 'hidden order' (HO) phases introduced before.

By various combinations of experimental methods (Sec.~\ref{sect:intro}), the most detailed understanding of HO has been achieved for the cubic heavy-fermion compound \CB. It  exhibits a second order transition at $T_\text{Q}=3.3$~K from paramagnetic phase~I into a HO phase~II and then into coexisting AFM phase~III at the lower N\'eel temperature $T_\text{N}=2.3$~K (Fig.~\ref{fig:CeB6_PDmod}). In zero field the HO phase~II does not lead to any new Bragg peaks in x-ray or neutron diffraction. The field dependence of $T_\text{Q}(H)$ has a large positive slope leading to an exceptional increase up to  $T_\text{Q}=10$~K at $H=35$~T. The compound is a prominent model system for HO because of the simplicity of 4$f$ states in that case. The $O_h$ symmetry (Fig.~\ref{fig:CeB6struc}) leads to a cubic CEF which splits the six $J=\frac{5}{2}$ Ce$^{3+}(4f^1)$ states into a $\Gamma_8$ ground state quartet and highly excited $\Gamma_7$ doublet at $\sim$530~K~\cite{zirngiebl:84, zirngiebl:85, loewenhaupt:86, YeKung19} (see Fig.~\ref{fig:REB6_CEF}), which may be neglected for all low-temperature phenomena. The quartet ground state is explicitly given by\vspace{-5pt}
\be
\bl
|+\ua\ket&=\sqrt{\textstyle\frac{5}{6}}|+{\textstyle\frac{5}{2}}\ket+\sqrt{\textstyle\frac{1}{6}}|-{\textstyle\frac{3}{2}}\ket; \;\;\;\;\;\;\;\;  |-\ua\ket=|{\textstyle+\frac{1}{2}}\ket,\\
|+\da\ket&=\sqrt{\textstyle\frac{5}{6}}|-{\textstyle\frac{5}{2}}\ket+\sqrt{\textstyle\frac{1}{6}}|+{\textstyle\frac{3}{2}}\ket; \;\;\;\;\;\;\;\;  |-\da\ket=|{\textstyle-\frac{1}{2}}\ket.
\label{eqn:CEF}
\el\vspace{5pt}
\ee
This may be thought of consisting of two orbitally inequivalent Kramers doublets with symmetry $\Gamma_7$ (left part) and $\Gamma_6$ (right part) that are forced into one quartet representation by the cubic symmetry.
Therefore it is suggestive to interpret $\sigma_z=\ua,\da$ as Kramers pseudo spin of each doublet and $\tau_z=\pm$ as orbital pseudo spin that distinguishes the two doublets \cite{ohkawa:85}. This may be formalized by introducing the representations for the two pseudo-spins $\boldtau$ and $\boldsigma$ \index{pseudo-spin} defined by
\bea
\boldtau=\fs\sum_{\tau\tau'\sigma}f^\dag_{\tau\sigma}\boldrho_{\tau,\tau'} f_{\tau'\sigma};\;\;
\boldsigma=\fs\sum_{\tau\sigma\sigma'}f^\dag_{\tau\sigma}\boldrho_{\sigma,\sigma'} f_{\tau\sigma'}
\eea
Here $\boldrho=(\rho_x,\rho_y,\rho_z)$ denotes the set of Pauli matrices. The $f^\dag_{\tau\sigma}$ one-electron fermion operators create the CEF states in
Eq.~(\ref{eqn:CEF}) according to $|\tau\sigma\ket=f^\dag_{\tau\sigma}|0\ket$ with $|0\ket$ denoting the empty $4f^0$ state.

\begin{table}[b!]
\tbl{Representations of 9 of the 15 multipoles of $\Gamma_8$ quartet: Stevens notation \index{Stevens representation} using total angular momentum \protect{\bJ} components or components of pseudospins \protect{$\boldsigma, \boldtau$}. The components of the total angular momentum are the linear combination $J_\alpha=\sum_n\lambda^n_\alpha X_n$ where coefficients $\lambda^n_\alpha$ can be read off in the last column. In the last row symmetrization (summation over all permutations of $x,y,z$) is denoted by a bar.\vspace{3pt}}
{\begin{tabular}{@{}cccc@{}}
\toprule
$O_h$ multipole$^\text{a}$ & rank & Stevens notation & pseudo-spin form\\
 (degeneracy) & $p$ & $J_\alpha$, $(\alpha=x,y,z)$ & $\sigma_\alpha,\tau_\alpha$\\
\midrule
$\Gamma^-_4(3)$ &1 (d)  &  \hphantom{0}  $J_x$   & $\frac{7}{6}[\sigma_x+\frac{2}{7}(-\tau_z\sigma_x+\sqrt{3}\tau_x\sigma_x)]$ \\
  \hphantom{0}   &           &  \hphantom{0}  $J_y$   & $\frac{7}{6}[\sigma_y+\frac{2}{7}(-\tau_z\sigma_y-\sqrt{3}\tau_x\sigma_y)]$\\
  \hphantom{0}   &          &  \hphantom{0}  $J_z$   & $\frac{7}{6}[\sigma_z+\frac{2}{7}(2\tau_z\sigma_z)]$\\
$\Gamma^+_3(2)$& 2 (q) & $O^0_2=\frac{1}{2}(2J_z^2-J_x^2-J_y^2)$     & $\tau_z=\frac{1}{8}O^0_2$\\
   \hphantom{0}  &          & $O^2_2=\frac{\sqrt{3}}{2}(J_x^2-J_y^2)$        & $\tau_x=\frac{1}{8}O^2_2$\\
$\Gamma^+_5(3)$ &2 (q) &  $O_{yz}=\frac{\sqrt{3}}{2}(J_yJ_z+J_zJ_y)$  & $\tau_y\sigma_x=\frac{1}{4}O_{yz}$\\
   \hphantom{0} &	   & $O_{zx}=\frac{\sqrt{3}}{2}(J_zJ_x+J_xJ_z)$   & $\tau_y\sigma_y=\frac{1}{4}O_{zx}$\\
   \hphantom{0} &	   & $O_{xy}=\frac{\sqrt{3}}{2}(J_xJ_y+J_yJ_x)$   & $\tau_y\sigma_z=\frac{1}{4}O_{xy}$\\
$\Gamma^-_2(1)$&3 (o)  & $T_{xyz}=\frac{\sqrt{15}}{6}\overline{J_xJ_yJ_z}$   & $\tau_y=\frac{\sqrt{5}}{45}T_{xyz}$\\
\bottomrule
\end{tabular}
}
\begin{tabnote}
$^\text{a}$d~=~dipole, q~=~quadrupole, o~=~octupole.\\
\end{tabnote}
\label{tbl:multipole}
\end{table}

\subsection{Pseudo-spin representation of $\Gamma_8$-quartet multipoles}
\label{sect:pseudospin}\index{pseudo-spin}\index{multipolar order parameters!pseudo-spin representation}

\begin{figure}[t]
\begin{center}
\includegraphics[width=\textwidth]{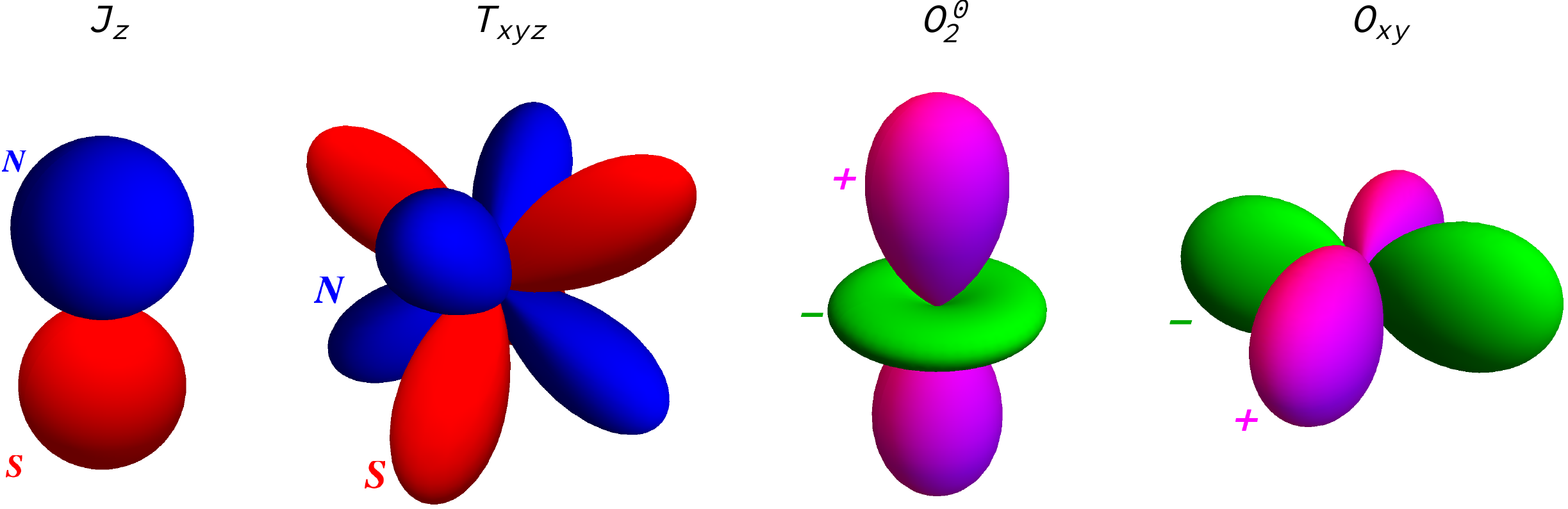}
\end{center}
\caption{Selected multipoles of $\Gamma_8$ quartet (degeneracy in parentheses): Dipolar $J_z$ ($\Gamma_4^-(3)$, rank 1), quadrupolar $O_2^0$ ($\Gamma_3^+(2)$, rank 2), $O_{xy}$ ($\Gamma_5^+(3)$, rank 2) and octupolar $T_{xyz}$ ($\Gamma_2^-(1)$, rank 3. Red/blue denote opposite signs of 4$f$ spin density and green/magenta opposite signs of 4$f$ charge densities.}
\label{fig:multipole}
\end{figure}

The 4$f$ electrons with orbital angular momentum \mbox{$l=3$} are arranged in a shell with orbital, spin and total angular momenta $(L,S,J)$ according to Hund's rules. For Ce$^{3+}$$(4f^1)$ and Yb$^{3+}$$(4f^{13})$ they are
equal to single electron or hole quantum numbers, respectively. The charge density and moment density operators of the 4$f$ shell may be expanded in terms of multipole basis functions consisting of  polynomials with rank $p\leq 2l$ associated with specific representations of the RE site symmetry. Their expectation values in a given 4$f$ state (e.g. the CEF ground state) correspond to the classical electrostatic and magnetostatic multipoles, as discussed extensively in \cite{kusunose:08}. With the help of the Wigner-Eckhard theorem for the states with total angular momentum $J$ the multipole operators of rank p may be expressed as combinations of  polynomials in ${\cal P}_p(J_x,J_y,J_z)$ of rank p belonging to cubic representation $(\Gamma\gamma)$ by using the Stevens operator technique \cite{hutchings:64}. \index{Stevens representation} Some of these operators up to rank 3 (octupoles) are listed in Table~\ref{tbl:multipole} and the symmetry of the corresponding real-space tesseral harmonics \cite{hutchings:64} in $(x,y,z)$ cartesian coordinates is shown in Fig.~\ref{fig:multipole}.

If we restrict further to the $\Gamma_8$ ground state in \CB\ the polynomials  ${\cal P}_p(J_x,J_y,J_z)$  may be mapped to the pseudospin algebra defined above by comparing their matrix elements within the quartet \cite{ohkawa:85,shiina:97}. The complete set of $\Gamma_8$ multipole operators in pseudospin basis is given by
\bea
\{X_n\}=\{\sigma_\alpha, \tau_\alpha, \sigma_\alpha\tau_\beta\}
\label{eqn:multipoleop}
\eea
with $n=1-15$ and $\alpha,\beta=x,y,z$. They constitute a basis set for the fifteen multipole moments, acting in the space of $\Gamma_8$ quartet states. The multipoles transforming as cubic representations in Table~\ref{tbl:multipole} are then generally linear combinations of the $X_n$, e.g. $J_\alpha=\sum_n\lambda^n_\alpha X_n$  where the $\lambda^n_\alpha$ may be read off from Table~\ref{tbl:multipole}.

\subsection{Multipole interaction model and symmetry breakings}
\label{sect:HOmulti}\index{multipole interaction model}

In the intermetallic RE compounds the small hybridization with conduction electrons leads to effective inter-site interactions between the multipoles on neighboring lattice sites. Those between the dipoles (rank 1) are commonly known as RKKY interactions, \index{RKKY interactions} but the concept may also be extended to higher-rank multipoles \cite{teitelbaum:76,schmitt:84,shiba:99} (Sec.\ref{sect:CEF}). In the same way as the RKKY terms lead to magnetic ordering of dipolar moments $\bJ$ at low temperature they may also induce 'hidden order' (HO) of multipoles with rank $p>1$. The  pseudo spin representation for \CB\ has been proposed \cite{ohkawa:85} and investigated in detail with respect to possible multipolar HO and excitations \cite{shiina:97,thalmeier:98,shiina:03}. A model Hamiltonian describing the effective intersite coupling of $\Gamma_8$ multipoles may be written as
\bea
{\cal H}&=&D\sum_{\langle ij\rangle}
[(\boldtau_i\cdot\boldtau_j)+(\boldsigma_i\cdot\boldsigma_j)+
4(\boldtau_i\cdot\boldtau_j)(\boldsigma_i\cdot\boldsigma_j)+\\
&&+\epsilon_\text{Q} 4 \tau_i^y\tau_j^y(\boldsigma_i\cdot\boldsigma_j)
+\epsilon_\text{O}\tau_i^y\tau_j^y]
-\frac{7}{6}g\mu_B\sum_i(\boldsigma_i+\frac{2}{7}\boldeta_i)\cdot\bH\non
\label{eqn:HAM}
\eea
where we introduced the multipole vector (cf. Table \ref{tbl:multipole}) $\boldeta=(-\tau_z\sigma_x+\sqrt{3}\tau_x\sigma_x,-\tau_z\sigma_y-\sqrt{3}\tau_x\sigma_y,2\tau_z\sigma_z)$ as an abbreviation. The first three terms describe a SU(4) 'supersymmetric' n.n. intersite interaction on the simple cubic lattice (sites $i,j$) of strength $D$ (coordination $z=6)$ which has no bias for any of the fifteen multipoles as primary order parameter.
The following two terms express the symmetry breaking that favors quadrupolar or octupolar order depending on the size of the parameters $\boldepsilon=(\epsilon_\text{Q},\epsilon_\text{O})$. They correspond to $\Gamma^+_5$-type quadrupoles and $\Gamma^-_2$ type octupole, respectively. This preference is concluded from the experimental evidence discussed below and from derivation of the effective ${\cal H}$ from a more fundamental Anderson-type Hamiltonian \cite{shiba:99, yamada:19}. The $\pm$ sign denotes even/odd behaviour under time reversal, i.e. the quadrupole corresponds to a charge and the octupole to a magnetic moment distribution (Fig.~\ref{fig:multipole}), in both cases with zero net charge or moment and therefore `'hidden''. The last term is the Zeeman energy in pseudospin representation.

\begin{figure}[b]
\includegraphics[width=\columnwidth]{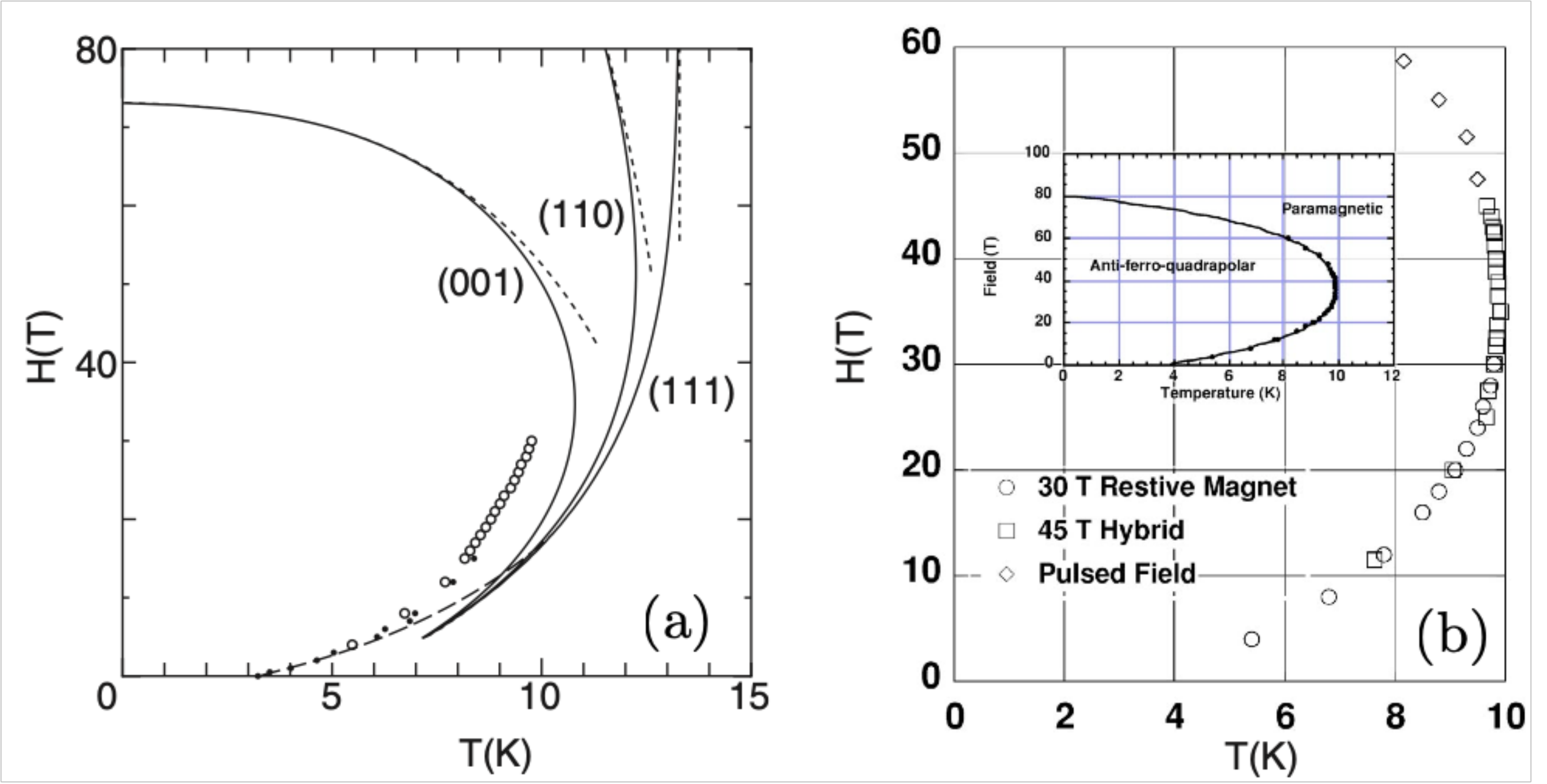}
\caption{(a)~High-field anisotropy of critical fields of AFQ phase~II for symmetry directions using $1/d$ expansion and compared to moderate field experiments (circles). At low fields critical field curves become isotropic in this method. Reproduced from Shiina \textit{et al.}~\cite{shiina:02}. (b) Experimental $[001]$ high field results. Inset shows quadratic fit that suggests a maximum critical field $H_{\rm c}(0)=80$~T. Reproduced from Goodrich \textit{et al.}~\cite{goodrich:04}.}
\label{fig:phaseBT}
\end{figure}

\subsection{Experimental identification of multipolar order parameters}
\label{sect:HOexp}

\index{multipolar order parameters!experimental identification}
In this section we present to some detail various experimental evidence to unravel the nature of HO in \CB, i.e. to identify which multipoles appear as order parameters below the transition temperature $T_\text{Q}$. Firstly it was observed that on approaching $T_\text{Q}$ from above certain elastic constants exhibit typical small anomalies, although no real softening \cite{nakamura:94}. This already indicated that the primary order should be of the  quadrupolar kind but not of the ferro-type. Therefore the  starting point of the model in Eq.~(\ref{eqn:HAM}) is appropriate. More direct evidence for the nature of HO comes from the following investigations:

\subsubsection*{\it a)~Anisotropy and slope of critical field curves}\index{CeB$_6$!critical fields|(}

\begin{figure}[b]
\begin{center}
\includegraphics[width=0.7\columnwidth]{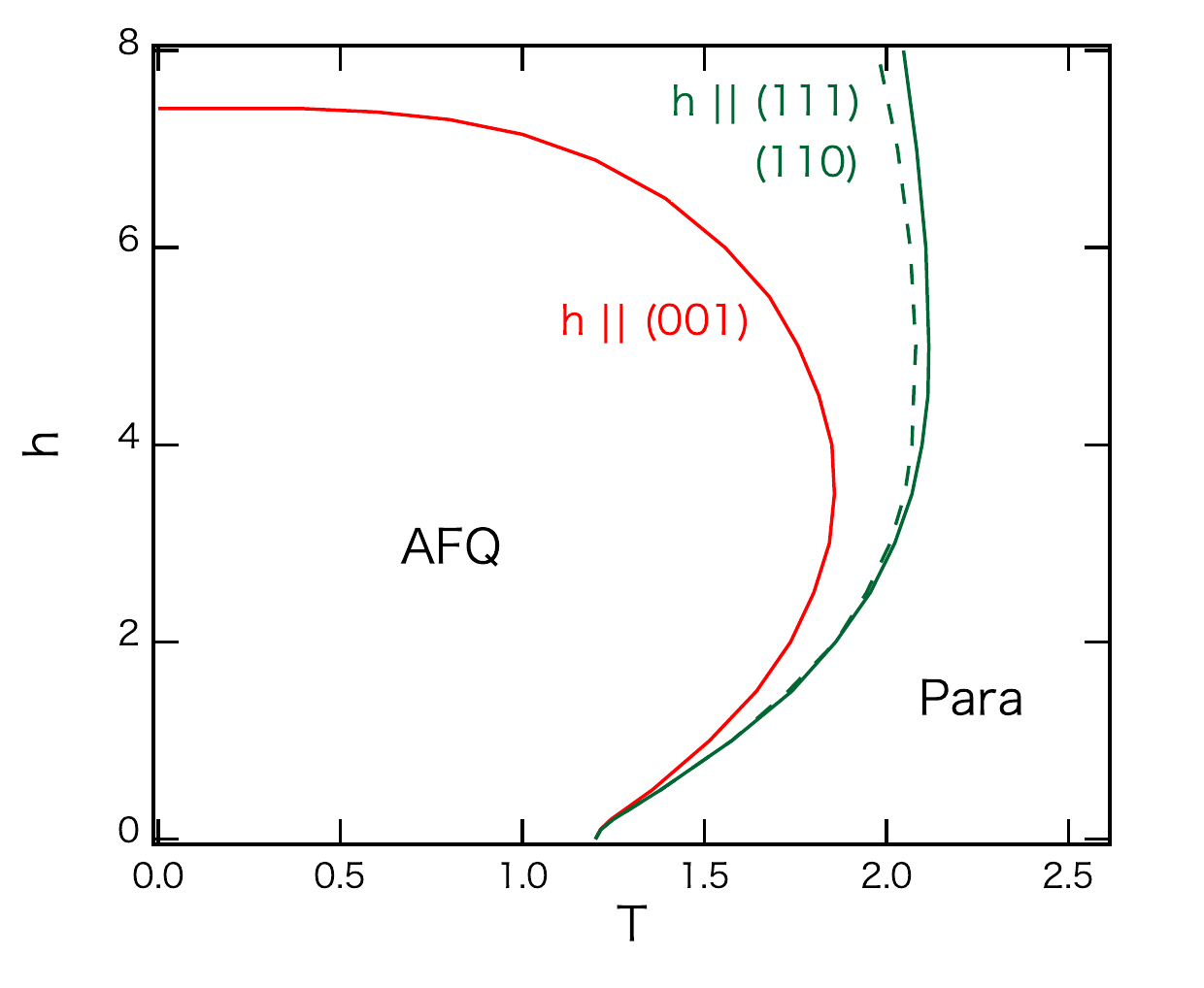}\vspace{-3pt}
\end{center}
\caption{Mean field $T_\text{Q}$ with $(\epsilon_\text{Q}, \epsilon_\text{O})=(0.2, 0.0)$ for magnetic fields along high symmetry axes (from \cite{shiina:98}). The field direction along [110] is given by the dashed line. The field and temperature are scaled by $T_0 (= 2Dz)$, the zero field $T_\text{Q}$ for $\epsilon_\text{Q}=\epsilon_\text{O}=0$.\vspace*{-5pt}}
\label{fig:fig1_8}
\end{figure}

The most striking property of $T_\text{Q}(H)$ is a very large positive slope of $(dT_\text{Q}(H)/dH)_0= 0.84$~K/T (Fig.~\ref{fig:phaseBT}). To understand this exceptional $B$-$T$ HO phase boundary a mean-field analysis of the model is required \cite{shiina:97}. Since we restrict to n.n. interactions and the quadrupolar order should not be of ferro type we start from a two sublattice $s=A,B, (\bar{s}=B,A)$ structure for the mean-field version of Eq.~(\ref{eqn:HAM}):
\be
H_{\rm mf}=-\sum_{s,i\in \bar{s}}h_s^n\hat{X}_{i\bar{s}}^n -NE(\bh),
\ee
where
\be
\nonumber
h_s^n=h^n-2zD\Lambda(n)x_s^n; \;\;\;
E(\bh)=2Dz\bx_a\cdot\boldLambda\cdot\bx_b ,
\ee
and $x_s=\bra\bX_s\ket$ is the mean-field value of the multipole vector $\bX=\{X_n\}$. It is also useful to define staggered $\bx_s=\fs(\bx_A-\bx_B)$ and uniform $\bx_f=\fs(\bx_A+\bx_B)$ order parameters. Furthermore we define the interaction model by setting
$\Lambda(n,n')= \Lambda(n)\delta_{n,n'}$ with $\Lambda(5) = 1+\epsilon_\text{O}$
($\Gamma^-_2$ octupole),  $\Lambda(8-10) = 1+\epsilon_\text{Q}$ ($\Gamma^+_5$ quadrupole) and
$\Lambda(n)=1$ else (all other multipoles). This singles out $\Gamma_5^+$ and $\Gamma_2^-$ as preferred HO parameters in accordance
with experiments discussed below. The components of the field vector $\bh=(h^n,n=1-15)$ can be read off by comparison with the last term in Eq.~(\ref{eqn:HAM}).

\begin{figure}
\mbox{\hspace{-0.015\columnwidth}\includegraphics[width=1.02\columnwidth]{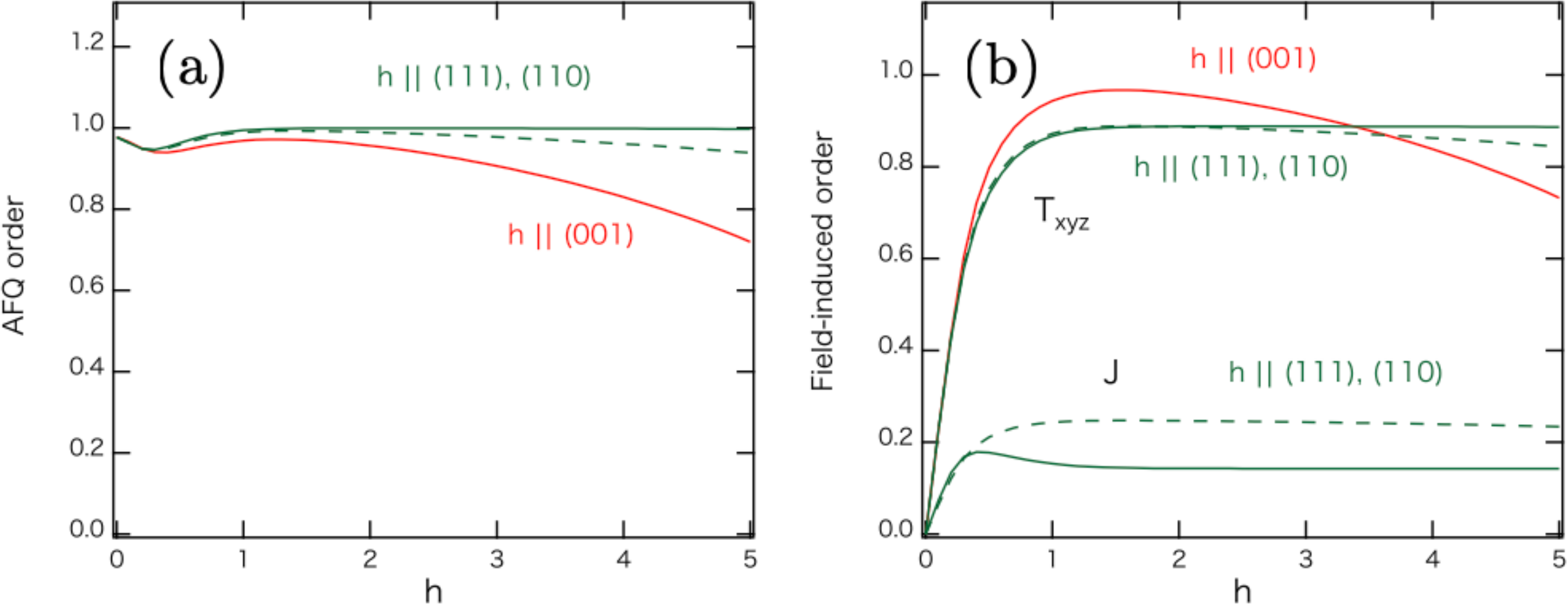}}
\caption{(a)~Staggered primary (quadrupolar $q_s$) order parameter at a fixed temperature $T=0.5$ as function of the magnetic fields along high symmetry axes. $q_s$ is scaled with the saturated value. The same scale for the field and the temperature is used as Fig.~\ref{fig:fig1_8}. (b)~Secondary (octupolar $o_s$ and dipolar $d_s$) order parameter as a function of the magnetic fields (from \cite{shiina:98}). $o_s$ is scaled with the saturated value, whereas $d_s$ is not scaled. The dashed lines in both figures are given by [110] field. For [001] field $d_s$ is not induced. See also Table~\ref{tbl:OPinduced} for the detailed components.
\index{multipolar order parameters!field dependence}}
\label{fig:fig1_9}
\end{figure}

The mean-field Hamiltonian for $A$, $B$ is a $4\times 4$ matrix in the quartet space. Its diagonalization leads to new split eigenstates (Fig.~\ref{fig:splitCEF}) from which the free energy may be obtained and minimized. For any field strength and direction the order parameters are then obtained. By the above choice of interaction the primary hidden order parameter is the threefold degenerate $\Gamma_5^+ (O_{yz},O_{zx},O_{xy})_s\equiv  4(\tau_y\sigma_x,\tau_y\sigma_y,\tau_y\sigma_z)_s$ at the staggered wave vector $\bQ'=(\fs\fs\fs)$ in r.l.u. $\frac{2\pi}{a}$.  The field dependent transition temperature $T_\text{Q}(H)$ is then obtained from the vanishing AFQ order, it is shown in Fig.~\ref{fig:fig1_8}. The transition temperature has a reentrant field dependence for all field directions. The strong increase of $T_\text{Q}(H)$ in Fig.~\ref{fig:fig1_8} up to intermediate fields has a simple origin in this model: For finite field, the octupolar staggered order is rapidly induced as {\it secondary} order parameter to its saturation value (Fig.~\ref{fig:fig1_9}). This stabilizes the AFQ phase and therefore leads to considerable increase in $T_\text{Q}(H)$. It should be noted that $T_\text{Q}$ is affected also by the direction of the field; $T_\text{Q}[111]$ and $T_\text{Q}[110]$ are much more enhanced than $T_\text{Q}[001]$ in high fields, originating from the anisotropic magnetization in the $\Gamma_8$ basis. This characteristic anisotropy is actually observed in the La-diluted system \cite{hiroi:98,akatsu:04}. On the other hand, the mean-field treatment has a deficiency as well: At zero field, due to the threefold degeneracy of the $\Gamma^+_5$ order parameter, fluctuations will be important and suppress  $T_\text{Q}^{\rm mf}(0)$ to the value $T_\text{Q}^{\rm exp}(0)=3.3$~K. Therefore the observed increase of $T_\text{Q}^{\rm exp}(H)$ starting from the zero field value up to the maximum $T_\text{Q}^{\rm exp}(H_\text{max})=10$~K at $H_\text{max}=35$~T will be even larger than predicted by the mean-field theory. This is shown in the experimental high-field phase diagram of Fig.~\ref{fig:phaseBT}. In fact, one can easily show that there is no chance for the mean-field $T_\text{Q}(H)$ to go beyond twice of the zero-field $T_\text{Q}$, irrespective of $(\epsilon_\text{Q}, \epsilon_\text{O})$. To improve the situation the contribution of thermal multipole fluctuations have been considered in low-field \cite{shiina:01} and high field case \cite{shiina:02} using a $1/d$ expansion in the spatial dimension $d$. The results in Fig.~\ref{fig:phaseBT}\,(a) which globally compares better with the experimental curve in Fig.~\ref{fig:phaseBT}\,(b). However in such expansion around the isotropic infinite dimensional limit the low field anisotropy information of  $T_\text{Q}(\bH)$ is lost.
\index{CeB$_6$!critical fields|)}

\subsubsection*{\it b)~Neutron diffraction in an applied magnetic field}\index{CeB$_6$!neutron diffraction|(}

This method gave the first clue on the symmetry of the quadrupolar HO \cite{effantin:85,erkelens:87}. Application of magnetic field along a symmetry direction, i.e. [001], [110], [111] reduces the symmetry group to lower than cubic. This has two effects: i) It selects a coherent superposition of the threefold degenerate primary HO representation $\Gamma_5^+$ that depends on field direction. (third row in Table~\ref{tbl:OPinduced} corresponds to experimental choice $[\bar{1}10]$). For a general field direction with unit vector $(\alpha\beta\gamma)$, the primary AFQ HO corresponds to  $\alpha O_{yz}+\beta O_{zx} +\gamma O_{xy}$ ii) Previously different $(O_h)$ representations may become mixed for finite field, leading to an {\it 'induced'} secondary order parameter with the same wave vector \bQ'\ whose amplitude becomes nonzero for finite field, in particular the dipole \bJ\ and octupole $T_{xyz}$. The homogeneous applied field breaks time reversal but preserves translational symmetry. Therefore, although the order parameter representations  become mixed and further secondary components with opposite time reversal symmetry are induced, their wave vector is identical to that of the primary order. Starting from the primary AFQ $\bigl[\Gamma^+_5,\bQ'=(\fs\fs\fs)\bigr]$ Table~\ref{tbl:OPinduced} lists the possible induced order for field along symmetry directions. The octupole $T_{xyz}$ is induced for all directions which explains the near isotropic enhancement of $T_\text{Q}(\bH)$. In particular it is seen that for \bH\ along $[\bar{1}10]$ a staggered dipolar moment along $[001]$ will be induced. This configuration corresponds to the ND experiment in \cite{effantin:85, erkelens:87} (third row in Table~\ref{tbl:OPinduced}) and leads to the observation of induced magnetic Bragg peaks from $J_z$ at \bQ'. Then, in reverse this observation may be interpreted as evidence for the underlying primary quadrupolar $O_{yz}-O_{zx}$ order.

\begin{table}[t]
\tbl{Primary quadrupolar (q) $\Gamma^+_5$ order induces secondary (d,o) order parameters of odd time reversal symmetry, depending on \bH\ direction. All order parameters are staggered with wave vector $\bQ'(\fs\fs\fs)$. Secondary induced moments (d) appear as AFM Bragg peaks at \bQ'.\vspace{3pt}}
{\begin{tabular}{@{}ccccc@{}}
\toprule
\bH\ direction & $^a$ primary $\Gamma^+_5$ (q) & induced $\Gamma^-_4$ (d) &  induced $\Gamma^-_2$ (o) & symmetry \\
\midrule
$[001]$ &  $O_{xy}$                             &     -                          & $T_{xyz}$  &  $C_{4v}$  \\
$[110]$ &  $O_{yz}+O_{zx}$                 &  $J_z$                    & $T_{xyz}$  &  $C_{2v}$  \\
$[\bar{1}10]$ &  $O_{yz}-O_{zx}$            &  $J_z$                     & $T_{xyz}$  &  $C_{2v}$  \\
$[111]$ &  $O_{yz}+O_{zx} +O_{xy}$   &  $J_x+J_y+J_z$     & $T_{xyz}$  &   $C_{3v}$ \\
\botrule
\end{tabular}
}
\begin{tabnote}
$^\text{a}$ The field selects a combination from the triply degenerate $\Gamma^+_5$ manifold. For general field direction with unit vector $(\alpha\beta\gamma)$ the linear combination $\alpha O_{yz}+\beta O_{zx} +\gamma O_{xy}$ is selected.
\end{tabnote}
\protect\label{tbl:OPinduced}
\end{table}
\index{CeB$_6$!neutron diffraction|)}

\subsubsection*{\it c)~NMR experiments}\index{CeB$_6$!nuclear magnetic resonance|(}

\begin{figure}
\begin{center}
\includegraphics[width=1.02\columnwidth]{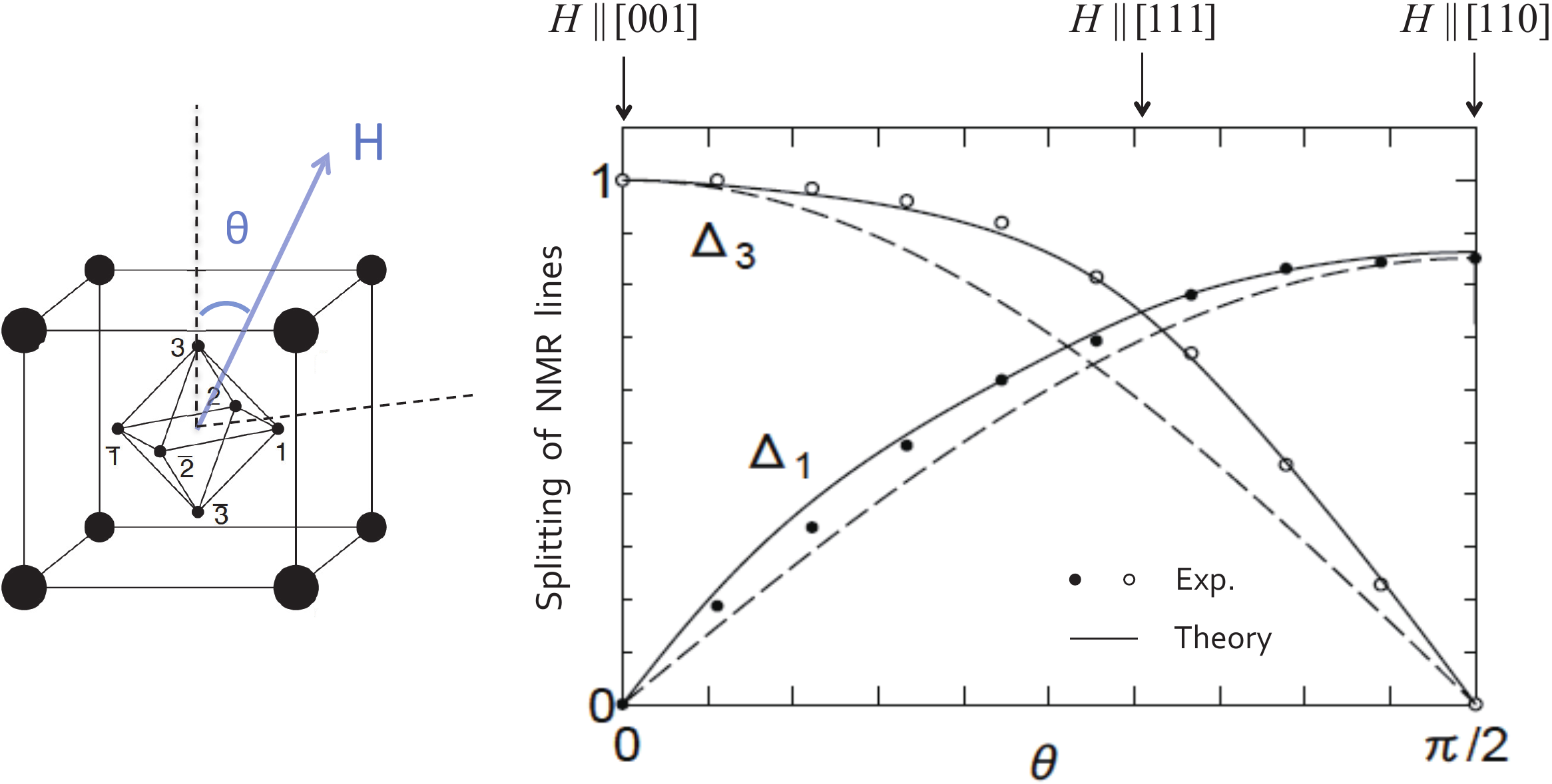}\vspace{-3pt}
\end{center}
\caption{NMR splittings $\Delta^{hf}_{1,3}(\theta)$ at inequivalent $^{11}B$ sites $1,3$ (see unit cell).
Circles are experimental data from \cite{takigawa:83}. Full line: model calculation
with Eq.~(\ref{eqn:hyperfine}). Broken line: octupolar part only.
The values are normalized to  $\Delta^{hf}_{3}(0)$ (from \cite{shiina:98}).}
\label{fig:CeB6_NMR}
\end{figure}

The dependence of NMR resonance lines of nuclear moments on applied field strength and direction contains important information on the underlying polarization of electronic magnetic moments, imprinted by the hyperfine interaction between the two types of moments. In this way analysis of NMR splittings as function of field strength and angle was used to infer the magnetic structure of induced moments in phase II of \CB\ \cite{takigawa:83}. Surprisingly the deduced structure did not agree with the structure obtained from ND results in a magnetic field \cite{effantin:85, erkelens:87}. In particular NMR lines of the $^{11}$B nucleus (site 3 in Figs.~\ref{fig:CeB6struc} and \ref{fig:CeB6_NMR}) show a clear splitting even when the field is oriented along the [001] direction. Now from Table~\ref{tbl:OPinduced} one observes that no dipolar moment $J_z$ is induced in this case. Assuming the standard hyperfine interaction where nuclear moments $I_z$ interact only with 4$f$ magnetic dipole moments $J_z$ one must conclude that there should be no NMR splitting for $\bH\parallel [001]$ under the assumption of an underlying AFQ structure as determined by ND, in clear contradiction to the observation \cite{takigawa:83}.

%
%
This discrepancy was solved by Shiina \textit{et al.} \cite{shiina:97,shiina:98} who showed that the local symmetry at the boron sites allows for a more general hyperfine interaction that couples the nuclear spin $\bI$ not only to the $4f$ dipolar moment but also to the octupolar moments. A simplified version \footnote{there are further contribution due to other induced octupoles \cite{shiina:97,shiina:98}} of the hyperfine Hamiltonian at the inequivalent $^{11}$B-sites (1,\,3 in Fig.~\ref{fig:CeB6struc}) is then given by
\be
\bl
H_1^{hf}&=a_1I_x\tilde{T}_{xyz}-b_1\bigl[I_y\tilde{J}_z(\bQ')+I_z\tilde{J}_y(\bQ')\bigr],
\\
H_3^{hf}&=a_3I_z\tilde{T}_{xyz}-b_3\bigl[I_x\tilde{J}_y(\bQ')+I_y\tilde{J}_x(\bQ')\bigr].
\label{eqn:hyperfine}
\el
\ee
Here the tilde denotes operators normalized to their maximum value and $a_{1,3}, b_{1,3}$ are hyperfine coupling constants of 4$f$ octupole and dipole moments for the two $^{11}$B-sites, respectively. The field is rotated in the diagonal plane containing $[001], [111], [110]$ axes with $\theta$ denoting the angle from [001]. Since the 4$f$ Zeeman energy  scale is much larger than the hyperfine energies, i.e. $g_J\mu_BH\gg a_{1,3},b_{1,3}$ the nuclear spins may simply be replaced by a classical vector that rotates with the field:
\bea
\bI=\frac{I}{\sqrt{2}}(\sin\theta,\sin\theta,\sqrt{2}\cos\theta).
\eea
Inserting this in Eq.~(\ref{eqn:hyperfine}) and using the mean-field solution for $\bra J_x\ket, \bra J_y\ket$ and $\bra T_{xyz}\ket$ leads to field-angle dependent hyperfine splittings $\Delta^{hf}_{1,3}$ of inequivalent $^{11}B$ sites that are  shown in Fig.~\ref{fig:CeB6_NMR}. Most importantly the  $\Delta^{hf}_{3}$ does not vanish for $\theta=0$ due to the octupolar  contribution in Eq.~(\ref{eqn:hyperfine}), in agreement with experiment.
This resolves the discrepancy with ND results. In fact the two methods are complementary: while ND determined the underlying AFQ structure via the induced magnetic dipoles, NMR identifies the existence of an induced strong octupolar component that was indirectly also inferred from the large positive slope of the critical field of phase II.
\index{CeB$_6$!nuclear magnetic resonance|)}

\subsubsection*{\it d)~Resonant x-ray diffraction (RXD) results}\index{CeB$_6$!resonant x-ray diffraction|(}

Although the previous methods concluded the existence and symmetry of HO from indirect evidence and its analysis, it would be reassuring to find direct evidence for quadrupoles and octupoles. In fact the more recent method of resonant x-ray diffraction is a useful new method to observe multipoles  up to fourth rank \cite{lovesey:05} directly. In \CB\ this has been carried out using the signals from (optical) dipolar E1 $(2p_{3/2}\ra 5$d$)$  and (optical) quadrupolar E2 $(2p_{3/2}\ra 4f)$ resonance transitions around the $L_3$ absorption edge. The transitions at $\omega_1=5724$~eV and at $\omega_2=5718$~eV differ by $\Delta\omega=6$~eV due to the larger binding energy of 4$f$ states and their line shapes overlap. The total intensity is given by $I(\omega,H)=|F_{E1}(\omega,H)+F_{E2}(\omega,H)|^2$
\begin{figure}
\begin{center}
\includegraphics[width=0.7\columnwidth]{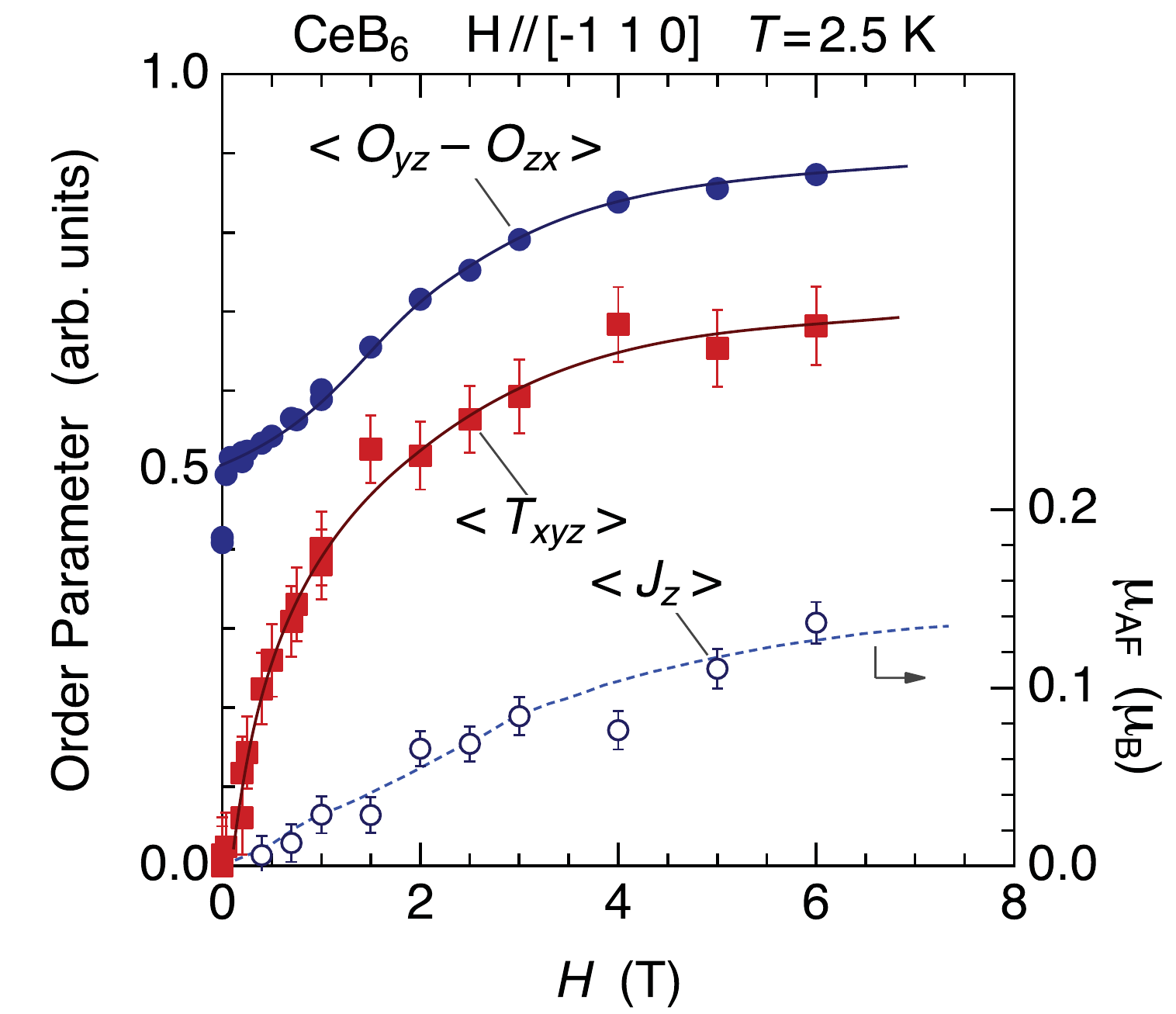}\vspace{-3pt}
\end{center}
\caption{Magnetic field dependence of the primary quadrupole HO $\bra O_{yz}-O_{zx}\ket$ and induced secondary dipolar $\bra J_z\ket$ and octupolar $\bra T_{xyz}\ket$ for $[\bar{1}10]$ field direction.
Symbols: RXD, dashed line: from ND experiments, full lines: guide to the eye. The momentum transfer is $(\frac{3}{2},\frac{3}{2},\fs)$ corresponding to \bQ' HO vector. Reproduced from Matsumura \textit{et al.}~\cite{matsumura:12}.}
\label{fig:CeB6_RXD}
\end{figure}
where $F_{E1}$,  $F_{E2}$ are the complex amplitudes for each process and H is the applied magnetic field (along the $[\bar{1}11]$ direction). These amplitudes contain contributions from electronic multipoles up to rank 2 $(E_1)$ and up to rank 4 $(E_2)$. To disentangle them the magnetic field reversal is an essential tool because the even and odd rank contributions in $F_{E1}$,  $F_{E2}$ behave even and odd under field reversal.  Let us define the average and difference intensities with respect to field reversal $H\ra -H$ by $I_{av}(\omega,H)=\fs[I(\omega,H)+I(\omega,-H)]$ and $\Delta I(\omega,H)=\fs[I(\omega,H)-I(\omega,-H)]$. From this approximate multipole order parameters may be extracted as \cite{matsumura:12}
\be
\bl
&\text{primary  (q):}\;\;\;\;\;\;\;\;\;\;\;  \bra O_{yz}-O_{zx}\ket_H \sim  \sqrt{I_{av}(\omega_1,H)};
\\
&\text{induced (d):}\;\;\;\;\;\;\;\;\;\;\;          \bra J _z\ket_H \sim  \Delta I(\omega_1)/\sqrt{I_{av}(\omega_1,H)};
\\
&\text{induced (o):}\;\;\;\;\;\;\;\;\;\;\;     \bra T_{xyz}\ket_H \sim   \Delta I(\omega_2)/\sqrt{I_{av}(\omega_1,H)}.
\el
\ee\smallskip
They are shown in Fig.~\ref{fig:CeB6_RXD}. The induced dipole agrees with the ND results (dashed line), and the field dependencies of all moments are qualitatively consistent with the theoretical results in Fig.~\ref{fig:fig1_9}.
The primary quadrupole $ \bra O_{yz}-O_{zx}\ket_H$ obtained from RXD in Fig.~\ref{fig:CeB6_RXD} (upper curve) shows considerable $H$ dependence much larger than predicted by the mean-field calculation (Fig.~\ref{fig:fig1_9}).  Again this is due to the neglect of fluctuations. Just as they suppress the experimental value of $T^{\rm exp}_\text{Q}(0)$ by a factor of two, they also suppress the size of the primary order parameter by a large factor $(\sim 1.5)$ as compared to the mean-field prediction. At considerably larger field they approach each other. The octupole has a pronounced convex bending which is a signature of the strong octupolar interaction. In fact, it is confirmed experimentally by investigating the dependence of the octupole-quadrupole ratio on the uniform magnetization. According to the theoretical study, this ratio is quite sensitive to the octupole interaction strength, irrespective of the fluctuation effect \cite{shiina:01}.

In summary, the conclusion from critical field anisotropy \cite{shiina:97}, field-induced neutron diffraction \cite{erkelens:87}, NMR results \cite{shiina:98}, and resonant x-ray scattering \cite{matsumura:09} indicate that HO may be well described as a primary antiferroquadrupole $\Gamma^+_5$ order with wave vector $\bQ'=(\fs\fs\fs)$ and a secondary strongly field induced octupolar $\Gamma^-_2$ order parameter in addition to a smaller induced dipole component, both at the same wave vector \bQ'. Semi-quantitative agreement with experiments may be achieved by choosing $\boldepsilon=(0.5,0.5)$ \cite{shiina:03} in the localized multipolar model of Eq.~(\ref{eqn:HAM}) and this should be considered as an appropriate set for \CB.\vspace{-2pt}
\index{CeB$_6$!resonant x-ray diffraction|)}

\section{The octupolar HO phase IV in diluted Ce$_{1-x}$La$_x$B$_6$}
\label{sect:CeLaB6}\index{CeB$_6$!La doped, Ce$_{1-x}$La$_x$B$_6$!octupolar order|(}

In the stoichiometric $R$B$_6$ compounds the 4$f$ element may easily be replaced by other rare earth species. The most interesting case is perhaps the series \CBL\ \mbox{$(0\leq x\leq 1)$} where the magnetic Ce$^{3+}$ sublattice is progressively diluted with nonmagnetic  La$^{3+}$ which has no 4$f$-electrons. This will have two main consequences: Firstly the inter-site multipole interactions will be progressively weakened \cite{lemmens:89} suppressing the tendency to multipolar order. Secondly the coherent heavy-fermion quasiparticle bands that exist for $x\simeq 1$ will gradually become site-incoherent and turn into narrow localized Kondo-resonance states. This can in fact be directly inferred from the change of resistivity $\rho(T)$ from correlated metal $A+BT^2$ behaviour to saturated unitary Kondo impurity resistivity at low temperatures \cite{sato:85}. In the present context we
focus on the evolution of the hidden order phase diagram with x. When the distance between the 4$f$ multipoles of Ce becomes larger and their interactions are reduced one has to ask how long the AFQ order will last. Simply extrapolating the mean-field solution of the concentrated compound does not give the correct answer. Because of the additional non-Kramers degeneracy of the $\Gamma_8$ ground state (expressed by $\boldtau$) the  single ion quadrupolar susceptibility $\chi^\text{Q}_{\Gamma_5^+}(T)$ [Eq.~(\ref{eqn:elastic})] has a Curie divergence $\sim 1/T$ for low $T$ \cite{thalmeier:91}. Therefore the mean-field approach would lead to a quadrupolar phase transition for an arbitrary dilute compound, although $T^{\rm mf}_\text{Q}(x)$ would approach zero for vanishing $x$. This is not the case and something rather more interesting is observed: The (zero-field) AFQ order vanishes rather rapidly with doping, and by $x\leq0.8$ is already replaced by a different phase~IV. Its nature has been investigated as intensely as that of the parent compound.

\subsection{Phase diagram and evidence for \emph{primary} octupolar order}
\label{sect:CeLaB6_PD}

\begin{figure}[b]
\includegraphics[width=\columnwidth]{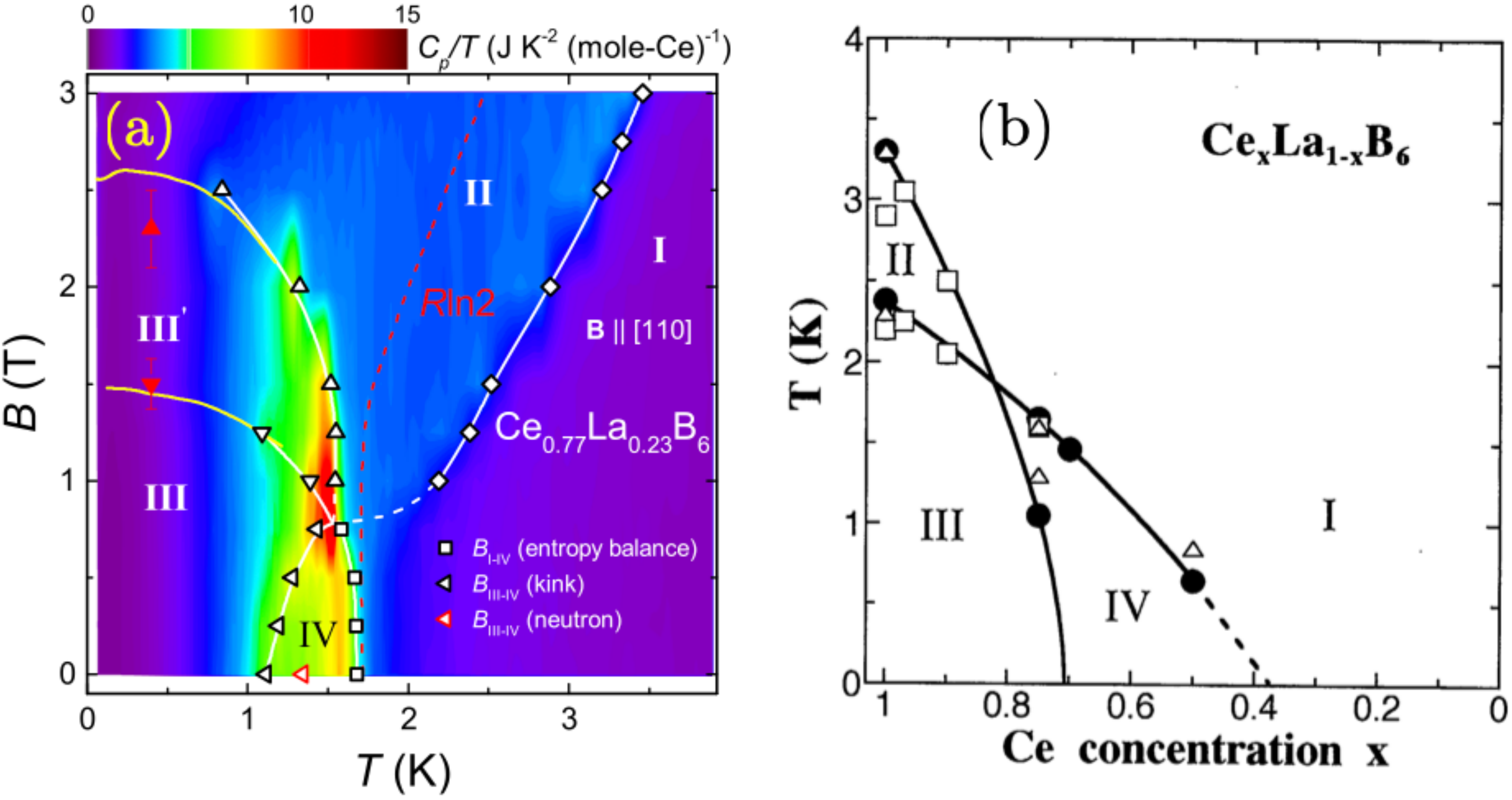}
\caption{(a) $B$-$T$ phase diagram of Ce$_{1-x}$La$_x$B$_6$ \index{CeB$_6$!La doped, Ce$_{1-x}$La$_x$B$_6$!$B$-$T$ phase diagram} for $x=0.23$ which shows the new antiferro-octupolar phase IV corner below $T_{\rm IV}=1.6$~K. AFQ phase II is no longer present for zero field. Reproduced from Jang \textit{et~al.}~\cite{jang:17}. (b)~$x$-$T$ phase diagram for phases I (para), phase II (AFQ), phase III (AFM) and phase IV (AFO). Reproduced from Tayama~\textit{et al.}~\cite{tayama:97}.}
\label{fig:phaseIV}
\end{figure}

An example of the global low-field phase diagram of \CBL\ \mbox{$(x=0.23)$} as obtained from magnetocaloric investigations \cite{jang:17} is shown in Fig.~\ref{fig:phaseIV} together with the $x$-$T$ phase diagram \cite{tayama:97}. Similar results were obtained from transport and magnetization experiments \cite{kobayashi:03,sera:18} and earlier from ultrasonic measurements \cite{suzuki:98,akatsu:03}. These and other macroscopic analysis
suggested the following basic properties of phase IV below \mbox{$T_\text{IV}\simeq 1.6$~K}: Contrary to phase~II of \CB\ at $T_\text{Q}$, a large specific heat jump is observed at $T_\text{IV}$. This indicates that the degeneracy of localized quartet states is completely lifted, different from AFQ order in phase~II where a twofold Kramers degeneracy remains in zero field (Fig.~\ref{fig:splitCEF}). Furthermore RXS gave clear evidence for a homogeneous \mbox{$(\bq=0)$} trigonal lattice distortion in phase IV along $[111]$ direction while none was observed in phase II (because the quadrupoles have a {\it staggered} order). In addition NMR and $\mu$SR experiments show the existence of an internal field below $T_\text{IV}$. Therefore phase IV breaks cubic crystal symmetry as well as time reversal symmetry. This requires a primary octupole order parameter belonging to $\Gamma^-_5$ as only plausible candidate \cite{kubo:04,kuramoto:09}. Irrespective of its translational character it will always induce a ferro-type quadrupole as secondary order parameter, already at {\it zero} field. This explains nicely the strong softening of elastic constants due to strain-quadrupole coupling immediately below $T_{\text IV}$ \cite{kubo:04}. The octupole order parameter of this symmetry and its induced quadrupoles are given in Table~\ref{tbl:octupole}.

\begin{table}[t]
\tbl{Octupolar order parameters for the phase IV, which is of antiferro-type with wave vector \bQ'. Secondary ferro-type quadrupolar order is induced already in zero field. The bar denotes symmetrization (summation over all permutations of $x,y,z$).\vspace{3pt}}
{\begin{tabular}{@{}ccccc@{}} \toprule
$O_h$ multipole & rank & Stevens notation & pseudo spin form & induced quadrupole\\
 (degeneracy) & p & $J_\alpha$, $(\alpha=x,y,z)$ & $\sigma_\alpha,\tau_\alpha$ & rank 2, $\Gamma_5^+(3)$
\\ \midrule
$\Gamma^-_5(3)$ &3 (o)    &   $T^\beta_x=\ftbj(\overline{J_xJ^2_y}-\overline{J^2_zJ_x})$
&  $\ftb(-\sqrt{3}\tau_z\sigma_x-\tau_x\sigma_x)$ & $O_{yz}$\\
&           &    $T^\beta_y=\ftbj(\overline{J_yJ^2_z}-\overline{J^2_xJ_y})$
 &    $\ftb(-\sqrt{3}\tau_z\sigma_y-\tau_x\sigma_y)$ & $O_{zx}$ \\
 &           &    $T^\beta_z=\ftbj(\overline{J_zJ^2_x}-\overline{J^2_yJ_z})$
 & $\ftb2\tau_x\sigma_z$  & $O_{xy}$ \\ \botrule
\end{tabular}
}
\begin{tabnote}
\end{tabnote}
\label{tbl:octupole}
\end{table}

\begin{figure}
\includegraphics[width=\textwidth]{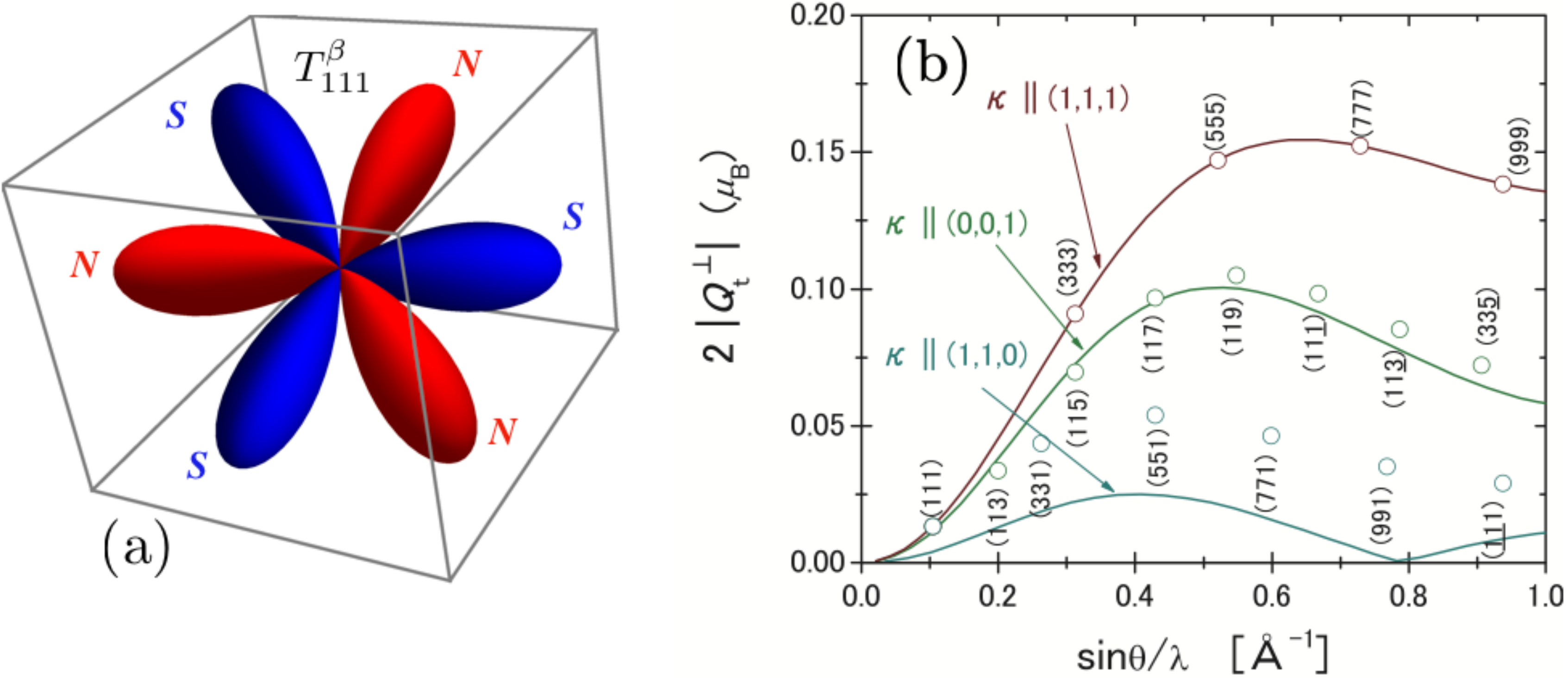}\vspace{-2pt}
\caption{(a)~A schematic plot of the symmetry of $T^\beta$ octupole for the (111) domain. (b)~Domain averaged magnetic form factor in the octupole order phase as function of the momentum transfer $|\boldkappa|/4\pi=\sin \theta/\lambda$ (from \cite{shiina:07}). Symbols $(lmn)$ corresponds to $\kappa = (l,m,n)/2$ and full lines to high symmetry directions. Underbars in the indices are defined as $\underline{n}= 10+n$.}
\label{fig:formfactor}
\end{figure}

In the supersymmetric part of the multipole intersite interaction Eq.~(\ref{eqn:HAM}) implicitly includes an isotropic term $\sim D\bT^\beta\cdot\bT^\beta$. One simple way to reproduce the octupolar order in Ce$_{1-x}$La$_x$B$_6$ is an appropriate enhancement of the octupole term in the interaction as $D \rightarrow D(1+\epsilon_{O'})$. Some variants of the model have been studied in the literature and provided a consistent mean-field picture to interpret the complex experimental results in phase IV \cite{kubo:04,sera:18}. However, it is not quite clear why the octupole interaction is selectively enhanced in the La-doped system. Another possible origin is the effect of random distribution of La that produces an additional CEF potential lower than cubic at Ce cites. Since the potential removes the non-Kramers degeneracy in the $\Gamma_8$ state, the randomness is expected to suppress the AFQ order more seriously. As a result, the AFO state surviving within the Kramers degeneracy gains a chance to overcome the quenched AFQ state \cite{sera:18,yamahara:19}. In fact, the existence of strong spatial disorder in phase IV of Ce$_{1-x}$La$_x$B$_6$ is inferred from broadening in the NMR spectra \cite{magishi:02}. Anyway the AFO mean-field ground state will be $\bra\bT^\beta\ket=\bra T^-_5\ket(\pm 1,\pm 1, \pm 1)$ corresponding to different domains. One of them, $(1,1,1)$ is illustrated in Fig.~\ref{fig:formfactor}\,(a). If we pick this domain the corresponding {\it homogeneous} induced $\Gamma^+_5$ (ferro-) quadrupole will be $\bra\bO\ket=\bra O_5^+\ket(1,1,1)$. Due to the coupling to homogeneous strains $\boldepsilon_{\Gamma_5}=(\epsilon_{yz},\epsilon_{zx},\epsilon_{xy})$ the crystal will distort with a trigonal strain $\epsilon_{\Gamma_5}(1,1,1)$ in accordance with conventional XD results \cite{inami:14}.

While these macroscopic symmetry considerations are consistent a direct proof by microscopic probes seems necessary. This was provided by analysis of angular dependent RXS \cite{mannix:05,kusunose:05,nagao:06} and large momentum transfer neutron diffraction \cite{kuwahara:07,shiina:07}.

Firstly, the former shows clearly that in phase IV the resonant scattering occurs at Bragg points corresponding to an AFO propagation vector $\bQ'=(\fs\fs\fs)$, the same as in the AFQ phase~II. Furthermore, the dependence of scattered intensity $I(\phi)$ on the azimuthal angle in the x-ray scattering plane shows a sixfold and threefold oscillation with $\phi$ in the full circle for $E2(\sigma\sigma)$ and  $E2(\sigma\pi)$ scattering channels, respectively. This can consistently be interpreted with an underlying octupolar-type hidden order \cite{mannix:05,matsumura:14}.

Secondly, although conventional neutron diffraction at low momentum transfer can only identify dipolar order, the scattering at very large momentum  transfer is sensitive to higher order (odd rank) multipoles. Shiina~\textit{et al.} \cite{shiina:07} have calculated the expected form factor for large momentum transfer $\boldkappa=\bk'-\bk$ of scattered neutrons for an underlying $\Gamma^-_5$ AFO order. It is shown in Fig.~\ref{fig:formfactor}\,(b). The fact that the form factor \index{magnetic form factor} vanishes for low $\boldkappa$ and shows a maximum at large $\boldkappa$ is a typical signature of multipolar order. The experiments have been performed \cite{kuwahara:07} and it was indeed observed on a few reflections that intensity {\it increases} with momentum transfer, in accordance with theoretical predictions. The anisotropy of the intensity for the momentum transfer can provide a further information on the form of the octupole, and its experimental identification is left for a future study.

Thus \CBL\ is one of the few confirmed cases of primary higher rank $r\geq 3$ multipolar order. The octupolar order in \CBL\ persists down to almost $x\approx 0.5$ \cite{jang:17}. For even lower $x$ the compounds are disordered at all temperature in zero fields and exhibit a Kondo impurity behavior. Another example for octupole ordering is NpO$_2$ \index{NpO$_2$}\cite{paixao:02,santini:09}, where the same component of octupole $T^\beta$ is believed to order with the triple-$q$ ordering vector. Note also that higher magnetic and nonmagnetic multipole states are proposed to resolve the nature of the famous HO phase of URu$_2$Si$_2$ \cite{ikeda:12,shibauchi:14,thalmeier:14}, but still remain controversial in that case.
\index{CeB$_6$!La doped, Ce$_{1-x}$La$_x$B$_6$!octupolar order|)}


\section{The collective excitations in the AFQ hidden order phase II of CeB$_6$}
\label{sect:CeB6excloc}\index{multipolar excitations}

Materials with CEF-split 4$f$ states show collective dispersive magnetic excitations, termed 'magnetic excitons' \cite{jensen:91} already in the paramagnetic phase. Their analysis leads to important information to build an exchange model \cite{haelg:86}. A magnetic phase transition may be preceded by a softening of excitons at the ordering wave vector \cite{jensen:91}. Below the transition their dispersion is modified due to the molecular field and additional collective Goldstone spin wave modes appear describing the order parameter dynamics.

\begin{figure}
\begin{center}
\includegraphics[width=0.55\columnwidth]{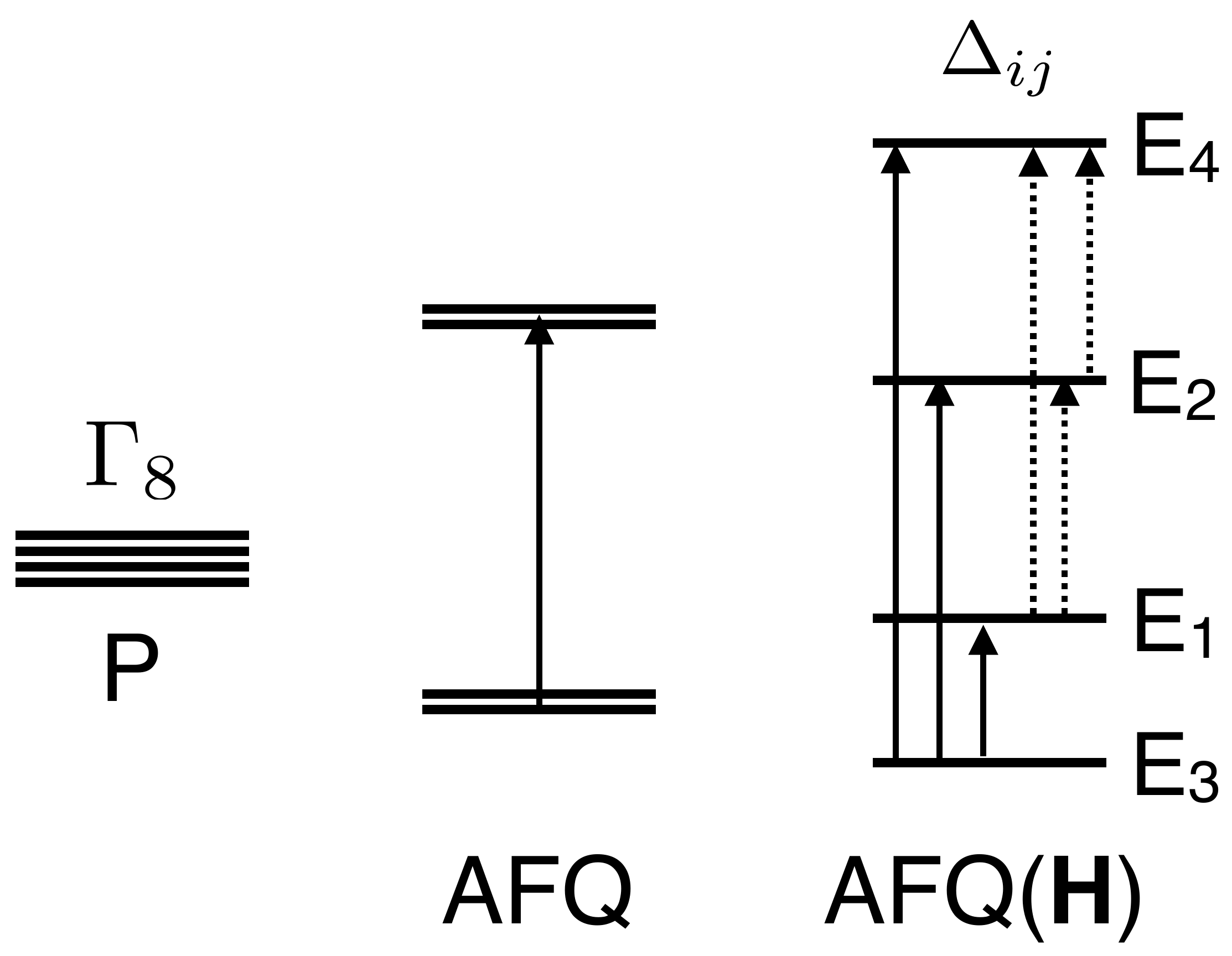}
\end{center}
\caption{Schematic splitting of $\Gamma_8$ multiplet in phase II into two Kramers doublets by AFQ molecular field and into four nondegenerate states due to action of AFQ, field \bH~and induced dipole and octupole molecular fields. Splittings are sublattice (s=A,B) dependent. For $T>0$, $\bH\neq 0$ all six excitations are possible. For low $T$ only three excitations from the ground state (full arrows) are important, leading to six dispersive modes in the interacting two-sublattice AFQ state (adapted from \cite{thalmeier:98}).}
\label{fig:splitCEF}
\end{figure}

In \CB\ we restrict ourselves to the fully degenerate $\Gamma_8$ (above $T_\text{Q}$) and ignore the very high energy $\Gamma_8-\Gamma_7$ excitation. Then we have only to consider the quasielastic excitations of the $\Gamma_8$ quartet. For $T<T_\text{Q}$ in the AFQ phase and by application of external fields the quartet splits as schematically shown in Fig.~\ref{fig:splitCEF} and the intersite coupling will then again lead to collective dispersive modes, this time on the energy scale of the $\Gamma_8$ splittings. However, in the present case their dispersion will not only be influenced by the dipolar exchange but by {\it all} multipolar interactions included in the intersite term of the Hamiltonian in Eq.~(\ref{eqn:HAM}). On the other hand their intensity appearing in INS is again be determined only by the {\it dipolar} dynamic structure function of these modes because neutrons (at low momentum transfer) do not directly couple to higher multipoles.

The magnetic excitation spectrum of the AFQ phase II of the model in Eq.~(\ref{eqn:HAM}) has been calculated with the complementary generalized Holstein-Primakoff approach \cite{shiina:03} or multipolar response function formalism in the RPA approach \cite{thalmeier:98,thalmeier:03}. Both include the full multipolar basis for calculation of the mode dispersions. For brevity we describe only the latter in this review but give a comparison of results from both methods for a typical case [Fig.~\ref{fig:disp}\,(b)]. As a first step (Sec~\ref{sect:HOexp}) one has to calculate the effective molecular fields (mf) $\langle X^n_A\rangle \pm \langle X^n_B\rangle$ (uniform and staggered) of each multipole basis operator which leads to the $\Gamma_8$ splitting (Fig.~\ref{fig:splitCEF}). Here $s=A,B$ denote the simple cubic sublattices of the antiferro-type HO defined by ordering vector $\bQ'$.

\subsection{Generalized multipolar RPA method}\index{random phase approximation}
\index{crystal electric field}\index{CeB$_6$!crystal electric field}

The CEF states are mixed by the molecular fields into new eigenstates $|\nu s\rangle_i$  with energies $E_\nu$ at every sublattice site $(s,i)$ as shown in Fig.~\ref{fig:splitCEF}. In terms of their standard basis operators a$^{si}_{\nu\mu}$= $|\nu s\rangle_i\langle\mu s|_i$  $(\nu,\mu=1,4)$ the mean-field approximation to Eq.~(\ref{eqn:HAM}) is given by
\be
H=
\sum_{\nu,si}E^s_\nu a^{si}_{\nu\nu}-\frac{1}{2}\sum_{\bra ij\ket ss'}\sum_{\nu\nu',\mu\mu'}
(\bM^s_{\nu\mu}\cdot\bD_{ss'}\cdot\bM^{s'}_{\nu'\mu'})
a^{si}_{\nu\mu}a^{s'j}_{\nu'\mu'}
\label{eqn:HAMMF}
\ee
where $\bM$ is a $n$-component vector $(n=1-15)$ of matrix elements for the multipole operators defined by $M^{ns}_{\nu\mu}$ = $\bra\nu,s|X^n_{is}|\mu,s\ket$ and the inter-sublattice multipole $n\times n$ diagonal interaction matrix is $\bD_{AB}(\bq)=\bD_{BA}(\bq)=-2zD\gamma_\bq\boldLambda$. Here $\boldLambda$ gives the relative interactions strengths of multipoles (Sec~\ref{sect:HOexp}), furthermore $\gamma_{\bq}=\frac{1}{3}(\cos q_x +\cos q_y +\cos q_z)$. With the thermal occupations of mf eigenstates given by $n_\nu=Z^{-1}\exp(-E_\nu/T)$ and $Z=\sum_\mu\exp(-E_\mu/T)$ the bare $n \times n$ multipolar susceptibility for each sublattice $s=A,B$ may be written as
\be
\chi^s_{0nl}(\omega)=\sum_{\nu\mu}
\frac{M^{ns}_{\nu\mu}M^{ls}_{\mu\nu}}{E_\mu-E_\nu-\omega +i\gamma_{\mu\nu}}
(n_\nu-n_\mu).
\label{eqn:baresus}
\ee
The $\gamma_{\mu\nu}$ line widths of $\Gamma_8$ transitions result from Landau damping due to conduction electrons \cite{thalmeier:03}. The collective response of all $15$ $\Gamma_8$ multipoles for the $2$ sublattices is then
described by the $30 \times 30$ RPA susceptibility matrix
\bea
\boldchi(\bq,\omega)=[\bun -\boldchi_0(\omega)\bD({\bq})]^{-1}
\boldchi_0(\omega)
\label{eqn:RPASUS}
\eea
where \bD\ consists of two anti-diagonal blocks $\bD_{AB}=\bD_{BA}$. The  elements of Eq.~(\ref{eqn:RPASUS}) may be used to construct the dipolar moment (\bJ) cartesian $(3\times 3)$ susceptibility matrix according to
\bea
\chi_{\alpha\beta}(\bq,\omega)=\sum_{ss',nm}
\lambda^n_\alpha\lambda^m_\beta\chi_{nm}^{ss'}(\bq,\omega)
\eea
where the $\lambda^n_\alpha$ are the coefficients of $J_\alpha$ in the pseudo spin representation (Table \ref{tbl:multipole}). Then the dynamical dipolar structure function, which is the only one observable in INS, may be written as
\bea
S(\bQ,\omega;\bH)&=&\frac{1}{\pi}[1-e^{\hbar\omega/kT}]\non\\
&&\times\sum_{\alpha\beta}
[\delta_{\alpha\beta}-\hat{\bQ}_\alpha\hat{\bQ}_\beta]
{\rm Im}\chi_{\alpha\beta}(\bq,\omega;\bH)
\label{eqn:strucfac}
\eea
It depends parametrically on field strength and direction. Here $\bQ = \bq+\bK$ is the total momentum transfer in INS with \bK\ denoting a reciprocal lattice vector and $\hat{\bQ} = \bQ/|\bQ|$. The structure function is proportional to the INS intensity and will be discussed for various field strengths and directions below.

\begin{figure}
\includegraphics[width=0.99\columnwidth]{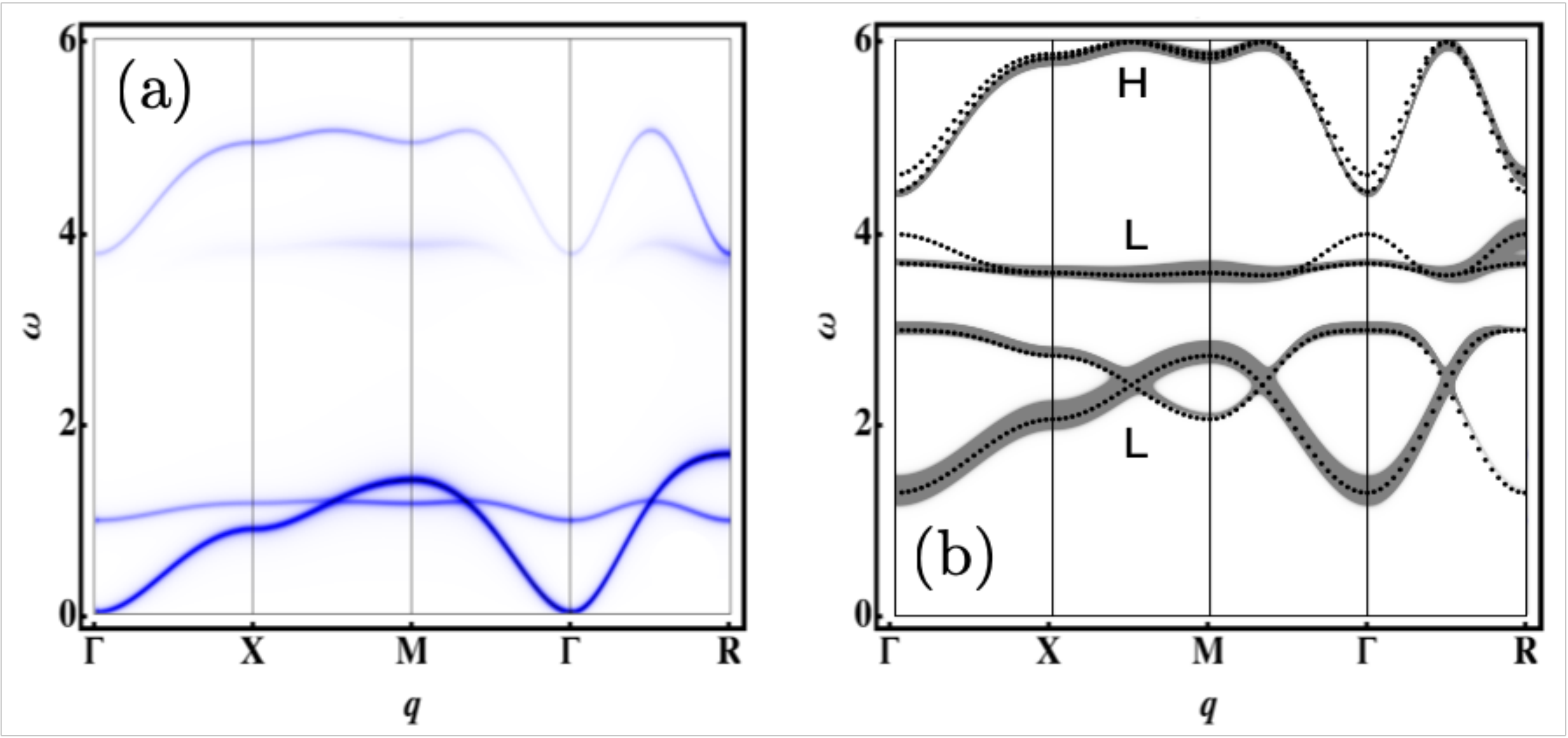}
\caption{Spectral function of magnetic excitations for zero field (a) and field along [001] axis (b) with $h'=1$ and $T=0.5T_0$ from HP \cite{shiina:03} (dots) and RPA \cite{thalmeier:03} (shading) approach. The momentum vector \bq\ is moving  along the Brillouin zone (BZ) path  $\varGamma$XM$\varGamma$R (here \bq\ is normalized to the length of each path segment to achieve  equal intervals). In this and all following similar figures the unit of mode energy $\omega$ is $T_0=2zD =0.41\; \mbox{meV} \; (4.74\; \mbox{K})$. Adapted from Ref.~\cite{thalmeier:03}.}
\label{fig:disp}
\end{figure}

The calculated dynamical RPA structure function $S(\bq,\omega)$ for \CB\ is shown in Fig.~\ref{fig:disp} for zero and finite field. Here we use model parameters $k_{\rm B} T_0=0.41~\text{meV}$ or $T_0=4.74 \; \mbox{K}$ and $\boldepsilon=(0.5,0.5)$ for \CB, also employed in Refs.~\cite{shiina:03,thalmeier:03}. Furthermore a dimensionless field strength is defined by $h'=0.672$H[T]/$T_0$[K]  with physical units for $H$ and $T_0$ (we will also use  $h=h'T_0$). Then $h'=1$ corresponds to $H=6.97 \; \mbox{T}$. There are locally six excitations between the molecular-field\,-- and Zeeman-split $\Gamma_8$ states, three from the ground state and three from the thermally excited states (Fig.~\ref{fig:splitCEF}), the latter are thermally suppressed except close to $T_\text{Q}$. Since there are two sublattices in the AFQ/AFO-type ordered phase six excitation branches will appear prominently that are mostly visible, e.g. in Fig.~\ref{fig:disp}\,(b) (for $h=0$ there are additional degeneracies). Roughly speaking, the six branches can be arranged in two groups: Firstly, two high energy branches (H) and secondly, four low energy branches (L), two of them almost degenerate and largely flat. The former are mostly stabilized at higher energy  by the octupolar molecular field while the latter are stabilized by the Zeeman term. When temperature approaches  $T^{\rm mf}_{Q}(h)$ from below, keeping the field constant the high energy modes collapse due to the reduced octupolar order while the low energy modes are less affected. Complementary, when temperature is kept constant much below the transition and the field is reduced to zero, the high-energy modes change little and the low energy modes are shifted downwards [cf. Fig.~\ref{fig:disp}\,(a,b)]. Due to the threefold degenerate $\Gamma^+_5$ order a Goldstone mode then appears at the $\varGamma$ point [Fig.~\ref{fig:disp}\,(a)]. It should be mentioned that the relative field independence of the higher modes is a consequence of the accidental degeneracy $\epsilon_{\rm Q}=\epsilon_{\rm O}$ assumed in the model \cite{thalmeier:03}. If we would choose $\epsilon_{\rm O}$ somewhat less than $\epsilon_{\rm Q}$ the octupolar order parameter $\langle \tau_z\rangle_\bQ$ would rapidly collapse at small fields and the two high energy modes with it, similar to the behaviour when temperature approaches the transition from below at zero field. The mode dispersions in the AFQ phase II of \CB\ from RPA calculations are in excellent agreement with the results from the HP approach, shown as dots in (b), including the intensities of modes as function of momentum \cite{shiina:03,thalmeier:03}. The latter method has later also been extended to the AFM phase III \cite{kusunose:01}. Experimental evidence for the multipolar mode dispersions in finite fields was found in \cite{bouvet:93} and \cite{jang:14, portnichenko:18}.

\subsection{Dependence of mode energies on field strength and field-angular rotation}\label{sect:CeB6fieldangular}
\index{CeB$_6$!$g$-factor anisotropy}\index{$g$-factor anisotropy}\index{CeB$_6$!magnetic resonant mode!field dependence}

The field dependence at constant \bq\ is complementary to the standard INS method where the full \bq-dispersion is determined for fixed field. In reality the latter may be difficult to carry out due to strong variation of
intensity in the BZ. In fact some of the excitations were mostly identified at the symmetry points of the BZ, in particular at $\varGamma$ and $R$, but also $X$, $M$. Therefore, for a comparison with theoretical results it may be a better strategy to keep the momentum transfer fixed at these symmetry points and vary the field strength and field direction. The mode frequencies are then recorded in radial plots in the field rotation plane.
This is a change of viewpoint as compared to the previous theoretical investigations \cite{shiina:03, thalmeier:03} which we discuss now in detail.

\begin{figure}[b]
\includegraphics[width=\columnwidth]{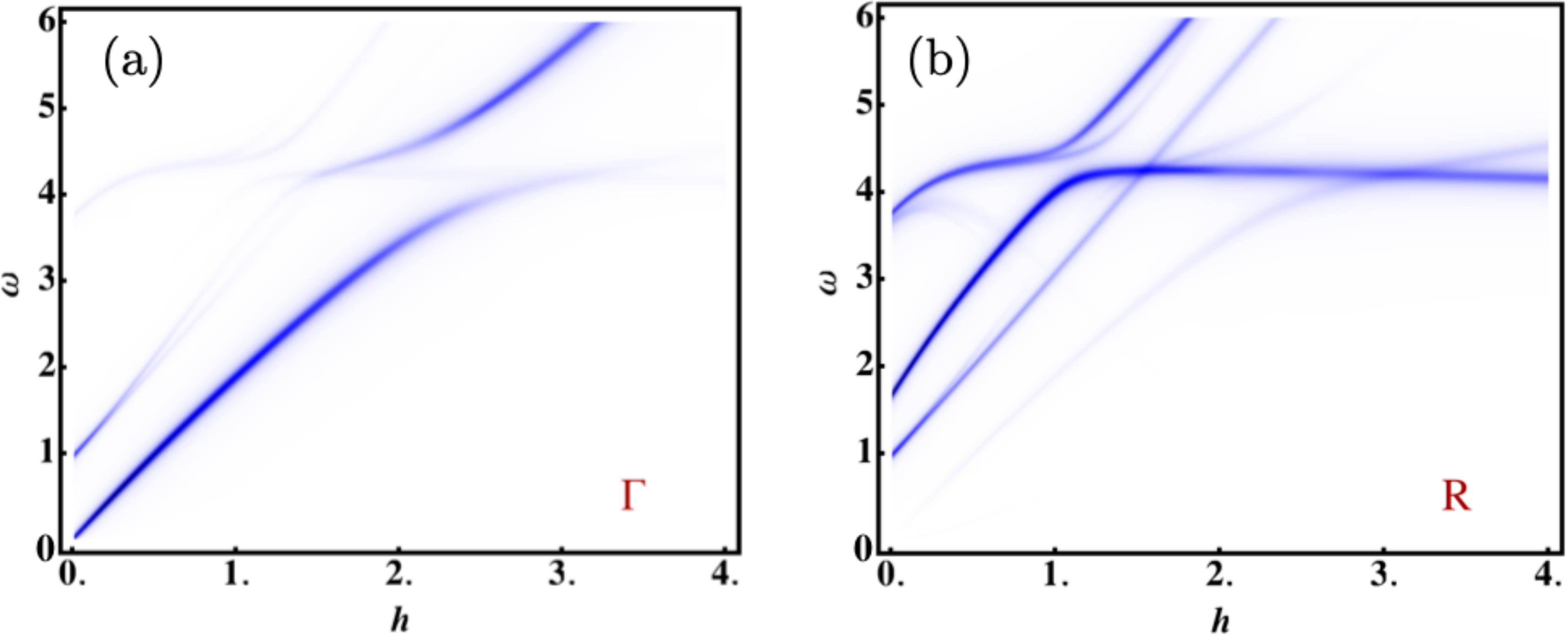}
\caption{$\bH \parallel [001]$ field dependence of mode frequencies for various BZ symmetry points $\varGamma(000)$ and $R(\fs\fs\fs)$. Low energy modes exhibit roughly linear Zeeman splitting. Field independent high energy modes are stabilized by the induced octupolar order parameter.\vspace{-3pt}}
\label{fig:fieldstrength}
\end{figure}

Without magnetic field the mean-field solution of Eq.~(\ref{eqn:HAM}) leads to a transition at $T_{\rm mf}=(1+\epsilon_\text{Q})T_0$ with the primary AFQ order parameter. At finite fields  a secondary dipole and octupole staggered order will be induced, depending on field direction (Fig.~\ref{fig:fig1_9}). Their associated molecular fields split and mix the local CEF energies and states. This information is encoded in the energy-denominator
and matrix elements of the RPA susceptibility in Eqs.~(\ref{eqn:baresus},\ref{eqn:RPASUS}) and hence the excitation spectrum of Eq.~(\ref{eqn:strucfac}) depends on order parameters, field strength and direction.
%
\begin{figure}[t]
\begin{center}
\includegraphics[width=\textwidth]{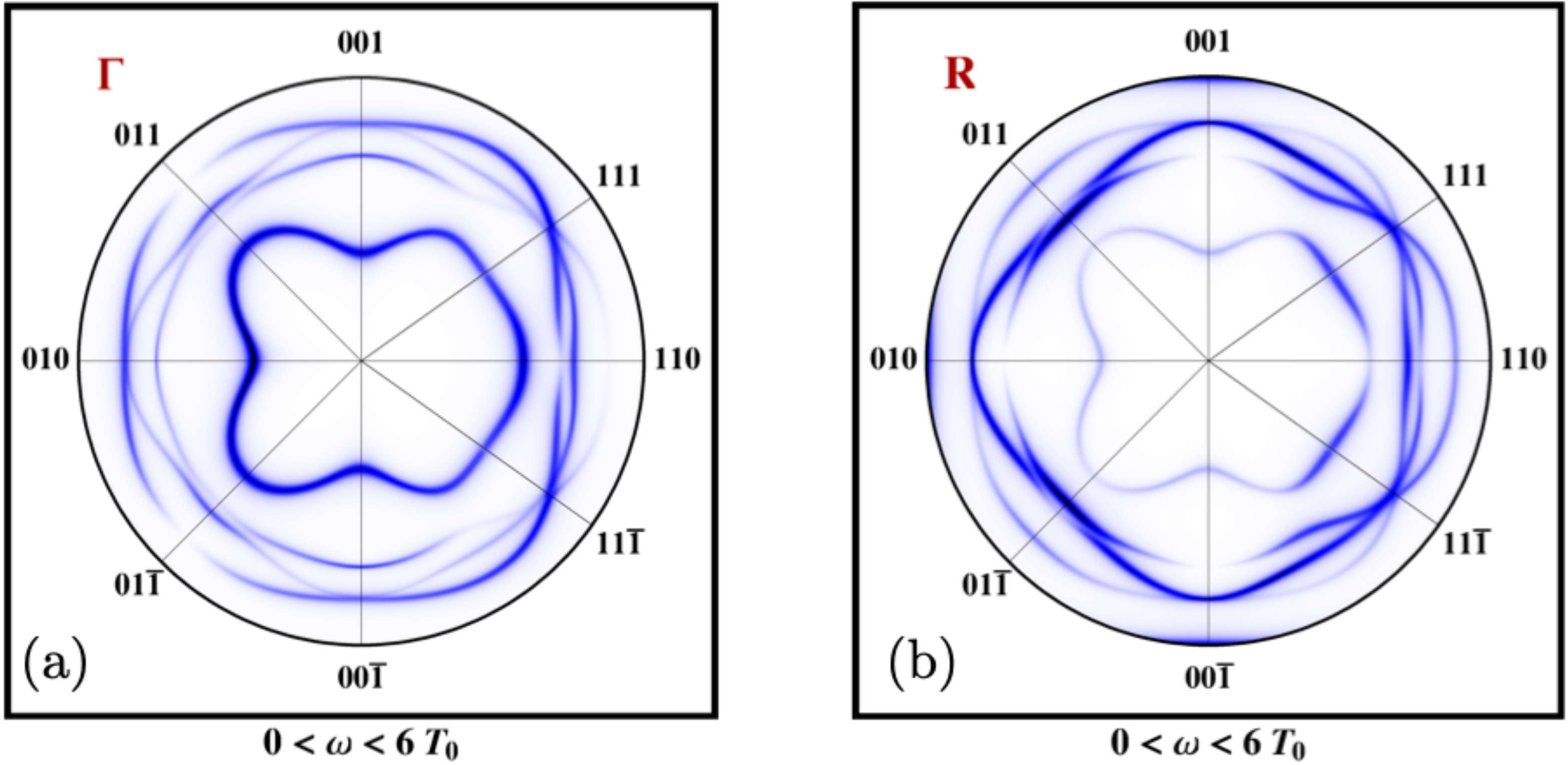}
\end{center}
\caption{Field-angular anisotropies of multipolar excitation branches at $\varGamma$ and $R$ points. A polar presentation is used where angle and radius correspond to field direction and mode energy, respectively ($\omega = 6T_0$ at black circle). Field strength is $h = 2T_0$ (14~T). For such large field angular anisotropy is pronounced and mode frequencies (radii) are more distinct. Some mode intensities are interchanged with $\varGamma\leftrightarrow R$.}
\label{fig:Hrot2}
\end{figure}
%
First we consider the continuous field strength dependence of mode frequencies at two symmetry points shown in Fig.~\ref{fig:fieldstrength} for field along the $[001]$ direction. In this case $O_{xy}$ is selected from the $\Gamma^+_5$ quadrupolar manifold as the primary order. As mentioned before, the high-energy modes $\omega\simeq 4T_0$ stabilized by the octupolar molecular field are hardly influenced by the field. On the other hand, the low-energy modes around  $\omega\simeq T_0$ stabilized directly by the Zeeman term show an approximately linear increase of frequency and splitting with the field. At the zone center $\varGamma$, the mode with highest intensity is the Goldstone mode starting at zero frequency, which then increases roughly linearly with field strength. Experimentally such linear field dependence of excitations at the $\varGamma$ point has been found both in neutron scattering \cite{portnichenko:16} and electron spin resonance \cite{DemishevSemeno06, DemishevSemeno08, DemishevSemeno09}, although the situation is complicated due to the additional AFM order at a different wave vector, which is not included in the present model.

The dependence of multipolar excitations on continuous field rotation at the $\varGamma$ and $R$ points is presented in the polar graphs of Fig.~\ref{fig:Hrot2} for field strength $h'=2$ ($H \simeq 14$~T). In these figures the radial coordinate represents the frequency of the excitation modes in the dipolar INS structure function of Eq.~(\ref{eqn:strucfac}). The angular coordinate  defines the field-rotation angle between the various cubic axes $[100],[110], [111]$ and their equivalents indicated at the outer boundary of the polar plots. The rotation is continuous and closed, however, in general not coplanar. The rotation sequence is chosen to facilitate comparison with present experiments on \CB\ carried out in the same geometry.

The field-angular mode anisotropies show various signatures worthwhile to look for in experiment. Firstly large fields increase the anisotropy in the field-angular variation of mode frequencies (Fig.~\ref{fig:Hrot2}). Furthermore the low energy (L) mode frequencies  (small polar radii) change more rapidly (expand) with increasing field strength than for the high energy modes (outer radii) in accordance with the previous discussion. We also note in both figures that the anisotropy pattern at the $\varGamma$ and $R$ points are quite similar. This is expected since both points have full cubic symmetry preserved. However, remarkably the relative intensity of low energy (small radii) and high energy (large radii) modes is interchanged when going from $\varGamma$ to $R$ and vice versa. At $X$ and $M$ the analogous behaviour is observed.

These polar anisotropy plots of multipolar excitations in \CB\ at BZ symmetry points present a compiled information on mode positions and intensities that may be very useful for comparing with experimental results and give guidance on where to look for the modes with largest intensity. If the model so far accepted for \CB\ with $\boldepsilon=(0.5,0.5)$ is reasonable, some features of the field anisotropy plots described above should be identified in future experiments.

\section{Resonant magnetic excitations in the itinerant CeB$_6$ Kondo lattice}
\label{sect:CeB6_itin}\index{spin exciton mode}\index{CeB$_6$!spin exciton mode|(}

The 4$f$ states in \CB, its ordered phases and excitations have sofar been treated in a completely localized 4$f$ picture. At first look this seems justified because  the valence of \CB\ (2.95+) obtained from photoemission \cite{kakizaki:95}  is very close to integer $3+$ corresponding to stable $4f^1$ configuration with a binding energy $|\epsilon_f|= 2.1\;\text{eV}$. The small deficiency is due to the Kondo resonance state formation above the Fermi level ($\epsilon_{\rm F}=0$) with a narrow width corresponding to a small Kondo temperature $T_\text{K}=4.5$~K \cite{horn:81}. In the lattice this temperature marks the onset of coherent heavy quasiparticle bands \index{heavy fermion metal!quasiparticle bands} with a width and (indirect) hybridization gap \index{hybridization gap}\index{heavy fermion metal!hybridization gap} of order $T_\text{K}$ and a corresponding large mass enhancement $m^*/m\simeq 20$ $(\gamma=250\;\text{mJ/mol K}^2)$ (Table~\ref{tbl:masses}). However, the quasiparticle band width given by $T_\text{K}$ is about the same as the ordering temperature $T_\text{Q}$ for AFQ hidden order.

Therefore it is questionable whether the fully localized approach to the HO phase and its excitations is sufficient. Indeed INS experiments \cite{friemel:12,jang:14} justify this question. They suggest that \CB\ exhibits a magnetic low  energy mode in zero field that has all the basic features of a collective itinerant spin exciton resonance within the hybridization and hidden order gaps: It sharply peaks at an energy $\omega_{\rm r}=0.5$~meV, it is narrowly confined at the simple cubic $R$ point with AFQ wave vector \bQ' and the temperature dependence of $\omega_{\rm r}$ and resonance intensity increase in an order-parameter like fashion with decreasing temperature. Simultaneously  the intensity for $\omega < \omega_{\rm r}$ is depressed, characteristic for a spin gap formation.

Such collective spin exciton modes are ubiquitous within the gap of {\it unconventional} superconductors, including high-$T_{\rm c}$ \cite{eschrig:06}, Fe-pnictide \cite{inosov:10,korshunov:08,onari:10} and heavy-fermion \cite{stock:08,eremin:08,thalmeier:16} superconductors. In this case the sign change of the unconventional superconducting gap $\Delta_{\bk+\bQ}=-\Delta_\bk$ at the SC resonance position \bQ\ is necessary. It ensures a finite coherence factor (matrix element of moment operator) at the gap threshold which results in a pronounced bound state peak in the collective magnetic response at an energy $\omega_{\rm r} < 2\Delta_0$ where $\Delta_0$ is the amplitude of the sign-changing gap function $\Delta_\bk$.

\subsection{Heavy quasiparticle band properties in the PAM}
\label{sect:PAM}\index{heavy fermion metal!quasiparticle bands|(}

Since \CB\ is in the normal state one must conjecture that the hybridization gap \index{hybridization gap}\index{heavy fermion metal!hybridization gap} and the additional gaps introduced by the orderings lead to the necessary singular behavior of the bare magnetic susceptibility to allow for a bound state \cite{thalmeier:16}. This may be described by the mean-field hybridization model of Eq.~(\ref{eq:HAMF}) supplemented by the effect of the molecular fields due to AFQ and AFM order which lead to the additional gapping of the mean-field quasiparticle spectrum \cite{akbari:12}. First we briefly outline the constrained mean-field theory of heavy electron bands in the conventional $\text{SU}(N_f)$ Anderson model \index{Anderson model} with $N_f=4$\,--\,fold degenerate conduction band and 4$f$ quartet ground state. The strong on-site Coulomb repulsion of $f$-electrons $U_{ff}$ eliminates double occupancy of $f$-electrons (Ce) or holes (Yb). This constraint is implemented using the auxiliary slave-boson field $b_i$ at each site that represents the empty (Ce) or full (Yb) 4$f$ shell. It requires the introduction of  a Lagrangian term $\lambda (\sum_mf_{im}^\dg f_{im}^{\phantom{\dagger}}+b_i^\dg b_i^{\phantom{\dagger}})$ with $1\leq m \leq N_f$. However, the constraint is enforced only on the average by performing the mean-field approximation $b=\bra b_i\ket$. The resulting mean-field Hamiltonian is described by
\begin{equation}
{\cal H}
=
\sum\limits_{{\bf k}m}
\varepsilon^c_{{\bf k}}c_{{\bf k}m}^{\dagger}c_{{\bf k}m }^{\phantom{\dagger}}+
\tilde{\varepsilon}^f_{\bf k} f_{{\bf k}m}^{\dagger}f_{{\bf k}m}^{\phantom{\dagger}}
+\tilde{V}_{{\bf k}}\left( c_{{\bf k}m }^{\dagger}f_{{\bf k}m}^{\phantom{\dagger}}
+ {\rm h.c.}\right)
+\lambda(r^2-1).
 \label{eq:HAMF}
\end{equation}\vspace{1pt}

\noindent For the conduction band $\varepsilon^c_{{\bf k}}$ a simple n.n. TB model is used which has the same main nesting \index{nesting}\index{Fermi surface!nesting properties} vector \bQ'~as the true Fermi surface \index{Fermi surface} of Fig.~\ref{fig:REB6_bands}. Furthermore $\lambda$ is a Lagrange parameter introduced to enforce the occupation constraint. It moves the effective $f$-electron level $ \tilde{\ve}^f_{\bk m}$ close to the Fermi level. Likewise the effective hybridization $\tilde{V}_{{\bf k}}$ is strongly reduced by the slave boson mean-field amplitude $\bra b_i\ket = r$. Together we obtain
\bea
\tilde{V}^2_\bk = r^2V^2_\bk = V^2_\bk(1-n_f)
;\;\;\;
\tilde{\ve}^f_{\bk m} &=& \ve^f_{\bk m}+\lam.
\eea
In the following discussion of magnetic response we use a simplified form of the Anderson model \index{Anderson model} that neglects the \bk\ dependence (but not necessarily orbital dependence) of hybridization. This means we are setting $\tilde{V}_\bk = \tilde{V}$. For our purpose this simplification is adequate, but it cannot always be used, e.g. for the derivation of electronic structure in the pseudogap Kondo insulator CeNiSn \cite{ikeda:96} \index{CeNiSn} or the topological insulators like SmB$_6$ (Sec.~\ref{sect:SmB6TI}) \cite{hanzawa:98,takimoto:11,dzero:10}.
\begin{figure}
\includegraphics[width=\textwidth]{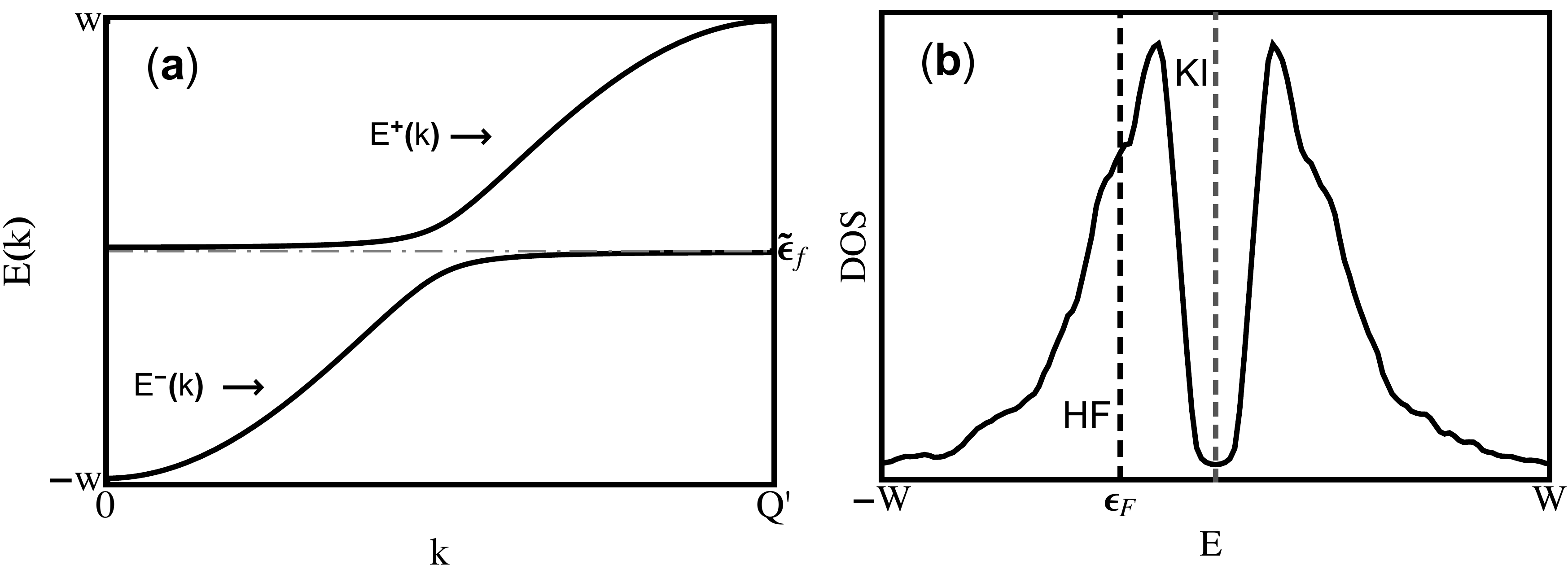}
\caption{(a) The hybridized quasiparticle bands around the renormalized $f$-level $\tilde{\epsilon}_f$ and (b) corresponding DOS for typical model parameters. Fermi level position for heavy-fermion metal (HF, like \CB) and Kondo insulator (KI, like \YB\ and \SB) is indicated. For KI $\epsilon_{\rm F}$ is inside the hybridization charge gap $\Delta_{\rm c}$ (from \cite{thalmeier:16}).}
\label{fig:Quasiband}
\end{figure}
The single particle type mean-field Hamiltonian may be diagonalized and then leads to the quasiparticle Hamiltonian
\be
\bl
{\cal H}_{MF}&= \sum\limits_{i,{\bf k},m,\alpha}\ve^{\alpha}_{{\bf k}}
a^{\dagger}_{\alpha,{\bf k}m}a_{\alpha,{\bf k}m}+\lambda(r^2-1),
\\
\ve^{\pm} _{\bf  k}&=\frac{1}{2}
\bigl[\epsilon^{c}_{{\bf  k}}+\tilde{\epsilon}^{f}_{{\bf  k}}\pm\sqrt{(\epsilon^{c}_{{\bf  k}}
-\tilde{\epsilon}^{f}_{{\bf  k}})^2+4\tilde{V}^2_{{\bf  k}}}\bigr].
\label{eq:HMFqp}
\el
\ee
Here $\ve^{\pm} _{\bf k}$ are the pair ($\alpha =\pm$) of hybridized quasiparticle ($a_{\alpha,{\bf k}m}$) bands, each $N_f$-fold degenerate. The indirect gap in Fig.~\ref{fig:Quasiband} has the size of the Kondo temperature: $\ve^+_0 -\ve^-_{\bQ'}\simeq T_\text{K}$ with $T_\text{K}=W\exp\bigl\{-2/[N_fJN_\text{c}(0)]\bigr\}$. Here $W$, $N_\text{c}(0)$ are conduction band width and DOS, respectively, $J=2V^2/|\ve_f|$ is the on-site exchange constant. The transformation to quasiparticle states is given by
\bea
f_{{\bf k}m }=u_{+, {\bf k}} a_{+,{\bf k}m }+u_{-, {\bf k}} a_{-,{\bf k}m };\;\;\;
c_{{\bf k}m }=u_{-, {\bf k}} a_{+,{\bf k}m }-u_{+, {\bf k}} a_{-,{\bf k}m }.\non\\
\eea
with the coefficients defined by
\bea
2u_{\pm, {\bf k}}^2 =
1\pm (\epsilon^{c}_{{\bf  k}}-\tilde{\epsilon}^{f}_{{\bf  k}})/\sqrt{(\epsilon^{c}_{{\bf  k}}
-\tilde{\epsilon}^{f}_{{\bf  k}})^2+4\tilde{V}^2_{{\bf  k}}}.
\label{eqn:qpcoeff}
\eea
They appear in the matrix elements in the numerator of the expression for the bare magnetic susceptibilities, possibly together with coherence factors of broken symmetry states (hidden order, magnetic or superconducting).
\index{heavy fermion metal!quasiparticle bands|)}

\subsection{Collective spin exciton modes}\label{sect:CeB6spinexc}
\index{spin exciton mode|(}

In a model for the AFQ and/or AFM ordered phases II and III of \CB\ further terms must be included which describe schematically the molecular fields due to orbital and (Kramers pseudo-) spin symmetry breaking due to AFQ and AFM order:
\bea
H_\text{AFQ}=\sum\limits_{{\bf k},\sigma=\ua\da}
\Delta_{{\bf Q}^\prime}(f_{{\bf k},+\sigma}^{\dagger}f_{{\bf k},+{\bf Q}^\prime-\sigma }+f_{{\bf k}, -\sigma}^{\dagger}
f_{{\bf k}+{\bf Q}^\prime,+\sigma}).
\label{eq:AFQ}
\eea
and
\bea
H_\text{AFM}=\sum\limits_{{\bf k}, \tau=\pm} \Delta_{\bf Q}
(f_{{\bf k}, \tau\uparrow }^{\dagger}f_{{\bf k}+{\bf Q}, \tau\downarrow }
+f_{{\bf k}, \tau\downarrow }^{\dagger}f_{{\bf k}+{\bf Q}, \tau\uparrow })
\label{eq:HAFQ-AFM}
\eea
\begin{table}[t]
\tbl{Overview of experimental spin resonance characteristics in heavy-fermion metals and Kondo insulators. Here $\Delta_{\rm c}$ denotes the quasiparticle charge gap equal to the hidden order or hybridization gap (for finite or vanishing $T_{\text{HO}}$), respectively. The resonance appears inside the charge gap  ($\frac{\omega_{\rm r}}{\Delta_{\rm c}} < 1$) around HO or characteristic FS vector \bQ'.\vspace{3pt}}
{\begin{tabular}{@{}lcccccc@{}} \toprule
compound & T$_{\text{HO}}$& $\Delta_{\rm c}$  & $\omega_{\rm r}$  &  $\frac{\omega_{\rm r}}{\Delta_{\rm c}}$  &  \bQ'  & Ref. \\
 & [K] & [meV] & [meV] & & [r.l.u.] & \\ \midrule
CeB$_{6}$  & 3.2 & 1.3  & 0.5  &  0.39  &  $(\fs\fs\fs)$   &  \cite{friemel:12, paulus:85,akbari:12}  \\
SmB$_{6}$  & - & 20  & 14  &  0.7  &  $(\fs 00)$   &  \cite{amorese:19,fuhrmann:15,fuhrmann:14}  \\
YbB$_{12}$ & - & 15 & 15  & $\sim$ 1&  $(\fs\fs\fs)$     & \cite{nemkovski:07,okamura:05,akbari:09}   \\
URu$_2$Si$_2$ & 17.8& 4.1& 1.86  & 0.45&  $(001)$    & \cite{bourdarot:10,aynajian:10,akbari:15}   \\ \botrule
\end{tabular}
}
\begin{tabnote}
\end{tabnote}
\protect\label{tbl:resonance}
\end{table}
Here $\tau=\pm$  is the pseudo-orbital and $\sigma=\ua\da$ the pseudo spin index of the $\Gamma_8$ quartet. For finding the magnetic excitations we first require the bare dipolar susceptibility
 $
 \chi_{0}^{ll^\prime}({\bf q},t)=-
 \theta(t)
 \bra
 T j_{{\bf q}}^{l}(t)j_{-{\bf q}}^{l^\prime}(0)
 \ket,
 $
 where
$
  j_{{\bf q}}^{l}=\sum\limits_{{\bf k}mm^\prime}
  f_{{\bf k}+{\bf q}m} ^{\dagger}
  {\hat M}^{l}_{mm^\prime}
  f_{{\bf k}m^\prime}
$
are  the physical magnetic dipole operators ($l,l^\prime=x,y,z$)
with $ {\hat M}^{z}= (7/6)\hat{\tau}_0 \otimes \hat{\sigma}_z$.
Due to cubic symmetry we can restrict to $ \chi_{0}^{zz}({\bf q},\omega)$ given by ($i\nu\rightarrow \omega+i 0^+$)
 \be
\chi^{zz}_{0}({\bf q},\omega)\propto
 \sum\limits_{\alpha \alpha^\prime{\bf k}m_1m_2}
   ( \hat{ \rho}_{{\bf  k},{\bf q} } ^{\alpha^\prime \alpha})^2
   \int d\omega^\prime
   {\hat G}^{0}_{ss}(i\nu+\omega^\prime)      {\hat G}^{0 }_{s^\prime s^\prime }(\omega^\prime)
 \ee
%
\begin{figure} [t]
\begin{center}
\mbox{\hspace{-2pt}\includegraphics[width=0.53\textwidth]{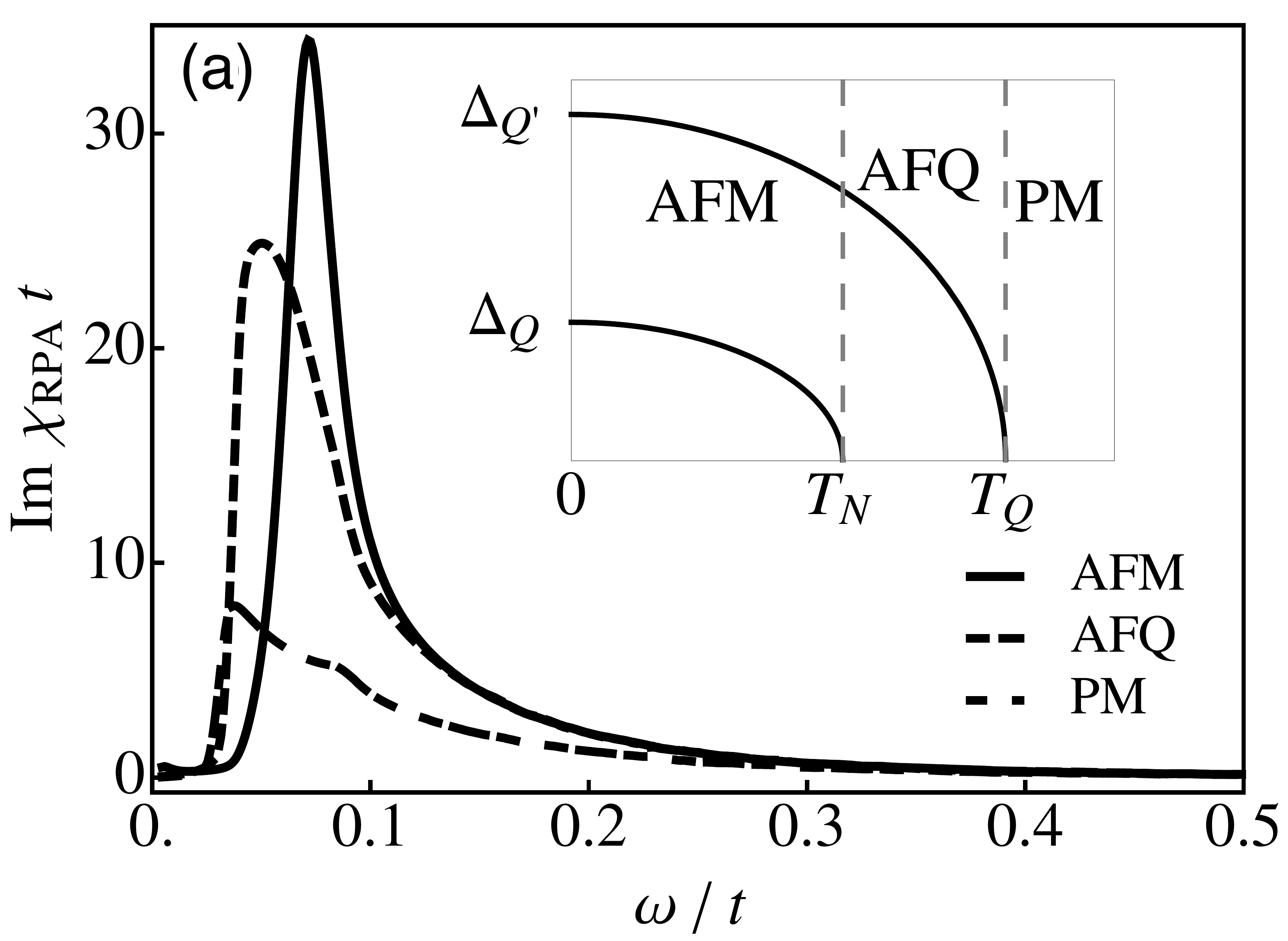}~~\includegraphics[width=0.47\textwidth]{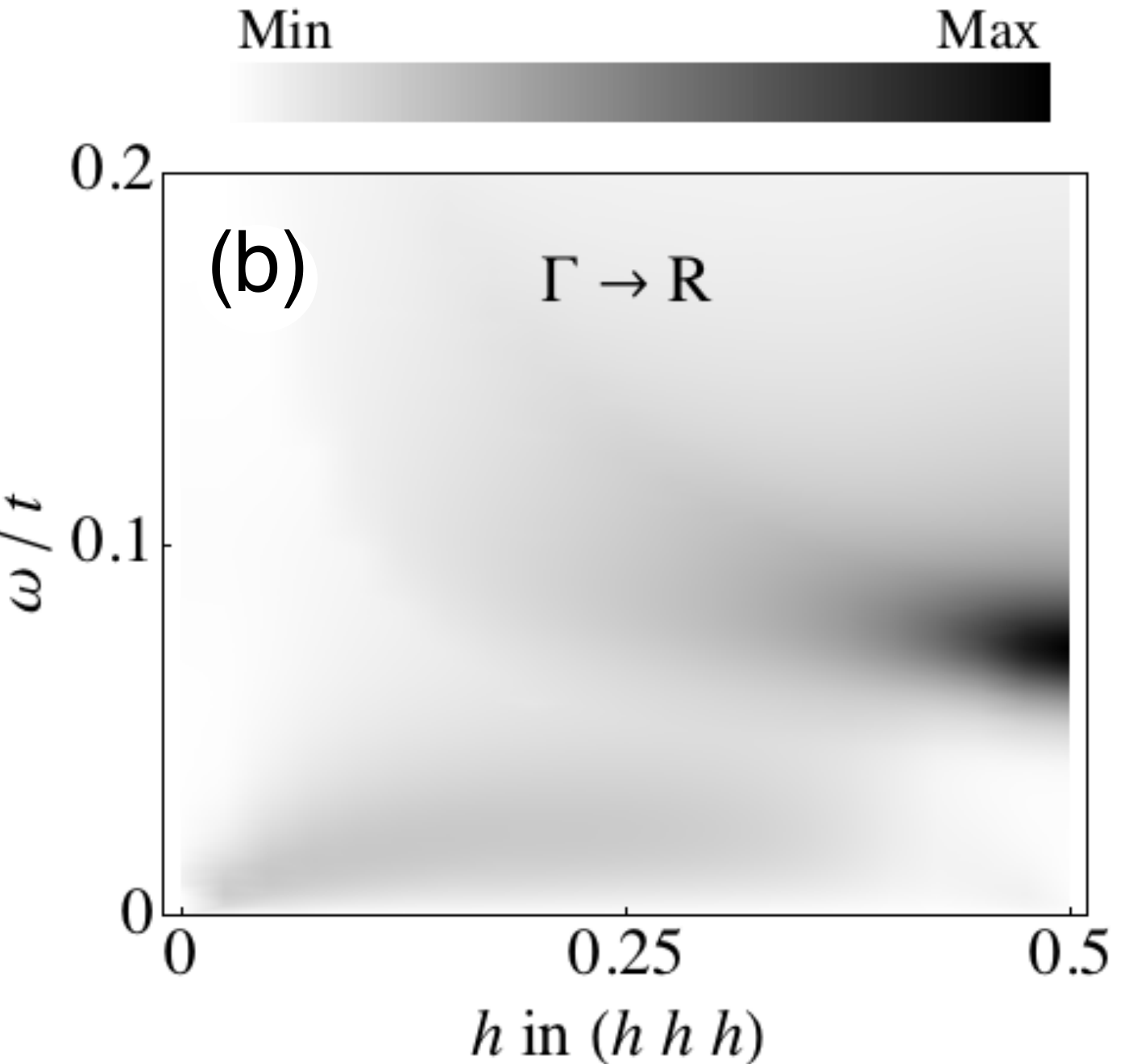}}\vspace{5pt}
\includegraphics[width=0.50\textwidth]{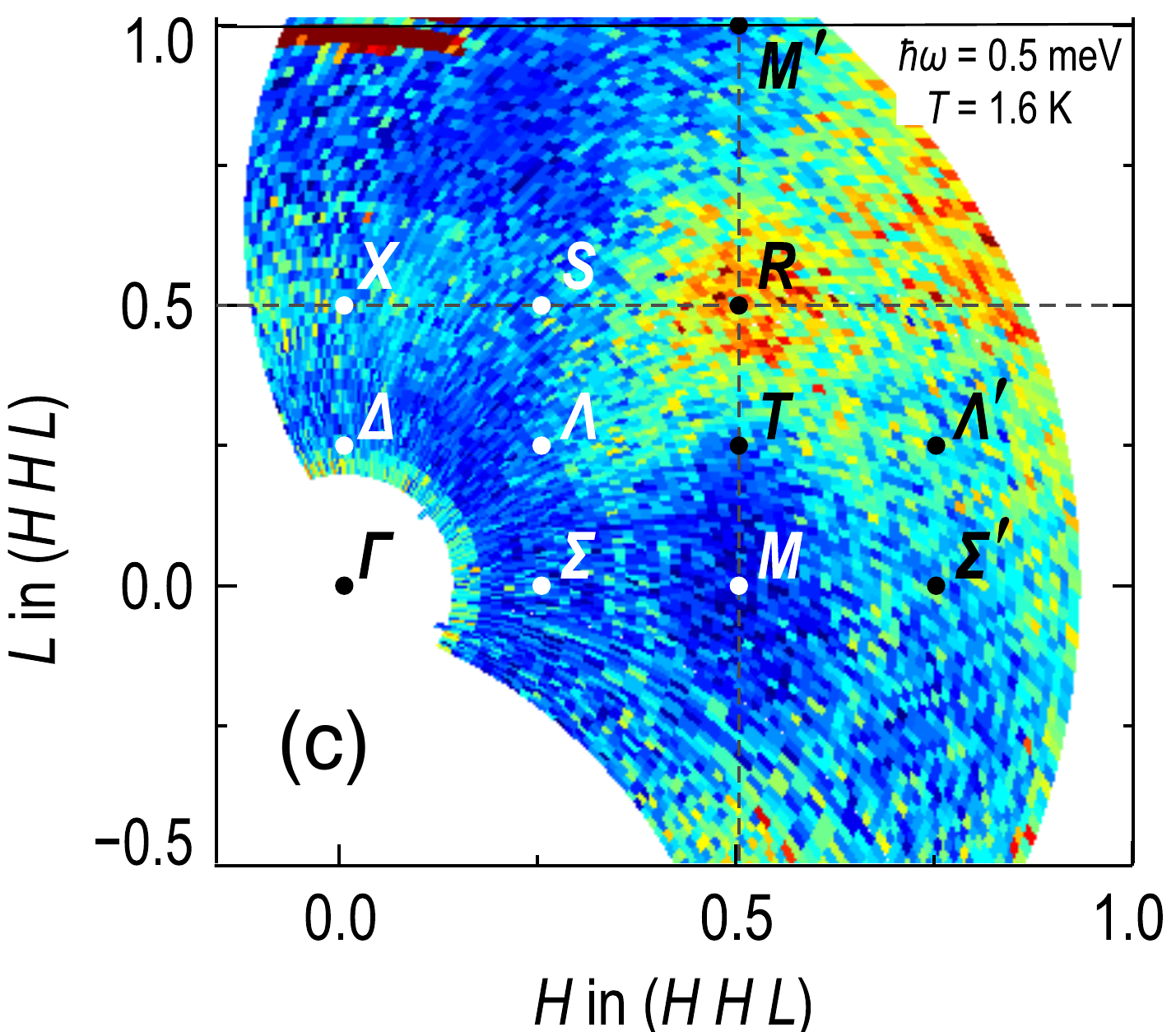}\vspace{-5pt}
\end{center}
\caption{(a)~Spin-exciton resonance peak in the RPA spectrum in the ordered phase developing at the $R$ point $\bQ'=(\frac{1}{2}\frac{1}{2}\frac{1}{2})$ (simple-cubic zone corner) due to AFQ and AFM gaps (inset shows the temperature dependence). $J_\bq$ is chosen as Lorentzian around $R$ point ($t = 22.4~\text{meV}$). (b)~Contour plot of $Im\chi_{\rm RPA}({\bf q},\omega)$ for \CB\ along the $\varGamma R$ direction. Localized resonance peak appears at $R$ for $\omega=\omega_{\rm r}$ and a spin gap develops below (adapted from \cite{akbari:12}). (c)~INS results at the experimental resonance frequency in $(hhl)$ plane. Intensity is narrowly confined at $R$ point resonance. Reproduced from Friemel~\textit{et al.}~\cite{friemel:12}.}
\label{fig:chiCeB6}
\end{figure}
%
where we abbreviate $s= (\alpha,{{\bf  k}+{\bf q}}, m_1 )$ and  $s^\prime=(\alpha^\prime,{{\bf  k}}, m_2)$.
It contains the effect of the modified quasiparticle energies in the Green's functions  ${\hat G}^{0 }_{s s}$ and matrix elements  $\hat{ \rho}_{{\bf  k},{\bf q} } ^{\alpha^\prime \alpha}$ containing the coefficients in Eq.~(\ref{eqn:qpcoeff}) reconstructed by the molecular fields of Eqs.~(\ref{eq:AFQ},\ref{eq:HAFQ-AFM}).
Due to the AFQ and AFM gap openings the  bare magnetic response described by $\chi^{zz}_{0}({\bf q},\omega)''$ is pushed to higher frequencies and the  real part is considerably enhanced. The collective RPA susceptibility,\index{random phase approximation} due to residual quasiparticle interactions described by $J_\bq$, is given by
\bea
 \chi_{\rm RPA}({\bf q},\omega)=
 [1- J_{{\bf q}}  \chi_{0}^{zz}({\bf q},\omega)]^{-1} \chi_{0}^{zz}({\bf q},\omega),
 \label{eq:RPA}
\eea
Once the enhanced $\chi_0^{zz}$ is large enough due to the influence of order parameters the  denominator of $ \chi_{\rm RPA}({\bf q},\omega)$ vanishes and a spin exciton bound state pole develops. This can be seen from Fig.~\ref{fig:chiCeB6}\,(a), where below $T_\text{Q}$ and in particular $T_\text{N}$ a sharp resonance appears around the AFQ ordering vector $\bq\approx \bQ'$ with an energy $\omega_{\rm r}/\Delta_{\rm c} =0.64$  ($T\rightarrow 0$). Here $\Delta_{\rm c}$ is the indirect hybridization charge gap (Table~\ref{tbl:resonance}) \index{heavy fermion metal!hybridization gap}\index{hybridization gap} determined by point contact spectroscopy \cite{paulus:85} and apparent in the DOS of Fig.~\ref{fig:Quasiband}\,(b). The momentum dependence of the spectrum along $[111]$ diagonal $\varGamma R$ line is presented in Fig.~\ref{fig:chiCeB6}\,(b). It demonstrates the confined resonance excitation at $\omega_{\rm r}$ and the signature of the spin gap \mbox{$(\omega\ll\omega_{\rm r})$}, both at the $R$ point. Away from the $R$ point the low energy spin fluctuations of the metallic state are still present. The complementary experimental \CB\ constant-$\omega$ INS intensity plot \cite{friemel:12} for \bq~in the $(hhl)$ scattering plane at $\omega=\omega_{\rm r}(0.5~\text{meV})$ is shown in Fig.~\ref{fig:chiCeB6}\,(c). It exhibits the pronounced accumulation of intensity confined narrowly at the resonance location $R$ corresponding to the peak formation in Fig.~\ref{fig:chiCeB6}\,(a). The above discussion of the dynamic response uses the RPA approach for the ordered phases of \CB. A theory beyond this approximation for the static response in the paramagnetic phase has been proposed in \cite{tazai:19}.
\index{CeB$_6$!spin exciton mode|)}

\section{Dispersive doublet spin exciton mode in the Kondo semiconductor \YB}\label{sect:YbB6}
\index{YbB$_{12}$!spin exciton mode|(}

In \CB\ the resonance is tied to the presence of  hidden and AFM order that enhance $\chi_{0}^{zz}({\bf q},\omega)$ (in experiment it appears only below $T_\text{N}$). This enables the existence of a pole in Eq.~(\ref{eq:RPA}). One might, however, expect that this is not always necessary and that under favorable conditions the resonance may appear already without the support of additional gapping due to order parameters. This case is realized in cubic \YB~\cite{akbari:09}. The compound is a model Kondo semiconductor with equal spin and charge gap of $\Delta_{\rm c} \sim 15\;\mbox{meV}$ \cite{nemkovski:07} and without any symmetry breaking. The 4$f$ hole in Yb$^{3+}$ has a lowest \mbox{$J=7/2$} multiplet which is split by the CEF into a $\Gamma_8^{(1)}$ ground state and two closeby doublets. The latter will be treated as another pseudo-quartet $\Gamma_8^{(2)}$.

\begin{figure} [t]
\begin{center}
\includegraphics[width=0.65\textwidth]{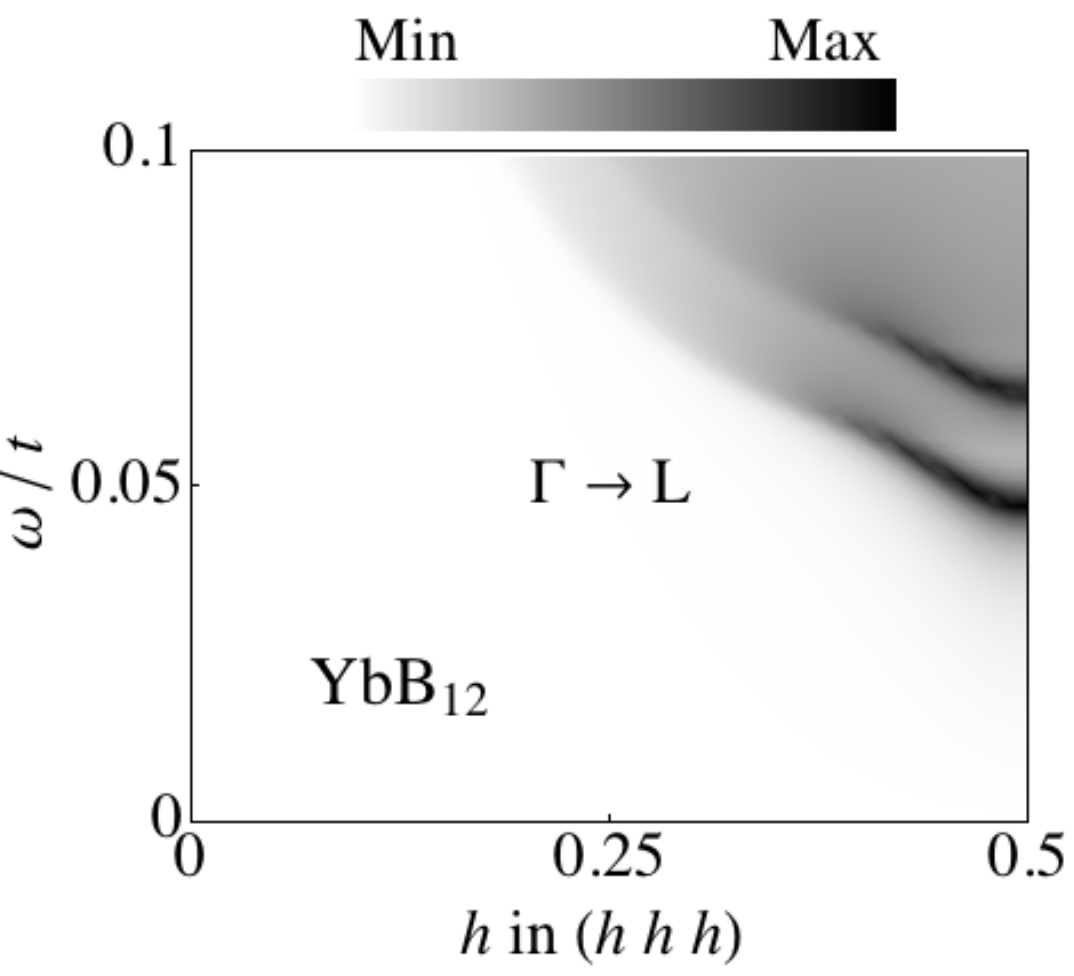}
\caption{Calculated dispersion of split resonance in \YB\ along $\varGamma L$ direction where $L$ is the fcc $(\frac{1}{2}\frac{1}{2}\frac{1}{2})$ point ($t=320$~meV). Dispersion stretches considerably into the BZ because $\omega_{\rm r}\simeq \Delta_{\rm c}$ (Table~\ref{tbl:resonance}), from \cite{akbari:09}.}
\label{fig:YbB12_speccont}
\end{center}
\end{figure}

Then the model in Eq.~(\ref{eq:HAMF}) must be slightly generalized replacing $\epsilon_f \rightarrow \epsilon_f+\Delta_\Gamma$
and $V\rightarrow V_\Gamma$ to include the CEF splitting $\Delta_2-\Delta_1$  of the two quartets and in particular their different average hybridization $V_\Gamma=\frac{1}{2}\bigl(\sum_m|V_{\Gamma m}|^2\bigl)^\frac{1}{2}$. Therefore we obtain two sets of quasiparticle bands with different size of the hybridization gap. \index{hybridization gap}\index{heavy fermion metal!hybridization gap} The associated bare susceptibility is then given by
\be
\chi_0^{\Gamma}({\bf  q},\omega)
=\sum_{{\bf  k},\pm}u^{\Gamma}_{\pm \bk+\bq} u^{\Gamma}_{\mp \bk}
\left[\frac{f\bigl(E^\pm_{\Gamma}({\bf  k}+{\bf  q})\bigr)-f\bigl(E^\mp_{\Gamma}({\bf k})\bigr)}
{E^\mp_{\Gamma}({\bf  k})-E^\pm_{\Gamma}({\bf  k}+{\bf  q})-\omega}\right],
\ee
This implies that the collective RPA susceptibility has contributions from the two sets of quasiparticle bands according to
\bea
\chi_{\rm RPA}({\bf  q},\omega)=\sum_{\Gamma}[1-{\it J}_{\Gamma}({\bf  q})
\chi_0^{\Gamma\Gamma}({\bf  q},\omega)]^{-1}\chi_0^{\Gamma\Gamma}({\bf  q},\omega).
\eea
Therefore one obtains two split collective modes with different energies if the resonance condition is fulfilled for each $\varGamma$. This is shown in Fig.~\ref{fig:YbB12_speccont} where two distinct resonance peaks appear at the $\bQ' =(\frac{1}{2}\frac{1}{2}\frac{1}{2})$ fcc $L$ point right on top of the single particle hybridization gap. (\YB\ has a different cubic structure with a fcc Yb sublattice). This wave vector corresponds to the low energy indirect interband ($\pm$) excitations across the hybridization gap (see illustration in Fig.\ref{fig:Quasiband}). When $|\bq|$ decreases away from the $L$ point the interband excitation energies increase from the indirect band gaps $\sim T_\text{K}$ at \bQ'~to the direct band gap $\sim 2\tilde{V}_\Gamma \gg T_\text{K}$ for $\bq\rightarrow 0$. This results in a decrease of ${\rm Re} \chi_0^{\Gamma\Gamma}({\bf q},\omega)$ for fixed $\omega$ and therefore the resonance condition becomes harder to fulfill for both modes. Ultimately at roughly one third into the BZ the intensity of the spin excitons vanishes (Fig.~\ref{fig:YbB12_speccont}). The upward dispersion of the split modes is again due to the behavior of ${\rm Re} \chi_0^{\Gamma\Gamma}({\bf  q},\omega)$ whose maximum in $\bq,\omega$ plane shifts to larger energies with decreasing $|\bq|$, this also results in larger energies of the two resonances (Fig.~\ref{fig:YbB12_speccont}). The model parameters have been adjusted to obtain the hybridization gap, the observed resonance energies \cite{nemkovski:07} (Table~\ref{tbl:resonance}) and the dispersive features. It is interesting to speculate what would happen if they could be tuned by pressure. A decrease of the gap or an increase in ${\it J}_{\Gamma}({\bf  q}$) might lead to a soft spin exciton mode at the $L$ point and result in an antiferromagnetic Kondo insulator.\index{spin exciton mode|)}\index{YbB$_{12}$!spin exciton mode|)} Although the Kondo semiconductor gap in YbB$_{12}$ is well documented, it shows anomalous transport properties that are not understood to date. Despite the insulating ground state, bulk Shubnikov\,--\,de~Haas (SdH) oscillations with a large effective mass are observed \cite{xiang:18}. Furthermore, thermal conductivity exhibits a linear term in $\kappa/T$ that is not of electronic origin \cite{sato:19}.

\section{Magnetic excitations, topological state in the mixed valent semiconductor SmB$_6$}\label{sect:SmB6}
\index{hybridization gap}\index{heavy fermion metal!hybridization gap}

\begin{figure}
\begin{center}
\includegraphics[width=0.82\textwidth]{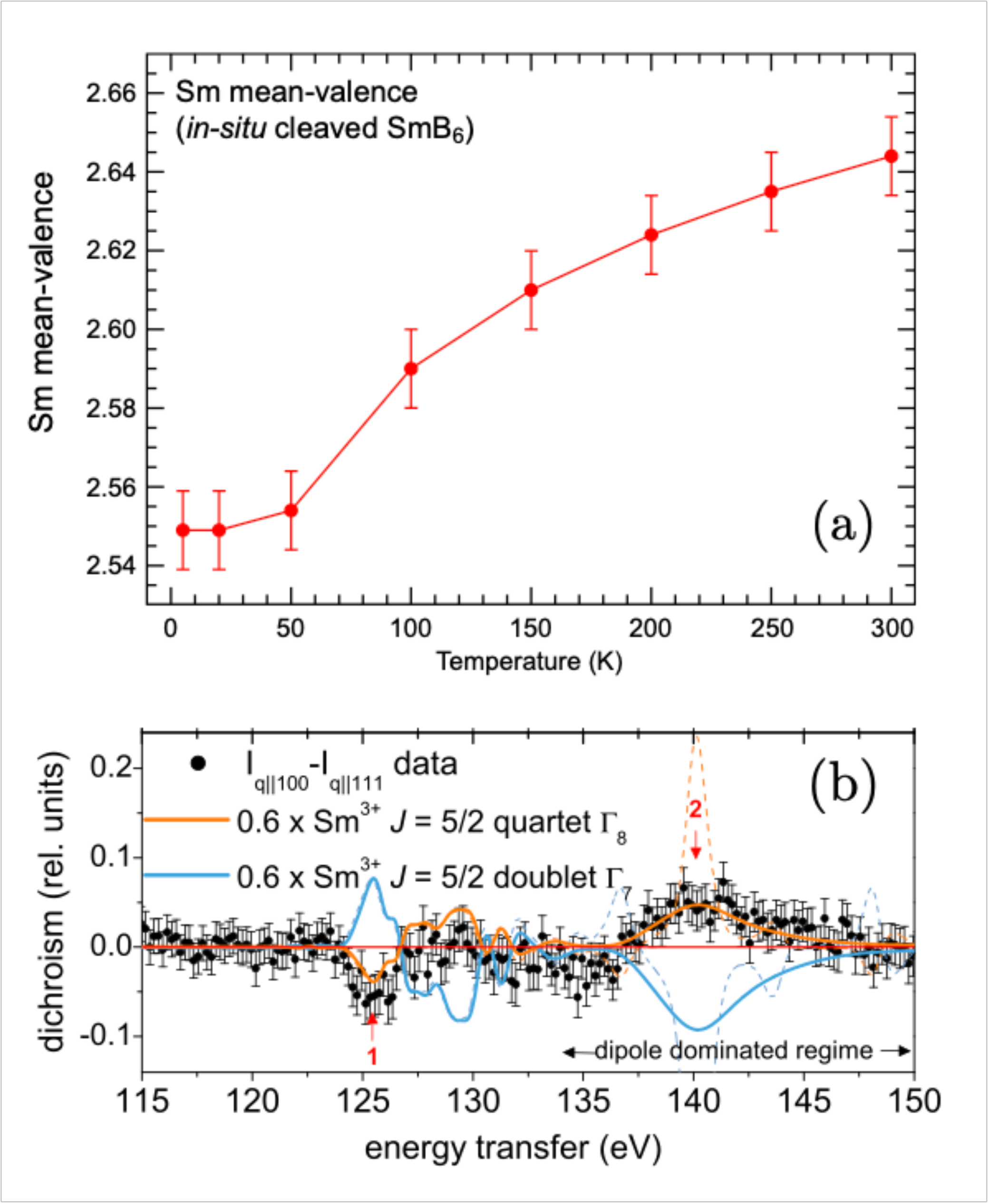}
\caption{(a)~Temperature dependence of Sm valence from HAXPES experiments (from \cite{utsumi:17}) (b) Dichroic spectrum $I_{\bq\parallel 100}-I_{\bq\parallel 111}$ (NIXS data: black dots) and comparison with simulations for $\Gamma_8$ and $\Gamma_7$ ground state. A scale factor 0.6 for the Sm$^{3+}$ part of the MV ground state is applied. Reproduced from Sundermann \textit{et~al.} \cite{sundermann:18}.}
\label{fig:SmB6_xray}
\end{center}
\end{figure}

Among the $R$B$_6$ series, \SB\ is the only strongly mixed valent (MV) compound \index{mixed valence} due to large hybridization of 4$f$ states and 5$d$ conduction states. This has been known for a long time, and the temperature and pressure dependence of the valence has been determined with various means, in particular XAS \cite{mizumaki:09,emi:18} and HAXPES \cite{utsumi:17} (see Fig.~\ref{fig:SmB6_xray}). At low temperature ($\simeq 5$~K) the valence is $v=2.55$, almost intermediate between the $2+$ and $3+$ configurations which have ground states \mbox{$J=0$} and $J=\frac{5}{2}$, respectively. The strong hybridization with conduction electrons which leads to this non-integer valence also generates a hybridization gap of the order $\Delta_{\rm c}=20$~meV schematically shown in Fig.~\ref{fig:Quasiband} within the Anderson model description.\index{Anderson model} Because the resistance shows an activated-type  increase when temperature decreases below $T\simeq 50$~K, somewhat below the gap  temperature scale, it was concluded that \SB\ is a ``Kondo-insulator'' where the Fermi energy $\epsilon_{\rm F}$ in Fig.~\ref{fig:Quasiband} lies inside the hybridization gap.\index{SmB$_6$!charge gap} This designation is now generally used although 'strongly correlated mixed valent semiconductor' would be more precise for \SB. Away from the hybridization region the 5$d$-like bulk conduction bands have dispersions centered around the $X$ points and are associated with constant-energy surfaces that qualitatively resemble much the 5$d$ Fermi surface \index{Fermi surface} in the reference compound LaB$_6$ (Fig.~\ref{fig:REB6_bands}).

\subsection{CEF and collective magnetic excitations}\label{sect:SmB6exc}
\index{SmB$_6$!magnetic excitations}

For describing the 5$d$-4$f$ hybridization in detail an idea about the Sm$^{3+}$ $J=\frac{5}{2}$ states and their CEF splitting is necessary. This question was settled by inelastic x-ray investigations \cite{sundermann:18,amorese:19} which have identified the Sm$^{3+}$ level scheme $\Gamma_8(0)-\Gamma_7(20~\text{meV})$. The symmetry of the ground state was concluded from x-ray dichroism spectra and comparison with simulation from a full multiplet calculation \cite{sundermann:18}. As can be seen in Fig.~\ref{fig:SmB6_xray} the $\Gamma_8$ ground state simulation fits very well to the data. The CEF splitting $\Delta=20\ \text{meV}$ of the upper $\Gamma_7$ level has been determined by RIXS experiments in an indirect manner via the CEF splitting of an excited $^4G^*_{5/2}$ term \cite{amorese:19}. Interestingly it is equal to the observed hybridization gap obtained, e.g, from optical conductivity \cite{gorshunov:99}. This explains also why the CEF splitting has not been found in INS experiments. The width of the quasielastic line in MV or Kondo compounds \index{mixed valence} is given by \cite{horn:81} $\Gamma(T)=T^*+A\sqrt{T}$ where $T^*$ is of order the hybridization gap $\Delta_{\rm c}$. Then naturally when the quasielastic line width is of the same order as the CEF splitting the latter cannot be identified in INS. However, INS did observe a pronounced spin exciton resonance located narrowly at the $X$-point at $\omega_{\rm r} = 14~\text{meV}$  \cite{fuhrmann:14,fuhrmann:15} and inside the hybridization gap. This collective magnetic mode is similar to the one found in the Kondo insulator \YB\ and in the HO/AFM state of the heavy-fermion metal \CB\ (Table~\ref{tbl:resonance}) discussed in the preceding sections.

\subsection{\SB\ as a strongly correlated topological insulator}\label{sect:SmB6TI}
\index{SmB$_6$!topological surface states}\index{topological insulator!strongly correlated|(}\index{Dirac surface states|(}

\begin{figure}
\begin{center}
\includegraphics[width=0.8\textwidth]{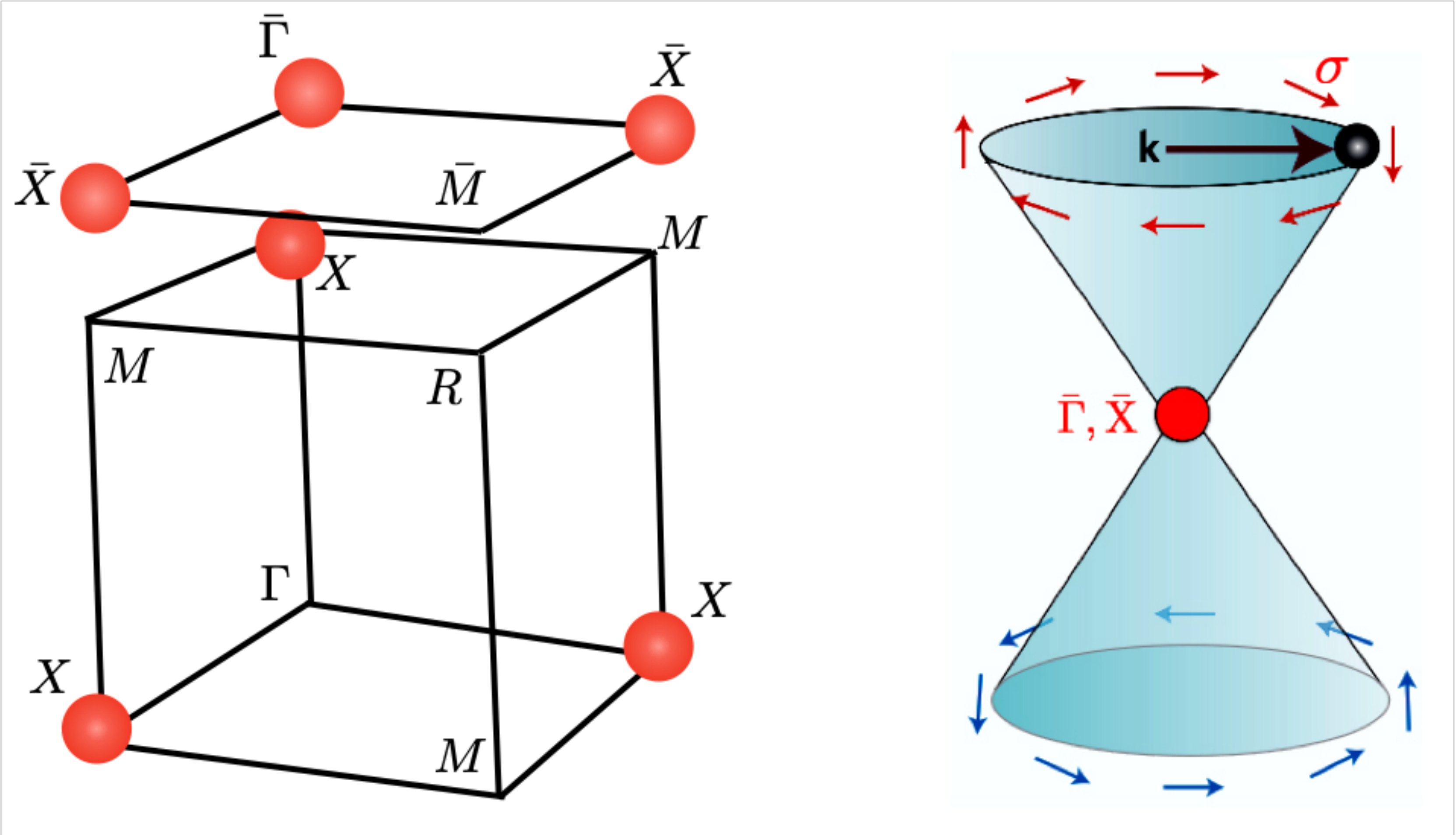}
\caption{3D BZ with TRI points $\varGamma(000)$, $X(\fs 00)$ etc., $M(\fs\fs 0)$ etc. and $R(\fs\fs\fs)$. Projected 2 BZ [along (001)] TRI points are denoted by bars \cite{takimoto:11}. The band inversion along $\varGamma X$ leads to an odd number $(3)$ of 2D isotropic Dirac cones at $\bar{\varGamma},\bar{X}$ in 2D surface projected BZ. Their dispersion $\pm v_F^s|\bk|$ is sketched and helical spin polarization with $\boldsigma\times\hat{\bk}=\pm 1$ indicated.}
\label{fig:TRIpoint}
\end{center}
\end{figure}

Early experiments on \SB\ had shown a puzzling feature: According to optical conductivity the hybridization gap should be well developed and lead to a complete suppression of dc transport. Although the resistance shows activated temperature behaviour  below $T\simeq 50$~K at even lower temperature  $T\leq 4$~K it abruptly saturates at a large but finite value \cite{menth:69,allen:79}. This behaviour was originally attributed to the formation of in-gap impurity bands although the saturation value does not increase with sample quality. However, with the advent of ideas on topological insulators \cite{fu:06,fu:07} it was realized \cite{takimoto:11,dzero:10,kim:13} that the resistance saturation in \SB~may have a more profound origin, namely being due to conducting topologically protected surface states.

The possibility of such states in non-interacting insulators was proposed in the ground-breaking work of Fu and Kane \cite{fu:06,fu:07}. For a 3D band insulator with spin orbit coupling in the presence of inversion $(I)$ and time reversal $(\Theta)$ symmetry the energy bands must be twofold (Kramers) degenerate at the time reversal invariant (TRI) points $\bk_m^*$ that are characterized by $\Theta\bk_m^*=-\bk_m^*+\bG$. Here \bG\ is a reciprocal lattice vector. There are $m=1$--8 such points in the simple cubic lattice of Fig.~\ref{fig:TRIpoint} (appropriate also for \SB\ when we restrict to Sm\,4$f$ and 5$d$ states). Due to inversion symmetry the $n$-th Bloch state at $\bk^*_m$ may be classified by its parity eigenvalue $\delta_m^n=\pm 1$. The set of products  for all {\it occupied} bulk states at a given TRI point $m$
\bea\textstyle
\delta_m=\prod_{(\ve_{n\bk}<\mu)}\delta_m^n=\pm 1
\eea
contains an important information on the dispersion of bands. If $\delta_m$ changes its sign from one TRI point to the next it means that there must have been one or an odd number of opposite-parity band crossings between them which are driven by the spin-orbit coupling. The product over all TRI points
\bea\textstyle
Z_2=\prod_m\delta_m=\pm 1
\eea
defines an Ising or $Z_2$ index \index{topological insulator!$Z_2$ index} that characterizes even $(Z_2=1)$ or odd $(Z_2=-1)$ overall number of band crossings. This 'strong' topological index signifies the absence or presence of topological order in the band insulator (the 'weak' indices associated with BZ surfaces will not be discussed here, see \cite{takimoto:11,legner:14,legner:15}). For $Z_2=1$ the bulk insulating state is designated trivial and non-trivial or topological for $Z_2=-1$. The latter corresponds to an internal 'twist' of the ground state wave function caused by the spin-orbit coupling generated band crossings. An adiabatic deformation of the lattice potential parameters cannot change it into a trivial band insulator, in particular not into the vacuum (loosely a trivial insulator with an arbitrary large gap). Therefore approaching the TI surface, i.e. the boundary to the vacuum the topologically distinct $(Z_2=\pm 1)$ regions can only be joined if the gap on the TI side vanishes at the surface. This implies the necessity of gapless surface states which have attractive features: They are characterized by a massless Dirac dispersion with spin-momentum locking leading to non-degenerate helical states. These surface states are robust and protected as long as the TI state of the bulk prevails. Many excellent realizations of weakly correlated TI \index{topological insulator!weakly correlated} materials and their surface states have been found by now \cite{ando:13} with Bi$_2$Se$_3$ and Bi$_2$Te$_3$ being most prominent examples \cite{hoefer:14}. Here the existence of protected surface states with Dirac dispersion and helical spin momentum locking which forbids backscattering have been verified in countless experiments like spin-resolved ARPES, QHE, SdH, magnetoresistance and STM-quasiparticle interference (QPI).

\begin{figure}
\begin{center}
\includegraphics[width=0.65\textwidth]{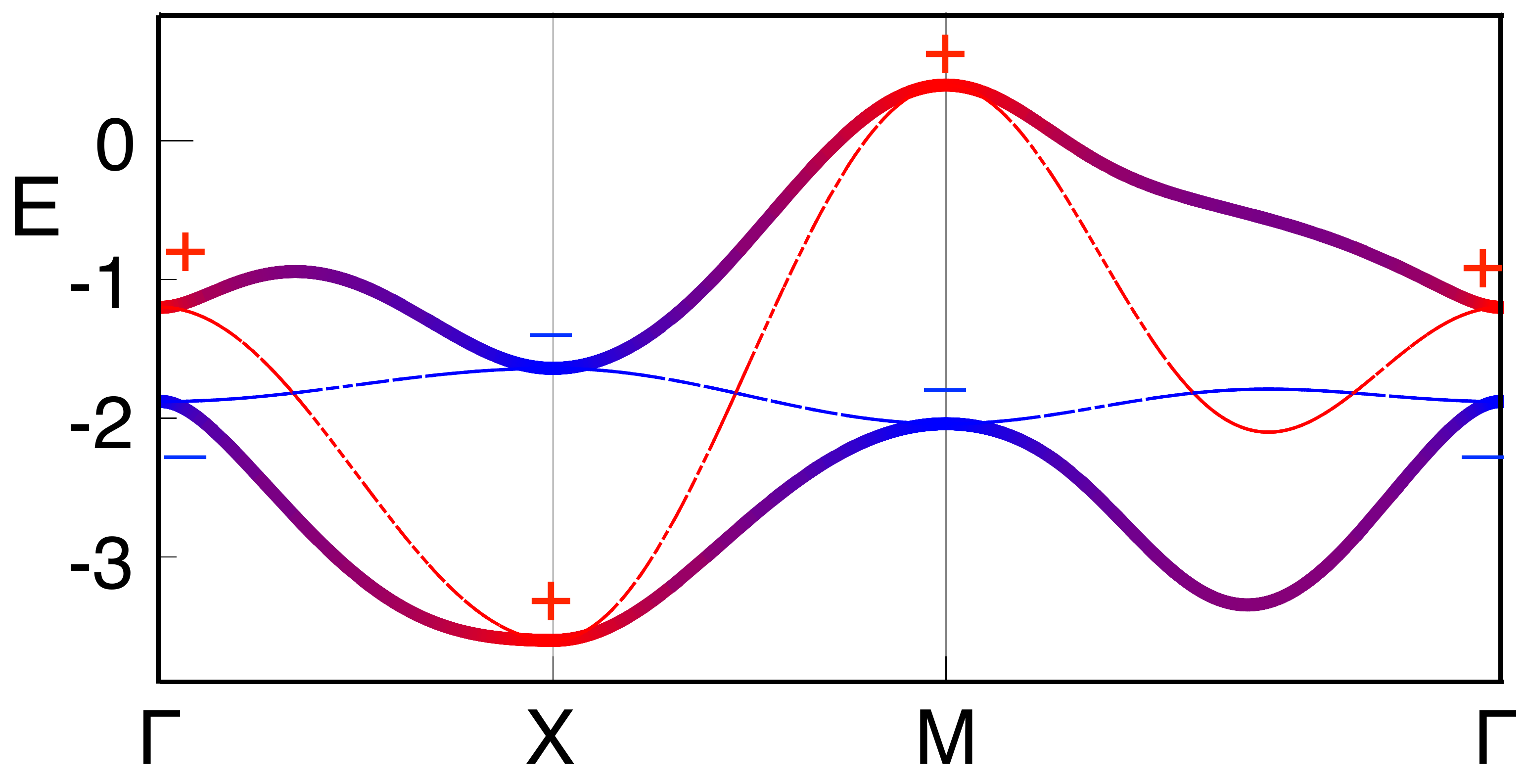}\hfill
\caption{Single $f$,\,$d$-orbital toy model band structure after Eq.~(\ref{eqn:SmB6qp}) corresponding qualitatively to the STI case $(Z_2=-1)$ of \SB. \index{SmB$_6$!toy model}
Parameters are $t_d=1$, $t_f=-0.1$, $t'_{d,f}=-0.4t_{d,f}$, $t''_{d,f}=0$, $\tilde{\epsilon}_f=-2$ and $\tilde{V}_{df}=0.5$. Band crossings (parity $\pm$ exchange) appear along $\varGamma X$ and $MX$ directions but not along $\varGamma M$. Adapted from Legner \textit{et~al.}~\cite{legner:14}.}
\label{fig:STI_disp}
\end{center}
\end{figure}

If this scenario should be applied to the strongly correlated case of the Kondo insulator \SB\ one has to invoke the question of topological classification in the presence of strong 4$f$ Coulomb repulsion $U_{ff}$. In various approximation schemes its effect may be described within the quasiparticle band picture, therefore the same procedure for obtaining  the topological index $Z_2$ may be used as defined above. The quasiparticle bands of \SB\ have been obtained by different approaches: Based on ab-initio LDA calculations \cite{antonov:02,kang:15}, LDA+DMFT \cite{kim:14} or LDA supplemented by Gutzwiller projection \cite{lu:13}. In these cases the full $\Gamma_8, \Gamma_7$ basis for the 4$f$ single particle states is included. Parameterized model calculations starting from the tight-binding picture have also been proposed, in decreasing level of complexity: A TB model based on the 4$f$ $\Gamma_8$ quartet only and including n.n. and n.n.n. hopping and hybridization \cite{takimoto:11} or n.n. processes only \cite{alexandrov:13}. This may be even more simplified by observing that only one of the orbital components of 5$d$ $e_g$ states hybridizes with the symmetry equivalent orbital component of $\Gamma_8$ \cite{kang:15}. Then the TB basis may be further reduced, involving only one (Kramers degenerate) orbital for 5$d$ conduction states as well as for 4$f$ states \cite{legner:14}. This toy model \index{SmB$_6$!toy model} for \SB\ is then similar to the basic Anderson model in the mean-field slave boson approximation (limit of infinite $U_{ff}$) discussed in Sec.~\ref{sect:PAM}. There is one important difference: For the topological index the quasiparticle band ordering at TRI points is essential. Therefore we cannot just assume a \bk-independent effective hybridization constant $\tilde{V}$ as would be appropriate for an {\it on-site} $p$-$f$ hybridization. Here, for $d$-$f$ {\it inter-site} hybridization $\tilde{V}_\bk$ must depend on momentum such that it vanishes at the TRI points due to inversion symmetry. With this modification the toy model leads to two (doubly Kramers degenerate) quasiparticle bands given by \cite{legner:14}:
\be
\bl
\ve^{\pm} _{\bf  k}=&\frac{1}{2}
\bigl[\epsilon^{c}_{{\bf  k}}+\tilde{\epsilon}^{f}_{{\bf  k}}\pm\sqrt{(\epsilon^{c}_{{\bf  k}}
-\tilde{\epsilon}^{f}_{{\bf  k}})^2+4\tilde{V}^2_{{\bf  k}}}\bigr]
\\
\epsilon^c_\bk=&-2t_d c_1(\bk)-4t'_dc_2(\bk)-8t_d''c_3(\bk)
\\
\tilde{\epsilon}^f_\bk=&\tilde{\epsilon}_f -2t_f c_1(\bk)-4t'_fc_2(\bk)-8t_f''c_3(\bk)
\\
\tilde{V}_\bk=&2\tilde{V}_{df}(s_x^2+s_y^2+s_z^2)^\fs
\label{eqn:SmB6qp}
\el
\ee
Here the definitions $s_\alpha=\sin k_\alpha$,  $c_\alpha=\cos k_\alpha$, $c_{\alpha\beta}=c_\alpha c_\beta$   $(\alpha, \beta=x,y,z)$ and furthermore $c_1(\bk)=c_x(\bk)+c_y(\bk)+c_z(\bk)$, $c_2(\bk)=c_{xy}(\bk)+c_{yz}(\bk)+c_{zx}(\bk)$ and $c_3(\bk)=c_x(\bk)c_y(\bk)c_z(\bk)$ were employed. The set of hopping parameters  $(t_d,t'_d,t''_d)$,  $(t_f,t'_f,t''_f)$ (up to 3rd n.n. and the renormalized $f$-orbital energy $\tilde{\epsilon}_f$ define the effective nonhybridized 5$d$ and 4$f$ bands and the {\it inter-site} n.n. hybridization $\tilde{V}_{df}$ mixes them to the two effective quasiparticle bands $\ve^{\pm} _{\bf  k}$. Obviously there is much freedom for parameter choice to model the bands. An example of both nonhybridized (thin dashed lines) and hybridized bands (full lines) is shown in Fig.~\ref{fig:STI_disp}. It demonstrates that i) there is an overall hybridization gap opening ii) there is a band inversion of opposite parity $(\pm)$ states along $\varGamma X$ and $MX$ directions but not along $\varGamma M$ and $\varGamma R$ (not shown) directions when moving in the BZ of Fig.~\ref{fig:TRIpoint}. Then we have $\delta_\varGamma=\delta_M =\delta_R=-1$ and $\delta_X=+1$ leading to $Z_2=\delta_\Gamma^{\phantom{3}}\delta_R^{\phantom{3}}\delta_X^3\delta_M^3=-1$. Therefore for the parameters in Fig.~\ref{fig:STI_disp} the model dispersion of \SB\ has an odd number of band inversions and corresponds to a strong topological insulator. This implies the existence of surface states \cite{hasan:10} which appear at the TRI points $\bar{\Gamma}$ and $\bar{X}$ of the 2D BZ which have 2D projected indices $\delta_{\bar{\Gamma}}=\delta_{\Gamma}\delta_{X}= -1$ and $\delta_{\bar{X}}=\delta_{X}\delta_{M}=-1$ (Fig.~\ref{fig:TRIpoint}). They are described by an effective 2D Hamiltonian leading to massless helical eigenstate
\bea
H_{2D}=\sum_\bk v_F^s(k_x\sigma_y-k_y\sigma_x); \;\;  \;\;
\ve^{\pm} _{s\bk}=\pm v_F^s|\bk|
\eea
where $v_F^s$ is the Fermi velocity (slope) of the Dirac-cone dispersion $\ve^{\pm}_{s\bk}$ which may be different for the $\bar{\Gamma}$ and $\bar{X}$ cones. The eigenstates have a helical spin-momentum locking with $\boldsigma\times\hat{\bk}=\pm 1$. The Dirac cones and the spin polarization of surface states is sketched in Fig.~\ref{fig:TRIpoint}.

As mentioned before the $Z_2$ index (and also the weak topological indices) \index{topological insulator!$Z_2$ index} depends on the model parameters of nonhybridized bands. Other situations are possible and the phase diagram \cite{tran:12,legner:14,legner:15}  in parameter space comprises trivial, weak  and strong topological insulators, only the latter was discussed here. For all TB models studied in this context two aspects have to be kept in mind: Firstly for the constituent bulk Kondo insulator gap to appear it is necessary to have non-vanishing hybridization $\tilde{V}_{df}$, but this is not yet sufficient: In addition the dispersion of nonhybridized $\tilde{\epsilon}^f_\bk$ along $\Gamma X$ must be {\it upward}  (thin blue line in Fig.~\ref{fig:STI_disp}) which poses constraints on the $f$-hopping parameters. Physically it is an effect of the closeby B~2$p$ valence band at $X$ (Fig.\ref{fig:REB6_bands}). Secondly, once this is satisfied the topological index $Z_2$ does {\it not} depend on the hybridization because always $ \tilde{V}_{\bk m}\equiv 0$ at the TRI points due to inversion symmetry. The $Z_2$ index, \index{topological insulator!$Z_2$ index} i.e. the question of trivial vs. topological insulator, which is simply the product of the $\delta_m={\rm sign}(\epsilon^c_{\bk^*_m}-\epsilon^f_{\bk^*_m})$, is then determined solely by the band ordering of the {\it nonhybridized} conduction $d$ band and effective $f$ band energies at the TRI points.

While the bulk TI nature of \SB\ due to 5$d$-4$f$ band crossing has been supported by ARPES \cite{denlinger:14,ramankutty:16} the observation and interpretation of surface state character \cite{jiang:13,xu:14} with this method is still controversial and therefore too early to review.  This may be partly due to materials problems because the \SB\ surfaces tend to  a diverse number of complicated reconstructions \cite{roessler:14,matt:18}. In particular as in YbB$_6$ the polar [001] \index{YbB$_6$!polar surface} surface may involve band bending effects that can lead to 2D (surface confined) states that are not due to the topologically non-trivial band crossing (see Ref.~\cite{ramankutty:16} and references therein). On the other hand, recent STM-QPI investigations \cite{pirie:18, matt:18} reported both the observation of bulk band crossing and hybridization gap formation as well as the existence of surface-state Dirac cones with a proposed large mass enhancement, thus supporting the topological Kondo insulator picture for \SB.
\index{Dirac surface states|)}\index{topological insulator!strongly correlated|)}

\section{Conclusions and outlook}
\label{sect:conclusion}

The cubic rare earth borides show a great variety of exotic ordered  states. This is enabled by the strong correlations due to 4$f$-electron Coulomb repulsion, the subtle influence of hybridization with conduction electrons and the possible high degeneracy of 4$f$ electron CEF ground states like quartets and triplets. Among the more exceptional cases are the heavy-fermion metal \CB\ and diluted \CBL\ with antiferro-type multipolar hidden order, PrB$_6$ with coexistence of magnetic and antiferro-quadrupolar order, the Jahn-Teller compounds DyB$_6$ and HoB$_6$ with ferroquadrupolar phases, the Kondo insulator \SB~with topological order, the spin-polarized semimetal EuB$_6$ and the 4$f$-5$d$ band semiconductor \YBB. The remaining $R$B$_6$ ($R$~=~Nd,\,Gd,\,Tb) exhibit antiferromagnetic order either with the canonical magnetic wave vector $(\fq\fq\fs)$ of the series or the simple type-I ordering vector $(00\fs)$ for the Nd case.

The new types of multipolar hidden order which involve higher than rank-1 (dipole) moments of the 4$f$ shell like quadrupoles and octupoles  cannot be easily identified with conventional diffraction methods. In this respect detailed NMR investigations and in particular the new technique of resonant x-ray scattering  have led to a real progress of understanding hidden order, most prominently in \CB\ and \CBL~but also in the 5$f$ actinide HO compounds.

The single strongly mixed valent compound of the $R$B$_6$ series, \SB\ is so far a unique example of a strongly correlated topological Kondo insulator. This is due to an intricate arrangement and crossing of 5$d$ and effective 4$f$ $\Gamma_8$-type quasiparticle bands in the BZ which leads to a non-trivial topological index $Z_2=-1$ that signifies the existence of massless helical Dirac surface states. While the evidence for band crossing and non-trivial $Z_2$ of bulk insulating state from theoretical investigation and photoemission experiments is convincing, the evidence for the topological helical surface states \cite{jiang:13} is controversial. Recent STM quasiparticle interference results have, however, given direct support for their existence \cite{pirie:18, matt:18}.

\begin{figure}
\begin{center}
\includegraphics[width=0.65\textwidth]{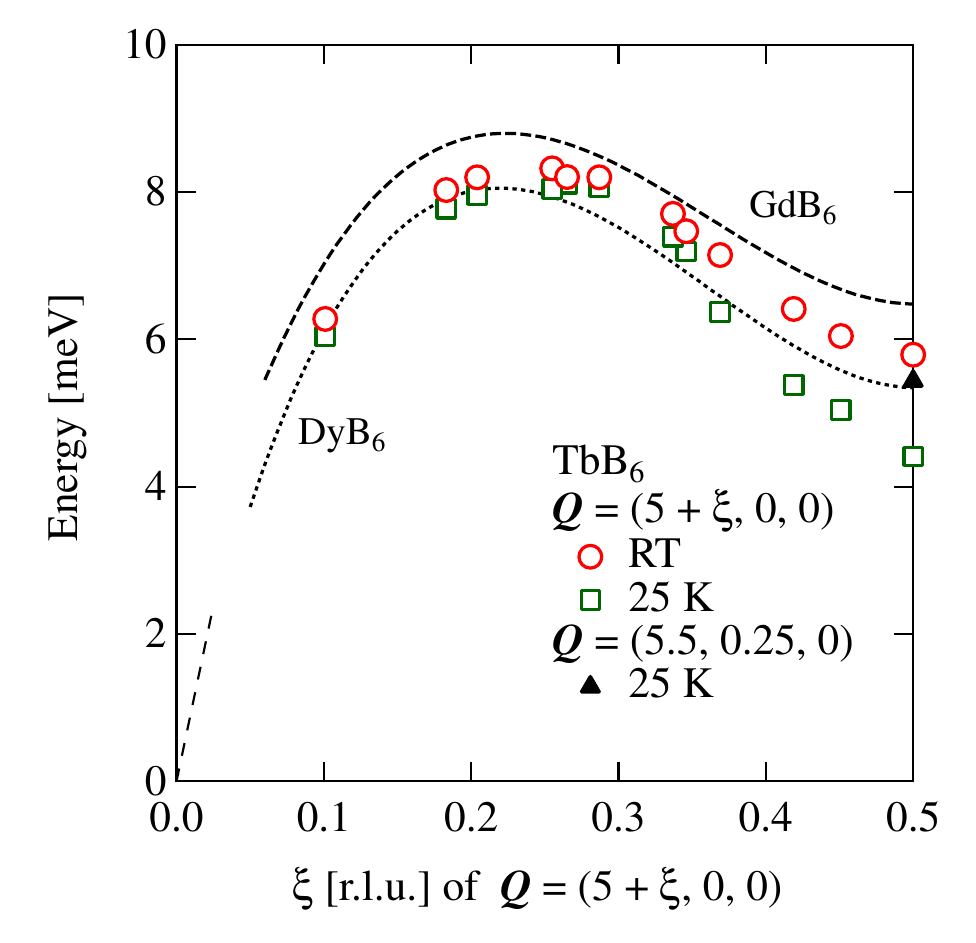}\index{rare-earth hexaborides!phonon anharmonicity}
\caption{Anharmonic phonon dispersion of $R$B$_6$ ($R$~=~Gd,\,Tb,\,Dy). Data for Tb show partial softening of the $L$-point $(\fs\fs\fs)$ (full fcc BZ) phonon with temperature and an essentially flat dispersion due to rattling motion in the oversized cages formed by B$_6$ octahedra. Lines indicate the experimental results for Gd, Dy at room temperature (RT). Reproduced from Iwasa~\textit{et~al.}~\cite{iwasa:14}.}
\label{fig:REB6_phonon}
\end{center}
\end{figure}

In those RE-boride compounds that show a well-defined 4$f$-5$d$ hybridization gap, another important signature of heavy quasiparticle physics \index{heavy fermion metal!quasiparticle bands} was found: The existence of spin-exciton resonances at sharp energies below the gap threshold and confined in \bq-space regions close to symmetry points of the BZ boundary. These resonances are well known from unconventional superconductors where, due to the sign change of the gap functions, a singular magnetic quasiparticle response appears at the gap threshold. This may also happen naturally for the normal state hybridization gap in Kondo compounds. In \CB, due to the metallic state, the assistance of AFQ/AFM order with additional gapping is necessary to stabilize the spin exciton bound state at the simple cubic $R(\fs\fs\fs)$ point. In the semiconducting state of \SB\ no local hidden order (other than topological) is necessary to create the resonance below the hybridization gap threshold $\omega_{\rm r}<\Delta_{\rm c}$ (Table~\ref{tbl:resonance}) at the simple cubic $X(\fs 0 0)$ point. A particular interesting case is cubic \YB: Because of two CEF $\Gamma_8$ (quasi-) quartets two orbitally distinct hybridization gaps lead to a twofold splitting of the spin resonance excitation that starts at the bcc $L$ point $(\fs\fs\fs)$ and disperses upward into the BZ. These spin resonances are beautiful manifestations of gapped heavy quasiparticles and their interactions in Kondo materials. \index{$\Gamma_8$ quartet|)}

In this review mostly electronic properties were considered with only an occasional mention of their coupling to lattice degrees of freedom. The latter would deserve a separate discussion. They have two aspects: Firstly the magnetoelastic coupling \index{magnetoelastic coupling} which leads to elastic constant anomalies, e.g. small effects for AFQ phase transitions as in \CB\ or full softening at Jahn-Teller transition with ensuing ferroquadrupolar order as in HoB$_6$. Secondly, as mentioned in the introduction the lanthanide contraction leads to oversized cages for the heavier RE ions formed by B$_6$ octahedra, starting with Gd (Fig.~\ref{fig:CeB6struc}). This enables a highly anharmonic 'rattling' motion in the cages whose signature are largely flat phonon modes. An example is shown in Fig.~\ref{fig:REB6_phonon}. It is not clear whether these strongly anomalous phonon dispersions are entirely due to the cage rattling effect, or whether strong-coupling of phonons to virtual CEF excitations plays an important role and this topic deserves further investigation.

\begin{figure}[t]
\begin{center}
\includegraphics[width=0.80\textwidth]{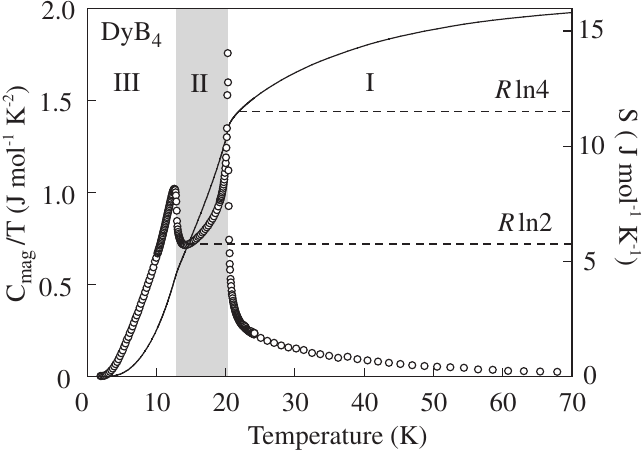}
\caption{Magnetic specific heat of DyB$_4$ identifies two transitions $T_{\rm c1}=20.3$~K and $T_{\rm c1}=12.7$~K. At I/II boundary $J_z$ Ising moment orders. In II J$_{x,y}$ moments and $O_{xy}$ quadrupoles still fluctuate as indicated by remaining entropy and large $c_{44}$ elastic constant softening. At II/III boundary they order and remove remaining entropy. This picture suggests that the two lowest Dy$^{3+}$ Kramers doublets form a quasi-quartet. Reproduced from Watanuki \textit{et~al.} \cite{watanuki:05}.}
\label{fig:DyB4_Cmag}
\end{center}
\end{figure}

With the exception of \YB~only the hexaboride $R$B$_6$ family was discussed in this work. There is, however another extended family of non-cubic $R$B$_4$ tetraborides which has attracted great attention, but for quite different reasons. Their layered structure has the RE ions arranged in quasi-2D sublattices with geometrically frustrated Shastry-Sutherland structure (a network of orthogonal dimers). The low RE site symmetry then always leads to (non-) Kramers CEF doublets, depending on $J$. It has been suggested that the in-plane $xy$ and perpendicular (Ising) $z$ components of the pseudo spin may be ordered/disordered independently due to the geometric frustration in the plane. The Ising component can exhibit metamagnetic behaviour as in \index{rare-earth tetraborides} TbB$_4$ \cite{inami:09,yoshii:08} \index{TbB$_4$}, HoB$_4$~\cite{matas:10} \index{HoB$_4$} and TmB$_4$~\cite{siemensmeyer:08}. \index{TmB$_4$} Of particular interest is DyB$_4$ \index{DyB$_4$} which has been considered as an example of geometrically frustrated dipoles and quadrupoles \cite{watanuki:05} that exhibit separate ordering of Ising and $xy$ magnetic moments while quadrupole moments are still disordered in between the transitions (Fig.~\ref{fig:DyB4_Cmag}). An alternative view has been put forward in \cite{matsumura:11}. As common with 2D geometric frustration effects \cite{schmidt:17} the collective behaviour of moments cannot be fully understood by simple analytic methods but needs to be supplemented by extensive numerical analysis \cite{huo:13}.

Besides the cubic binary rare earth borides, the tetragonal ternary RE borocarbides RB$_2$C$_2$ are another family that exhibit multipolar ordering. Because of their lower symmetry quartet degeneracy of 4$f$ states is already lifted and therefore exotic order is a priori less likely. Nevertheless spontaneous or field induced quadrupolar order has been identified for a few compounds. In particular DyB$_2$C$_2$\index{DyB$_2$C$_2$} shows an AFQ transition [$\bQ'=(00\fs)$] with a exceptionally large $T_\text{Q}=24.7$~K, followed by a magnetic transition at $T_\text{C} = 15.3$~K \cite{yamauchi:99}. The quadrupolar nature of the AFQ phase has been proven by resonant x-ray scattering \cite{hirota:00,matsumura:02}. Due to the large $T_\text{Q}$ and the total entropy release of $4R\ln 2$ one is led to assume that a quasi-quartet of two closeby Kramers doublets form the basis for the quadrupolar and magnetic order. AFQ order has also been suggested for TbB$_2$C$_2$ \cite{kaneko:03}\index{TbB$_2$C$_2$} and HoB$_2$C$_2$ \cite{yanagisawa:05}\index{HoB$_2$C$_2$}, but in contrast to the Dy compound it is only of the field-induced type while at zero field magnetic order prevails.

\section*{Acknowledgments}

\mbox{R.~S.} and \mbox{P.~T.} gratefully acknowledge collaboration with O.~Sakai and H.~Shiba on part of this work. A.~A. and P.~T. also thank D.~S. Inosov and L.~H. Tjeng and his group for fruitful discussions. \mbox{A.~A.} acknowledges financial support through National Research Foundation (NRF) funded by the Ministry of Science of Korea (Grants No.~2017R1D1A1B03033465 and No.~2019R1H1A2039733).

\clearpage

\section*{List of Acronyms}
\begin{tabbing}
Physical Acronyms \= Meaning of Acronyms \= \kill
AFM/AFQ/AFO \> antiferromagnetic/-quadrupolar/-octupolar \\
ARPES \> angle-resolved photoemission spectroscopy \\
BZ   \>  Brillouin zone\\
CEF \> crystal electric field \\
dHvA \> de~Haas\,--\,van~Alphen \\
DMFT \> dynamical mean-field theory \\
DOS \> density of states \\
FM/FQ \> ferromagnetic/ferroquadrupolar\\
FS    \> Fermi surface \\
HAXPES \> hard x-ray photoelectron spectroscopy \\
HF   \> heavy fermion \\
HFM  \> heavy-fermion metal \\
HO \> hidden order \\
HP  \> Holstein-Primakoff \\
IC \> incommensurate \\
INS   \> inelastic neutron scattering \\
JT \> Jahn-Teller \\
LDA \> local density approximation \\
MV  \> mixed valence \\
n.n./n.n.n. \> nearest/next-nearest neighbor \\
ND   \> neutron diffraction \\
NIXS/RIXS \> non-/resonant inelastic x-ray scattering \\
NMR/ESR  \> nuclear magnetic/electron spin resonance \\
PAM \> periodic Anderson model\\
QHE  \> quantum Hall effect \\
QPI \> quasiparticle interference \\
RE \> rare earth \\
RKKY \>  Ruderman-Kittel-Kasuya-Yosida \\
RPA \> random phase approximation \\
RXD \> resonant x-ray diffraction \\
SdH \> Shubnikov\,--\,de~Haas \\
STM  \> surface tunneling microscopy \\
TI \> topological insulator \\
TRI \> time reversal invariant\\
XAS/XD \> x-ray absorption spectroscopy/x-ray diffraction
\end{tabbing}


\bibliography{Chapter_Thalmeier}     

\begin{thebibliography}{100}
\newcommand{\enquote}[1]{``#1''}
\providecommand{\url}[1]{\texttt{#1}}
\providecommand{\urlprefix}{URL }
\providecommand{\eprint}[2][]{\url{#2}}

\bibitem{kusunose:08}
Kusunose, H.; \enquote{Description of multipole in $f$-electron systems};
  \emph{J.~Phys. Soc. Jpn.} \textbf{77}, 064710 (2008).

\bibitem{santini:09}
Santini, P., Carretta, S., Amoretti, G., Caciuffo, R., Magnani, N., and Lander,
  G.~H.; \enquote{Multipolar interactions in $f$-electron systems: The paradigm
  of actinide dioxides}; \emph{Rev. Mod. Phys.} \textbf{81}, 807 (2009).

\bibitem{erkelens:87}
Erkelens, W. A.~C., Regnault, L.~P., Burlet, P., and Rossat-Mignod, J.;
  \enquote{Neutron scattering study of the antiferroquadrupolar ordering in
  {CeB$_6$} and {Ce$_{0.75}$La$_{0.25}$B$_6$}}; \emph{J.~Magn. Magn. Mater.}
  \textbf{63--64}, 61 (1987).

\bibitem{kuwahara:07}
Kuwahara, K., Iwasa, K., Kohgi, M., Aso, N., Sera, M., and Iga, F.;
  \enquote{Detection of neutrons scattering from phase {IV} of
  {Ce$_{0.7}$La$_{0.3}$B$_6$}: a confirmation of the octupole order};
  \emph{J.~Phys. Soc. Jpn.} \textbf{76}, 093702 (2007).

\bibitem{nagao:01}
Nagao, T. and Igarashi, J.; \enquote{Resonant x-ray scattering from the
  quadrupolar ordering phase of {CeB$_6$}}; \emph{J.~Phys. Soc. Jpn.}
  \textbf{70}, 2892 (2001).

\bibitem{matsumura:09}
Matsumura, T., Yonmura, T., Kunimori, K., Sera, M., and Iga, F.;
  \enquote{Magnetic field induced 4f octupole in {CeB$_6$} probed by resonant
  X-ray diffraction}; \emph{Phys. Rev. Lett.} \textbf{103}, 017203 (2009).

\bibitem{nagao:06}
Nagao, T. and Igarashi, J.; \enquote{Electric quadrupole contribution to
  resonant x-ray scattering: application to multipole ordering phases in
  {Ce$_{1-x}$La$_{x}$B$_6$}}; \emph{Phys. Rev.~B} \textbf{74}, 104404 (2006).

\bibitem{nakamura:96}
Nakamura, S., Goto, T., and Kunii, S.; \enquote{Ultrasonic investigation of
  quadrupolar response in {Kondo} system {Ce$_x$La$_{1-x}$B$_6$}};
  \emph{Physica~B} \textbf{219--220}, 89 (1996).

\bibitem{yanagisawa:18}
Yanagisawa, T., Mombetsu, S., Hidaka, H., Amitsuka, H., Cong, P.~T., Yasin, S.,
  Zherlitsyn, S., Wosnitza, J., Huang, K., Kanchanavatee, N., Janoschek, M.,
  Maple, M.~B., and Aoki, D.; \enquote{Search for multipolar instability in
  {URu$_2$Si$_2$} studied by ultrasonic measurements under pulsed magnetic
  field}; \emph{Phys. Rev.~B} \textbf{97}, 155137 (2018).

\bibitem{takigawa:83}
Takigawa, M., Yasuoka, H., Tanaka, T., and Ishizawa, Y.; \enquote{{NMR} study
  on the spin structure of {CeB$_6$}}; \emph{J.~Phys. Soc. Jpn.} \textbf{52},
  728 (1983).

\bibitem{shiina:97}
Shiina, R., Shiba, H., and Thalmeier, P.; \enquote{Magnetic-field effects on
  quadrupolar ordering in a {$\Gamma_8$}-quartet system {CeB$_6$}};
  \emph{J.~Phys. Soc. Jpn.} \textbf{66}, 1741 (1997).

\bibitem{shiina:07}
Shiina, R., Sakai, O., and Shiba, H.; \enquote{Magnetic form factor of elastic
  neutron scattering expected for octupolar phases in
  {Ce$_{0.7}$La$_{0.3}$B$_6$} and {NpO$_2$}}; \emph{J.~Phys. Soc. Jpn.}
  \textbf{76}, 094702 (2007).

\bibitem{ikeda:12}
Ikeda, H., Suzuki, M.-T., Arita, R., Takimoto, T., Shibauchi, T., and Matsuda,
  Y.; \enquote{Emergent rank-5 nematic order in {URu$_2$Si$_2$}}; \emph{Nat.
  Phys.} \textbf{8}, 528 (2012).

\bibitem{shibauchi:14}
Shibauchi, T., Ikeda, H., and Matsuda, Y.; \enquote{Broken symmetries in
  {URu$_2$Si$_2$}}; \emph{Philos. Mag.} \textbf{94}, 3747 (2014).

\bibitem{thalmeier:14}
Thalmeier, P., Takimoto, T., and Ikeda, H.; \enquote{Itinerant multipolar order
  in {URu$_2$Si$_2$} and its signature in magnetic and lattice properties};
  \emph{Phil. Mag.} \textbf{32--33}, 3863 (2014).

\bibitem{akbari:15}
Akbari, A. and Thalmeier, P.; \enquote{Collective spin resonance excitation in
  the gapped itinerant multipole hidden order phase of {URu$_2$Si$_2$}};
  \emph{Phys. Rev.~B} \textbf{92}, 094512 (2015).

\bibitem{kuramoto:09}
Kuramoto, Y., Kusunose, H., and Kiss, A.; \enquote{Multipole orders and
  fluctuations in strongly correlated electron systems}; \emph{J.~Phys. Soc.
  Jpn.} \textbf{78}, 072001 (2009).

\bibitem{takimoto:06}
Takimoto, T.; \enquote{Antiferro-hexadecapole scenario for metal-insulator
  transition in {PrRu$_4$P$_{12}$}}; \emph{J.~Phys. Soc. Jpn.} \textbf{75},
  034714 (2006).

\bibitem{shiina:13}
Shiina, R.; \enquote{Theory of metal-insulator transition and unconventional
  magnetic ordering in {SmRu$_4$P$_{12}$}}; \emph{J.~Phys. Soc. Jpn.}
  \textbf{82}, 083713 (2013).

\bibitem{onimaru:16}
Onimaru, T. and Kusunose, H.; \enquote{Exotic quadrupolar phenomena in
  non-{Kramers} doublet systems}; \emph{J.~Phys. Soc. Jpn.} \textbf{85}, 082002
  (2016).

\bibitem{zerec:04}
Zerec, I., Keppens, V., McGuire, M.~A., Mandrus, D., Sales, B.~C., and
  Thalmeier, P.; \enquote{Four-well tunneling states and elastic response of
  clathrates}; \emph{Phys. Rev. Lett.} \textbf{92}, 185502 (2004).

\bibitem{iwasa:11}
Iwasa, K., Igarashi, R., Saito, K., Laulhe, C., Orihara, T., Kunii, S.,
  Kuwahara, K., Nakao, H., Murakami, Y., Iga, F., Sera, M., Tsutsui, S.,
  Uchiyama, H., and Baron, A. Q.~R.; \enquote{Motion of the guest ion as
  precursor to the first-order phase transition in the cage system {GdB$_6$}};
  \emph{Phys. Rev.~B} \textbf{B\,84}, 214308 (2011).

\bibitem{iwasa:14}
Iwasa, K., Iga, F., Yonemoto, A., Otomo, Y., Tsutsui, S., and Baron, A. Q.~R.;
  \enquote{Universality of anharmonic motion of heavy rare-earth atoms in
  hexaborides}; \emph{J.~Phys. Soc. Jpn.} \textbf{83}, 094604 (2014).

\bibitem{serebrennikov:16}
Serebrennikov, D.~A., Clementyev, E.~S., and Alekseev, P.~A.; \enquote{Analysis
  of the crystal lattice instability for cage-cluster systems using superatom
  model}; \emph{J. Exp. Theor. Phys.} \textbf{123}, 452 (2016).

\bibitem{luethi:84}
L\"uthi, B., Blumenr\"oder, S., Hillebrands, B., Zirngiebl, E., G\"untherodt,
  G., and Winzer, K.; \enquote{Elastic and magnetoelastic effects in
  {CeB$_6$}}; \emph{Z.~Phys.~B} 31 (1984).

\bibitem{ramankutty:16}
Ramankutty, S.~V., de~Jong, N., Huang, Y.~K., Zwartsenberg, B., Massee, F.,
  Bay, T.~V., Golden, M.~S., and Frantzekakis, E.; \enquote{Comparitive study
  of rare earth hexaborides using high resolution angle-resolved
  photoemission}; \emph{J.~Electron Spectrosc. Relat. Phenom.} \textbf{208}, 43
  (2016).

\bibitem{tan:15}
Tan, B.~S., Hsu, Y.-T., Zeng, B., Hatnean, M.~C., Harrison, N., Zhu, Z.,
  Hartstein, M., Kiourlappou, M., Srivastava, A., Johannes, M.~D., Murphy,
  T.~P., Park, J.-H., Balicas, L., Lonzarich, G.~G., Balakrishnan, G., and
  Sebastian, S.~E.; \enquote{Unconventional {Fermi} surface in an insulating
  state}; \emph{Science} \textbf{349}, 287 (2015).

\bibitem{onuki:89}
Onuki, Y., Komatsubara, T., Reinders, P. H.~P., and Springford, M.;
  \enquote{Fermi surface and cyclotron mass of {CeB$_6$}}; \emph{J.~Phys. Soc.
  Jpn.} \textbf{58}, 3698 (1989).

\bibitem{kubo:93}
Kubo, Y., Asano, S., Harima, H., and Yanase, A.; \enquote{Electronic structure
  and the {Fermi} surfaces of antiferromagnetic {NdB$_6$}}; \emph{J.~Phys. Soc.
  Jpn.} \textbf{62}, 205 (1993).

\bibitem{joss:87}
Joss, W., van Ruitenbeek, J.~M., Crabtree, G.~W., Tholence, J.~L., van Deursen,
  A. P.~J., and Fisk, Z.; \enquote{Observation of the magnetic field dependence
  of the cyclotron mass in the {Kondo} lattice {CeB$_6$}}; \emph{Phys. Rev.
  Lett.} \textbf{59}, 1609 (1987).

\bibitem{hutchings:64}
Hutchings, M.~T.; \emph{{\rm ``Point-charge calculations of energy levels of
  magnetic ions in crystalline electric fields''}}; in F. Seitz and D. Turnbull
  (Eds.), \textit{Solid State Physics}, vol.~16, p.~227,  (Academic Press, New
  York,~1964).

\bibitem{lea:62}
Lea, K.~R., Leask, M. J.~M., and Wolf, W.~P.; \enquote{The raising of angular
  momentum degeneracy of $f$-electron terms by cubic crystal fields}; \emph{J.
  Phys. Chem. Solids} \textbf{23}, 1381 (1962).

\bibitem{sundermann:18}
Sundermann, M., Yavas, H., Chen, K., Kim, D.~J., Fisk, Z., Kasinathan, D.,
  Haverkort, M.~W., Thalmeier, Severing, A., and Tjeng, L.~H.; \enquote{4$f$
  crystal field ground state of the strongly correlated topological insulator
  {SmB$_6$}}; \emph{Phys. Rev. Lett.} \textbf{120}, 016402 (2018).

\bibitem{hamamoto:17}
Hamamoto, S., Fujioka, S., Kanai, Y., Yamagami, K., Nakatani, Y., Nakagawa, K.,
  Fujiwara, H., Kiss, T., Higashiya, A., Yamasaki, A., Kadano, T., Imada, S.,
  Tanaka, A., Tamasaku, K., Yabashi, M., Ishikawa, T., Matsumoto, K.~T.,
  Onimaru, T., Takabatake, T., and Sekiyama, A.; \enquote{Linear dichroism in
  angle-resolved core-level photoemission spectra reflecting 4$f$ ground-state
  symmetry of strongly correlated cubic {Pr} compounds}; \emph{J.~Phys. Soc.
  Jpn.} \textbf{86}, 123703 (2017).

\bibitem{amorese:19}
Amorese, A., Stockert, O., Kummer, K., Brookes, N.~B., Kim, D.-J., Fisk, Z.,
  Haverkort, M.~W., Thalmeier, P., Tjeng, L.~H., and Severing, A.; \enquote{A
  RIXS investigation of the crystal-field splitting of {Sm$^{3+}$} in
  {SmB$_6$}}; \emph{arXiv:1901.10808}  (2019).

\bibitem{loewenhaupt:86}
Loewenhaupt, M. and Prager, M.; \enquote{Crystal fields in {PrB$_6$} and
  {NdB$_6$}}; \emph{Z.~Phys.~B} \textbf{62}, 195 (1986).

\bibitem{teitelbaum:76}
Teitelbaum, H.~H. and Levy, P.~M.; \enquote{Indirect multipole interactions in
  metallic rare-earth compounds}; \emph{Phys. Rev.~B} \textbf{14}, 3058 (1976).

\bibitem{schmitt:84}
Schmitt, D. and Levy, P.~M.; \enquote{Ab initio calculation of indirect
  multipolar interactions in {DyZn}}; \emph{Phys. Rev.~B} \textbf{29}, 2850
  (1984).

\bibitem{schlottmann:00}
Schlottmann, P.; \enquote{{RKKY} interaction between {Ce} ions in
  {Ce$_6$La$_{1-x}$B$_6$}}; \emph{Phys. Rev.~B} \textbf{62}, 10067 (2000).

\bibitem{kuramoto:02}
Kuramoto, Y. and Kubo, K.; \enquote{Interpocket polarization model for magnetic
  structures in rare-earth boride compounds}; \emph{J.~Phys. Soc. Jpn.}
  \textbf{71}, 2633 (2002).

\bibitem{yamada:19}
Yamada, T. and Hanzawa, K.; \enquote{Derivation of {RKKY} interaction between
  multipole moments in {CeB$_6$} by the effective {Wannier} model based on the
  bandstructure calculation}; \emph{J.~Phys. Soc. Jpn.} \textbf{88}, 084703
  (2019).

\bibitem{shiba:99}
Shiba, H., Sakai, O., and Shiina, R.; \enquote{Nature of {Ce-Ce} interaction in
  {CeB$_6$} and its consequences}; \emph{J.~Phys. Soc. Jpn.} \textbf{68}, 1988
  (1999).

\bibitem{koitzsch:16}
Koitzsch, A., Herming, N., Knupfer, M., B\"uchner, B., Portnichenko, P.~Y.,
  Dukhnenko, A.~V., Shitsevalova, N.~Y., Filipov, V.~B., Lev, L.~L., Strocov,
  V.~N., Ollivier, J., and Inosov, D.~S.; \enquote{Nesting-driven multipolar
  order in {CeB$_6$} from photoemission tomography}; \emph{Nat. Commun.}
  \textbf{7} (2016).

\bibitem{thalmeier:91}
Thalmeier, P. and L\"uthi, B.; \emph{{\rm ``The electron-phonon interaction in
  intermetallic compounds''}}; chap.~96 in K. A. Gschneidner Jr. and LeRoy
  Eyring (Eds.), \textit{Handbook on the Physics and Chemistry of Rare Earths},
  vol.~14, pp. 225--341,  (North-Holland, Amsterdam,~1991).

\bibitem{pofahl:87}
Pofahl, G., Zirngiebl, E., Blumenr\"oder, S., Brenten, H., and G\"untherodt,
  G.; \enquote{Crystalline-electric field level scheme of {NdB$_6$}};
  \emph{Z.~Phys.~B} \textbf{66}, 339 (1987).

\bibitem{kobayashi:01}
Kobayashi, S., Sera, M., Hiroi, M., Nishizaki, T., Kobayashi, N., and Kunii,
  S.; \enquote{Anisotropic magnetic phase diagram of {PrB$_6$} dominated by the
  {O$_{xy}$} antiferro-quadrupolar interaction}; \emph{J.~Phys. Soc. Jpn.}
  \textbf{70}, 1721 (2001).

\bibitem{kuromaru:02}
Kuromaru, T., Kusunose, H., and Kuramoto, Y.; \enquote{Multipolar ordering in
  {PrB$_6$}}; \emph{J.~Phys. Soc. Jpn.} \textbf{71}, 130 (2002).

\bibitem{nakamura:94}
Nakamura, S., Goto, T., Kunii, S., Iwashita, K., and Tamaki, A.;
  \enquote{Quadrupole-strain interaction in Rare Earth hexaborides};
  \emph{J.~Phys. Soc. Jpn.} \textbf{63}, 623 (1994).

\bibitem{awaji:99}
Awaji, S., Kobayashi, N., Sakatsume, S., Kunii, S., and Sera, M.;
  \enquote{Metamagnetic transition in {NdB$_6$} with a small magnetic
  anisotropy in low magnetic fields}; \emph{J.~Phys. Soc. Jpn.} \textbf{68},
  1518 (1999).

\bibitem{yonemura:09}
Yonemura, T., Tanida, H., Sera, M., and Iga, F.; \enquote{Competition between
  the quadrupole interaction and crystalline electric field effect in the
  antiferromagnetic ordered phase of {NdB$_6$}}; \emph{J.~Phys. Soc. Jpn.}
  \textbf{78}, 114705 (2009).

\bibitem{denlinger:02}
Denlinger, J.~D., Clack, J.~A., Allen, J.~W., Gweon, G.-H., Poirier, D.~M.,
  Olson, C.~G., Sarrao, J.~L., Bianchi, A.~D., and Fisk, Z.; \enquote{Bulk band
  gaps in divalent hexaborides}; \emph{Phys. Rev. Lett.} \textbf{89}, 157601
  (2002).

\bibitem{kim:08}
Kim, J., Kim, Y.-J., Kunes, J., Cho, B.~K., and Choi, E.~J.; \enquote{Optical
  spectroscopy and electronic band structure of ferromagnetic {EuB$_6$}};
  \emph{Phys. Rev.~B} \textbf{78}, 165120 (2008).

\bibitem{kreissl:05}
Kreissl, M. and Nolting, W.; \enquote{Electronic properties of {EuB$_6$} in the
  ferromagnetic regime: Half-metal versus semiconductor}; \emph{Phys. Rev.~B}
  \textbf{72}, 245117 (2005).

\bibitem{pohlit:18}
Pohlit, M., R\"ossler, S., Ohno, Y., Ohno, H., v.~Molnar, S., Fisk, Z.,
  M\"uller, J., and Wirth, S.; \enquote{Evidence for ferromagnetic clusters in
  the colossal magnetoresistance material {EuB$_6$}}; \emph{Phys. Rev.~B}
  \textbf{120}, 257201 (2018).

\bibitem{wigger:04}
Wigger, G.~A., Monnier, R., Ott, H.~R., Young, D.~P., and Fisk, Z.;
  \enquote{Electronic transport in {EuB$_6$}}; \emph{Phys. Rev.~B} \textbf{69},
  125118 (2004).

\bibitem{amara:10}
Amara, M., Galera, R.-M., Aviani, I., and Givord, F.; \enquote{Macroscopic and
  microscopic investigation of the antiferromagnetic phase of {TbB$_6$}};
  \emph{Phys. Rev.~B} \textbf{82}, 224411 (2010).

\bibitem{iwasa:18}
Iwasa, K., Iga, F., Moyoshi, T., Nakao, A., and Ohhara, T.;
  \enquote{Magnetic-ordering propagation vectors of terbium hexaboride
  revisited}; \emph{J.~Phys. Soc. Jpn.} \textbf{87}, 064705 (2018).

\bibitem{goto:00}
Goto, T., Nemoto, Y., Nakano, Y., Nakamura, S., Kajitani, T., and Kunii, S.;
  \enquote{Quadrupolar effect of {HoB$_6$} and {DyB$_6$}}; \emph{Physica B}
  \textbf{281--282}, 586 (2000).

\bibitem{sera:19}
Sera, M., Yonemura, T., Itamochi, K., Matsumura, T., Hiroi, M., and Takahashi,
  K.; \enquote{Not a simple ferro-quadrupole order in {DyB$_6$}};
  \emph{J.~Phys. Soc. Jpn.} \textbf{88}, 054703 (2019).

\bibitem{luethi:73}
L\"uthi, B., Mullen, M.~E., Andres, K., Bucher, E., and Maita, J.~P.;
  \enquote{Experimental investication of the cooperative {Jahn-Teller} effect
  in {TmCd}}; \emph{Phys. Rev.~B} \textbf{8}, 2639 (1973).

\bibitem{mullen:74}
Mullen, M.~E., L\"uthi, B., Wang, P.~S., Bucher, E., Longinotti, L.~D., Maita,
  J.~P., and Ott, H.~R.; \enquote{Magnetic-ion-lattice interaction: Rare-earth
  antimonides}; \emph{Phys. Rev.~B} \textbf{10}, 186 (1974).

\bibitem{kang:16}
Kang, C.-J., Denlinger, J.~D., Allen, J.~W., Min, C.-H., Reinert, F., Kang,
  B.~Y., Cho, B.~K., Kang, J.~S., Shim, J.~H., and Min, B.~I.;
  \enquote{Electronic structure of {YbB$_6$}: Is it a topological insulator or
  not?}; \emph{Phys. Rev. Lett.} \textbf{116}, 116401 (2016).

\bibitem{nakamura:95}
Nakamura, S., Goto, T., and Kunii, S.; \enquote{Magnetic phase diagrams of the
  dense {Kondo} compounds {CeB$_6$} and {Ce$_{0.5}$La$_{0.5}$B$_6$}};
  \emph{J.~Phys. Soc. Jpn.} \textbf{64}, 3941 (1995).

\bibitem{onuki:04}
Onuki, Y., Settai, R., Sugiyama, K., Takeuchi, T., Kobayashi, T.~C., Haga, Y.,
  and Yamamoto, E.; \enquote{Recent advances in the magnetism and
  superconductivity of heavy fermion systems}; \emph{J.~Phys. Soc. Jpn.}
  \textbf{73}, 769 (2004).

\bibitem{thalmeier:05}
Thalmeier, P. and Zwicknagl, G.; \emph{{\rm ``Unconventional superconductivity
  and magnetism in lanthanide and actinide intermetallic compounds''}};
  chap.~219 in \textit{Handbook on the Physics and Chemistry of Rare Earths},
  vol. 34, pp. 135--287,  (Elsevier, Amsterdam,~2005).

\bibitem{thalmeier:05a}
Thalmeier, P., Zwicknagl, G., Stockert, O., Sparn, G., and Steglich, F.;
  \emph{{\rm ``Superconductivity in heavy fermion compounds''}}; in A.~V.
  Narlikar (Ed.), \textit{Frontiers in {S}uperconducting {M}aterials}, pp.
  109--182,  (Springer, Berlin Heidelberg,~2005).

\bibitem{zirngiebl:84}
Zirngiebl, E., Hillebrands, B., Blumenr\"oder, S., G\"untherodt, G.,
  Loewenhaupt, M., Carpenter, J.~M., Winzer, K., and Fisk, Z.;
  \enquote{Crystal-field excitations in {CeB$_6$} studied by {Raman} and
  neutron spectroscopy}; \emph{Phys. Rev.~B} \textbf{30}, 4052 (1984).

\bibitem{zirngiebl:85}
Zirngiebl, E., Hillebrands, B., Blumenr\"oder, S., and G\"untherodt, G.;
  \enquote{New crystal-field level scheme of CeB$_6$ deduced from Raman and
  neutron spectroscopy}; \emph{J. Appl. Phys.} \textbf{57}, 3769 (1985).

\bibitem{YeKung19}
Ye, M., Kung, H.-H., Rosa, P. F.~S., Bauer, E.~D., Fisk, Z., and Blumberg, G.;
  \enquote{Raman spectroscopy of $f$-electron metals: An example of
  ${\mathrm{CeB}}_{6}$}; \emph{Phys. Rev. Materials} \textbf{3}, 065003 (2019).

\bibitem{ohkawa:85}
Ohkawa, F.~J.; \enquote{Orbital antiferromagnetism in {CeB$_6$}};
  \emph{J.~Phys. Soc. Jpn.} \textbf{54}, 3909 (1985).

\bibitem{thalmeier:98}
Thalmeier, P., Shiina, R., Shiba, H., and Sakai, O.; \enquote{Theory of
  multipolar excitations in {CeB$_6$}}; \emph{J.~Phys. Soc. Jpn.} \textbf{67},
  2363 (1998).

\bibitem{shiina:03}
Shiina, R., Shiba, H., Thalmeier, P., Takahashi, A., and Sakai, O.;
  \enquote{Dynamics of multipoles and neutron scattering spectra in quadrupolar
  ordering phase of {CeB$_6$}}; \emph{J.~Phys. Soc. Jpn.} \textbf{72}, 1216
  (2003).

\bibitem{shiina:02}
Shiina, R.; \enquote{Remark on high-field phase diagram of {CeB$_6$}};
  \emph{J.~Phys. Soc. Jpn.} \textbf{71}, 2257 (2002).

\bibitem{goodrich:04}
Goodrich, R.~G., Young, D.~P., Hall, D., Balicas, L., Fisk, Z., Harrison, N.,
  Betts, J., Migliori, A., Woodward, F.~M., and Lynn, J.~W.; \enquote{Extension
  of the temperature-magnetic field phase diagram of {CeB$_6$}}; \emph{Phys.
  Rev.~B} \textbf{69}, 054415 (2004).

\bibitem{shiina:98}
Shiina, R., Sakai, O., Shiba, H., and Thalmeier, P.; \enquote{Interplay of
  field-induced multipoles in {CeB$_6$}}; \emph{J.~Phys. Soc. Jpn.}
  \textbf{67}, 941 (1998).

\bibitem{hiroi:98}
Hiroi, M., Kobayashi, S., Sera, M., Kobayashi, N., and Kunii, S.;
  \enquote{Reentrant behavior and strong anisotropy of the phase boundary
  between antiferro-quadrupolar ordered and paramagnetic phases in
  {Ce$_x$La$_{1-x}$B$_6$} in high magnetic fields}; \emph{Phys. Rev. Lett.}
  \textbf{81}, 2510 (1998).

\bibitem{akatsu:04}
Akatsu, M., Goto, T., Suzuki, O., Nemoto, Y., Nakamura, S., Kunii, S., and
  Kido, G.; \enquote{Magnetic anisotropy of the antiferroquadrupole phase in
  {Ce$_{0.50}$La$_{0.50}$B$_6$}}; \emph{Phys. Rev. Lett.} \textbf{93}, 156409
  (2004).

\bibitem{shiina:01}
Shiina, R.; \enquote{Quadrupolar phase transition and field-dependent
  multipolar fluctuation in {CeB$_6$}}; \emph{J.~Phys. Soc. Jpn.} \textbf{70},
  2746 (2001).

\bibitem{effantin:85}
Effantin, J.~M., Rossat-Mignod, J., Burlet, P., Bartholin, H., Kunii, S., and
  Kasuya, T.; \enquote{Magnetic phase diagram of {CeB$_6$}}; \emph{J.~Magn.
  Magn. Mater.} \textbf{47--48}, 145 (1985).

\bibitem{lovesey:05}
Lovesey, S.~W., Balcar, E., Knight, K.~S., and Rodriguez, J.~F.;
  \enquote{Electronic properties of crystalline materials observed in x-ray
  diffraction}; \emph{Phys. Rep.} \textbf{411}, 233 (2005).

\bibitem{matsumura:12}
Matsumura, T., Yonemura, T., Kunimori, K., Sera, M., Iga, F., Nagao, T., and
  Igarashi, J.; \enquote{Antiferroquadrupole order and magnetic field induced
  octupole in {CeB$_6$}}; \emph{Phys. Rev.~B} \textbf{85}, 174417 (2012).

\bibitem{lemmens:89}
Lemmens, P., Ewert, S., Thalmeier, P., Lenz, D., and Winzer, K.;
  \enquote{Elastic constants and quadrupolar interactions in the {(La,Ce)B$_6$}
  series}; \emph{Z.~Phys.~B} \textbf{76}, 501 (1989).

\bibitem{sato:85}
Sato, N., Sumiyama, A., Kunii, S., Nagano, H., and Kasuya, T.;
  \enquote{Interaction between {Kondo} states and the {Hall} effect of dense
  {Kondo} system {Ce$_x$La$_{1-x}$B$_6$}}; \emph{J.~Phys. Soc. Jpn.}
  \textbf{54}, 1923 (1985).

\bibitem{jang:17}
Jang, D., Portnichenko, P.~Y., Cameron, A.~S., Friemel, G., Dukhnenko, A.,
  Shitsevalova, N.~Y., Filipov, V.~B., Schneidewind, A., Ivanov, A., Inosov,
  D.~S., and Brando, M.; \enquote{Large positive correlation between the
  effective electron mass and the multipolar fluctuation in the heavy
  fermion-metal {Ce$_{1-x}$La$_x$B$_6$}}; \emph{npj Quantum Mater.} \textbf{2},
  62 (2017).

\bibitem{tayama:97}
Tayama, T., Sakakibara, T., Tenya, K., Amitsuka, H., and Kunii, S.;
  \enquote{Magnetic phase diagram of {Ce$_x$La$_{1-x}$B$_6$} studied by static
  magnetization measurement at very low temperatures}; \emph{J.~Phys. Soc.
  Jpn.} \textbf{66}, 2268 (1997).

\bibitem{kobayashi:03}
Kobayashi, S., Yoshino, Y., Tsuji, S., Tou, H., Sera, M., and Iga, F.;
  \enquote{Appearance of the phase {IV} in {Ce$_x$La$_{1-x}$B$_6$} at {$x\sim
  0.8$}}; \emph{J.~Phys. Soc. Jpn.} \textbf{72}, 2947 (2003).

\bibitem{sera:18}
Sera, M., Kunimori, K., Matsumura, T., Kondo, A., Tanida, H., Tou, H., and Iga,
  F.; \enquote{Appearance of the octupole ordered phase {IV} in
  {Ce$_x$La$_{1-x}$B$_6$}}; \emph{Phys. Rev.~B} \textbf{97}, 184417 (2018).

\bibitem{suzuki:98}
Suzuki, O., Goto, T., Nakamura, S., Matsumura, T., and Kunii, S.;
  \enquote{Magnetic phase diagrams of {Kondo} compounds
  {Ce$_{0.75}$La$_{0.25}$B$_6$} and {Ce$_{0.6}$La$_{0.4}$B$_6$}};
  \emph{J.~Phys. Soc. Jpn.} \textbf{67}, 4243 (1998).

\bibitem{akatsu:03}
Akatsu, M., Goto, T., Nemoto, Y., Suzuki, O., Nakamura, S., and Kunii, S.;
  \enquote{Trigonal lattice distortion and ferro-quadrupole ordering in phase
  IV of {Ce$_x$La$_{1-x}$B$_6$} (x=0.75 and 0.70)}; \emph{J.~Phys. Soc. Jpn.}
  \textbf{72}, 205 (2003).

\bibitem{kubo:04}
Kubo, K. and Kuramoto, Y.; \enquote{Octupole ordering model for the phase {IV}
  of {Ce$_{x}$La$_{1-x}$B$_6$}}; \emph{J.~Phys. Soc. Jpn.} \textbf{73}, 216
  (2004).

\bibitem{yamahara:19}
Yamahara, D. and Shiina, R.; \emph{unpublished}  (2019).

\bibitem{magishi:02}
Magishi, K., Kawakami, M., Saito, T., Koyama, K., Mizuno, K., and Kunii, S.;
  \enquote{{$^{11}$B} {NMR} study of {Ce$_x$La$_{1-x}$B$_6$}}; \emph{Z.
  Naturforsch. A} \textbf{57a}, 441 (2002).

\bibitem{inami:14}
Inami, T.; \enquote{Large ferroquadrupole moment induced in the
  octupole-ordered {Ce$_{0.7}$La$_{0.3}$B$_6$} revealed by high-resolution
  x-ray diffraction}; \emph{Phys. Rev.~B} \textbf{90}, 041108 (2014).

\bibitem{mannix:05}
Mannix, D., Tanaka, Y., Carbone, D., Bernhoeft, N., and Kunii, S.;
  \enquote{Order parameter segregation in {Ce$_{0.7}$La$_{0.3}$B$_6$}: 4$f$
  octopole and 5$d$ dipole magnetic order}; \emph{Phys. Rev. Lett.}
  \textbf{95}, 117206 (2005).

\bibitem{kusunose:05}
Kusunose, H. and Kuramoto, Y.; \enquote{Evidence for octupole order in
  {Ce$_{0.7}$La$_{0.3}$B$_6$} from resonant x-ray scattering}; \emph{J.~Phys.
  Soc. Jpn.} \textbf{74}, 3139 (2005).

\bibitem{matsumura:14}
Matsumura, T., Michimura, S., Inami, T., Otsubo, T., Tanida, H., Iga, F., and
  Sera, M.; \enquote{Evidence for hidden quadrupolar fluctuations behind the
  octupole order in {Ce$_{0.7}$La$_{0.3}$B$_6$} from resonant x-ray diffraction
  in magnetic fields}; \emph{Phys. Rev.~B} \textbf{89}, 014422 (2014).

\bibitem{paixao:02}
Paixao, J.~A., Detlefs, C., Longfield, M.~J., Caciuffo, R., Santini, P., and
  Bernhoeft, N.; \enquote{Triple-{{\bf q}} octupolar ordering in {NpO$_2$}};
  \emph{Phys. Rev. Lett.} \textbf{89}, 187202 (2002).

\bibitem{jensen:91}
Jensen, J. and Mackintosh, A.~R.; \emph{{\rm ``Rare Earth magnetism and
  excitations''}},  (Clarendon Press, Oxford,~1991).

\bibitem{haelg:86}
H\"alg, B. and Furrer, A.; \enquote{Anisotropic exchange and spin dynamics in
  the {type-I (-IA)} antiferromagnets {CeAs}, {CeSb}, and {USb}: A neutron
  study}; \emph{Phys. Rev.~B} \textbf{34}, 6258 (1986).

\bibitem{thalmeier:03}
Thalmeier, P., Shiina, R., Shiba, H., Takahashi, A., and Sakai, O.;
  \enquote{Temperature and field dependence of multipolar excitations in
  {CeB$_6$}}; \emph{J.~Phys. Soc. Jpn.} \textbf{72}, 3219 (2003).

\bibitem{kusunose:01}
Kusunose, H. and Kuramoto, Y.; \enquote{Spin-orbital wave excitations in
  orbitally degenerate exchange model with multipolar interactions};
  \emph{J.~Phys. Soc. Jpn.} \textbf{70}, 3076 (2001).

\bibitem{bouvet:93}
Bouvet, A.; \emph{{\rm \'Etude par diffusion in\'elastique de neutrons de
  propri\'et\'es magn\'etiques de borures de terre rare: {CeB$_6$, PrB$_6$ et
  YbB$_{12}$}}}; Ph.D. thesis; L'Universit\'e Joseph Fourier, Grenoble (1993).

\bibitem{jang:14}
Jang, H., Friemel, G., Ollivier, J., Dukhnenko, A.~V., Shitsevalova, N.~Y.,
  Filipov, V.~B., Keimer, B., and Inosov, D.~S.; \enquote{Intense low-energy
  ferromagnetic fluctuations in the antiferromagnet heavy-fermion metal
  {CeB$_6$}}; \emph{Nat. Mater.} \textbf{13}, 682 (2014).

\bibitem{portnichenko:18}
Portnichenko, P.~Y.; \emph{{\rm ``Magnetic dynamics in heavy-fermion systems
  with multipolar ordering''}}; Ph.D. thesis; Technische Universit\"at Dresden
  (2018).

\bibitem{portnichenko:16}
Portnichenko, P.~Y., Demishev, S.~V., Semeno, A.~V., Ohta, H., Cameron, A.~S.,
  Surmach, M.~A., Jang, H., Friemel, G., Dukhnenko, A.~V., Shitsevalova, N.~Y.,
  Filipov, V.~B., Schneidewind, A., Ollivier, J., Podlesnyak, A., and Inosov,
  D.~S.; \enquote{Magnetic field dependence of the neutron spin resonance in
  {CeB$_6$}}; \emph{Phys. Rev.~B} \textbf{94}, 035144 (2016).

\bibitem{DemishevSemeno06}
Demishev, S., Semeno, A., Bogach, A., Paderno, Y., Shitsevalova, N., and
  Sluchanko, N.; \enquote{Magnetic resonance in cerium hexaboride caused by
  quadrupolar ordering}; \emph{J. Magn. Magn. Mater.} \textbf{300}, e534 --
  e537 (2006).

\bibitem{DemishevSemeno08}
Demishev, S.~V., Semeno, A.~V., Ohta, H., Okubo, S., Paderno, Y.~B.,
  Shitsevalova, N.~Y., and Sluchanko, N.~E.; \enquote{High-frequency study of
  the orbital ordering resonance in the strongly correlated heavy fermion metal
  CeB$_6$}; \emph{Appl. Magn. Reson.} \textbf{35}, 319--326 (2008).

\bibitem{DemishevSemeno09}
Demishev, S.~V., Semeno, A.~V., Bogach, A.~V., Samarin, N.~A., Ishchenko,
  T.~V., Filipov, V.~B., Shitsevalova, N.~Y., and Sluchanko, N.~E.;
  \enquote{Magnetic spin resonance in CeB$_6$}; \emph{Phys. Rev. B}
  \textbf{80}, 245106 (2009).

\bibitem{kakizaki:95}
Kakizaki, A., Harasawa, A., Ishii, T., Kashiwakura, T., Kamata, A., and Kunii,
  S.; \enquote{Electronic structure of {CeB$_6$ studied by 3$d$ XPS and
  high-resolution 4$d$-4$f$ resonant photoemission}}; \emph{J.~Phys. Soc. Jpn.}
  \textbf{64}, 302 (1995).

\bibitem{horn:81}
Horn, S., Steglich, F., Loewenhaupt, M., Scheuer, H., Felsch, W., and Winzer,
  K.; \enquote{The magnetic behaviour of {CeB$_6$}: Comparison between elstic
  and inelastic neutron scattering, intial susceptibility and high-field
  magnetization}; \emph{Z.~Phys.~B} \textbf{42}, 125 (1981).

\bibitem{friemel:12}
Friemel, G., Li, Y., Dukhnenko, A.~V., Shitsevalova, N.~Y., Sluchanko, N.~E.,
  Ivanov, A., Filipov, V.~B., Keimer, B., and Inosov, D.~S.; \enquote{Resonant
  magnetic exciton mode in the heavy-fermion antiferromagnet {CeB$_6$}};
  \emph{Nat. Commun.} \textbf{3}, 830 (2012).

\bibitem{eschrig:06}
Eschrig, M.; \enquote{The effect of collective spin-1 excitations on electronic
  spectra in high-$T_{\rm c}$ superconductors}; \emph{Adv. Phys.} \textbf{55},
  47 (2006).

\bibitem{inosov:10}
Inosov, D.~S., Park, J.~T., Bourges, P., Sun, D.~L., Sidis, Y., Schneidewind,
  A., Hradil, K., Haug, D., Lin, C.~T., Keimer, B., and Hinkov, V.;
  \enquote{Normal-state spin dynamics and temperature-dependent spin-resonance
  energy in optimally doped {BaFe$_{1.85}$Co$_{0.15}$As$_2$}}; \emph{Nat.
  Phys.} \textbf{6}, 178 (2010).

\bibitem{korshunov:08}
Korshunov, M.~M. and Eremin, I.; \enquote{Theory of magnetic excitations in
  iron-based layered superconductors}; \emph{Phys. Rev.~B} \textbf{78},
  140509(R) (2008).

\bibitem{onari:10}
Onari, S., Kontani, H., and Sato, M.; \enquote{Structure of Neutron-Scattering
  Peak in both $s_{++}$ wave and $s_{+-}$ wave states of an iron pnictide
  superconductor}; \emph{Phys. Rev.~B} \textbf{81}, 060504(R) (2010).

\bibitem{stock:08}
Stock, C., Broholm, C., Hudis, J., Kang, H.~J., and Petrovic, C.; \enquote{Spin
  resonance in the d-wave superconductor {CeCoIn$_5$}}; \emph{Phys. Rev. Lett.}
  \textbf{100}, 087001 (2008).

\bibitem{eremin:08}
Eremin, I., Zwicknagl, G., Thalmeier, P., and Fulde, P.; \enquote{Feedback spin
  resonance in superconducting {CeCu$_2$Si$_2$} and {CeCoIn$_5$}}; \emph{Phys.
  Rev. Lett.} \textbf{101}, 187001 (2008).

\bibitem{thalmeier:16}
Thalmeier, P. and Akbari, A.; \emph{{\rm ``Resonant spin excitations in
  unconventional heavy fermion superconductors and Kondo lattice compounds''}};
  in J.~Jedrzejewski (Ed.), \textit{Quantum Criticality in Condensed Matter},
  p.~44,  (World Scientific, Singapore,~2016).

\bibitem{akbari:12}
Akbari, A. and Thalmeier, P.; \enquote{Spin exciton formation inside the hidden
  order phase of {CeB$_6$}}; \emph{Phys. Rev. Lett.} \textbf{108}, 146403
  (2010).

\bibitem{ikeda:96}
Ikeda, H. and Miyake, K.; \enquote{A theory of anisotropic semiconductor of
  heavy fermions}; \emph{J.~Phys. Soc. Jpn.} \textbf{65}, 1769 (1996).

\bibitem{hanzawa:98}
Hanzawa, K.; \enquote{Theory of intermediate-valence states in {Sm} compounds};
  \emph{J.~Phys. Soc. Jpn.} \textbf{67}, 3151 (1998).

\bibitem{takimoto:11}
Takimoto, T.; \enquote{{SmB$_6$}: A promising candidate for a topological
  insulator}; \emph{J.~Phys. Soc. Jpn.} \textbf{80}, 123710 (2011).

\bibitem{dzero:10}
Dzero, M., Sun, K., Galitski, V., and Coleman, P.; \enquote{Topological {Kondo}
  insulators}; \emph{Phys. Rev. Lett.} \textbf{104}, 106408 (2010).

\bibitem{paulus:85}
Paulus, E. and Voss, G.; \enquote{Point contact spectra of cerium compounds};
  \emph{J.~Magn. Magn. Mater.} \textbf{47}, 539 (1985).

\bibitem{fuhrmann:15}
Fuhrmann, W.~T., Leiner, J., Nikolic, P., Granroth, G.~E., Stone, M.~B.,
  Lumsden, M.~D., DeBeer-Schmitt, L., Alekseev, P.~A., Koohpayeh, J.-M. M.
  S.~M., Cottingham, P., Phelan, W.~A., Schoop, L., McQueen, T.~M., and
  Broholm, C.; \enquote{Interaction driven subgap spin exciton in the {Kondo}
  insulator {SmB$_6$}}; \emph{Phys. Rev. Lett.} \textbf{114}, 036401 (2015).

\bibitem{fuhrmann:14}
Fuhrmann, W.~T. and Nicolic, P.; \enquote{In-gap collective mode spectrum of
  the topological {Kondo} insulator {SmB$_6$}}; \emph{Phys. Rev.~B}
  \textbf{90}, 195144 (2014).

\bibitem{nemkovski:07}
Nemkovski, K.~S., Mignot, J.-M., Alekseev, P.~A., Ivanov, A.~S., Nefeodova,
  E.~V., Rybina, A.~V., Regnault, L.-P., Iga, F., and Takabatake, T.;
  \enquote{Polarized-neutron study of spin dynamics in the {Kondo} insulator
  {YbB$_{12}$}}; \emph{Phys. Rev. Lett.} \textbf{99}, 137204 (2007).

\bibitem{okamura:05}
Okamura, H., Michizawa, T., Nanba, T., i.~Kimura, S., Iga, F., and Takabatake,
  T.; \enquote{Indirect and direct energy gaps in {Kondo} semiconductor
  {YbB$_{12}$}}; \emph{J.~Phys. Soc. Jpn.} \textbf{74}, 1954 (2005).

\bibitem{akbari:09}
Akbari, A., Thalmeier, P., and Fulde, P.; \enquote{Theory of Spin Exciton in
  the {Kondo} Semiconductor {YbB$_{12}$}}; \emph{Phys. Rev. Lett.}
  \textbf{102}, 106402 (2009).

\bibitem{bourdarot:10}
Bourdarot, F., Hassinger, E., Raymond, S., Aoki, D., Taufour, V., Regnault,
  L.-P., and Flouquet, J.; \enquote{Precise study of the resonance at {{\bf
  Q}$_0$=$(1,0,0)$} in {URu$_2$Si$_2$}}; \emph{J.~Phys. Soc. Jpn.} \textbf{79},
  064719 (2010).

\bibitem{aynajian:10}
Aynajian, P., da~Silva~Neto, E.~H., Parker, C.~V., Huang, Y., Pasupathy, A.,
  Mydosh, J., and Yazdani, A.; \enquote{Visualizing the formation of the
  {Kondo} lattice and the hidden order in {URu$_2$Si$_2$}}; \emph{Proc. Nat.
  Acad. Sci.} \textbf{107}, 10383 (2010).

\bibitem{tazai:19}
Tazai, R. and Kontani, H.; \enquote{Multipole fluctuation theory for heavy
  fermion systems: Application to multipole orders in {CeB$_6$}};
  \emph{arXiv:1901.06213}  (2019).

\bibitem{xiang:18}
Xiang, Z., Kasahara, Y., Asaba, T., Lawson, B., Tinsman, C., Chen, L.,
  Sugimoto, K., Kawaguchi, S., Sato, Y., Li, G., Yao, S., Chen, Y.~L., Iga, F.,
  Singleton, J., Matsuda, Y., and Li, L.; \enquote{Quantum oscillations of
  electrical resistivity in an insulator}; \emph{Science} \textbf{362}, 65
  (2018).

\bibitem{sato:19}
Sato, Y., Xiang, Z., Kasaara, Y., Taniguchi, T., Kasahara, S., Chen, L., Asaba,
  T., Tinsman, C., Murayama, H., Tanaka, O., Mizukami, Y., Shibauchi, T., Iga,
  F., Singleton, J., and Li, L.; \enquote{Unconventional thermal metallic state
  of charge-neutral fermions in an insulator}; \emph{Nat. Phys.} (in press),
  doi: 10.1038/s41567--019--0552--2 (2019).

\bibitem{utsumi:17}
Utsumi, Y., Kasinathan, D., Ko, K.-T., Agrestini, S., Haverkort, M.~W., Wirth,
  S., Wu, Y.-H., Tsuei, K.-D., Kim, D.-J., Fisk, Z., Tanaka, A., Thalmeier, P.,
  and Tjeng, L.~H.; \enquote{Bulk and surface electronic properties of
  {SmB$_6$}: A hard x-ray photoelectron spectroscopy study}; \emph{Phys.
  Rev.~B} \textbf{96}, 155130 (2017).

\bibitem{mizumaki:09}
Mizumaki, M., Tsutsui, S., and Iga, F.; \enquote{Temperature dependence of {Sm}
  valence in {SmB$_6$} studied by X-ray absorption spectroscopy};
  \emph{J.~Phys.: Conf. Ser.} \textbf{176}, 012034 (2009).

\bibitem{emi:18}
Emi, N., Kawamura, N., Mizumaki, M., Koyama, T., Ishimatsu, N., Pristas, G.,
  Kagayama, T., Shimizu, K., Osanai, Y., Iga, F., and Mito, T.;
  \enquote{Kondo-like behavior near the magnetic instability in {SmB$_6$}:
  Temperature and pressure dependences of the {Sm} valence}; \emph{Phys.
  Rev.~B} \textbf{97}, 161116 (2018).

\bibitem{gorshunov:99}
Gorshunov, B., Sluchanko, N., Dressel, A. V.~M., Knebel, G., Loidl, A., and
  Kunii, S.; \enquote{Low-energy electrodynamics of {SmB$_6$}}; \emph{Phys.
  Rev.~B} \textbf{59}, 1808 (1999).

\bibitem{menth:69}
Menth, A., Buehler, E., and Geballe, T.~H.; \enquote{Magnetic and
  semiconducting properties of {SmB$_6$}}; \emph{Phys. Rev. Lett.} \textbf{22},
  295 (1969).

\bibitem{allen:79}
Allen, J.~W., Batlogg, B., and Wachter, P.; \enquote{Large low-temperature Hall
  effect and resistivity in mixed-valent {SmB$_6$}}; \emph{Phys. Rev.~B}
  \textbf{20}, 4807 (1979).

\bibitem{fu:06}
Fu, L. and Kane, C.~L.; \enquote{Time reversal polarization and a {Z$_2$}
  adiabatic spin pump}; \emph{Phys. Rev.~B} \textbf{74}, 195312 (2006).

\bibitem{fu:07}
Fu, L. and Kane, C.~L.; \enquote{Topolocigal insulators with inversion
  symmetry}; \emph{Phys. Rev.~B} \textbf{76}, 045302 (2007).

\bibitem{kim:13}
Kim, D.~J., Thomas, S., Grant, T., Botimer, J., Fisk, Z., and Xia, J.;
  \enquote{Surface {Hall} effect and nonlocal transport in {SmB$_6$}: evidence
  for surface conduction}; \emph{Sci. Rep.} \textbf{3}, 3150 (2013).

\bibitem{legner:14}
Legner, M., R\"uegg, A., and Sigrist, M.; \enquote{Topological invariants,
  surface states, and interaction-driven phase transitions in correlated
  {Kondo} insulators with cubic symmetry}; \emph{Phys. Rev.~B} \textbf{89},
  085110 (2014).

\bibitem{legner:15}
Legner, M., R\"uegg, A., and Sigrist, M.; \enquote{Surface-state spin textures
  and mirror {Chern} numbers in topological {Kondo} insulators}; \emph{Phys.
  Rev. Lett.} \textbf{115}, 156405 (2015).

\bibitem{ando:13}
Ando, Y.; \enquote{Topological insulator materials}; \emph{J.~Phys. Soc. Jpn.}
  \textbf{82}, 102001 (2013).

\bibitem{hoefer:14}
Hoefer, K., Becker, C., Rata, D., Swanson, J., Thalmeier, P., and Tjeng, L.~H.;
  \enquote{Intrinsic conduction through topological surface states of
  insulating {Bi$_2$Te$_3$} epitaxial thin films}; \emph{PNAS} \textbf{111},
  14979 (2014).

\bibitem{antonov:02}
Antonov, V.~N., Harmon, B.~N., and Yaresko, A.~N.; \enquote{Electronic
  structure of mixed-valence semiconductors in the {LSDA+U} approximation.
  {II.} {SmB$_6$} and {YbB$_{12}$}}; \emph{Phys. Rev.~B} \textbf{66}, 165209
  (2002).

\bibitem{kang:15}
Kang, C.-J., Kim, J., Kim, K., Kang, J., Denlinger, J.~D., and Min, B.~I.;
  \enquote{Band symmetries of mixed-valence topological insulator: {SmB$_6$}};
  \emph{J.~Phys. Soc. Jpn.} \textbf{84}, 024722 (2015).

\bibitem{kim:14}
Kim, J., Kim, K., Kang, C.-J., Kim, S., Choi, H.~C., Kang, J.-S., Denlinger,
  J.~D., and Min, B.~I.; \enquote{Termination-dependent surface in-gap states
  in a potential mixed-valent topological insulator: {SmB$_6$}}; \emph{Phys.
  Rev.~B} \textbf{90}, 075131 (2014).

\bibitem{lu:13}
Lu, F., Zhao, J.~Z., Weng, H., Fang, Z., and Dai, X.; \enquote{Correlated
  topological insulator with mixed valence}; \emph{Phys. Rev. Lett.}
  \textbf{110}, 096401 (2013).

\bibitem{alexandrov:13}
Alexandrov, V., Dzero, M., and Coleman, P.; \enquote{Cubic topological {Kondo}
  insulators}; \emph{Phys. Rev. Lett.} \textbf{111}, 226403 (2013).

\bibitem{hasan:10}
Hasan, M.~Z. and Kane, C.~L.; \enquote{Colloquium: Topological insulators};
  \emph{Rev. Mod. Phys.} \textbf{82}, 3054 (2010).

\bibitem{tran:12}
Tran, M.-T., Takimoto, T., and Kim, K.-S.; \enquote{Phase diagram for a
  topological {Kondo} insulating system}; \emph{Phys. Rev.~B} \textbf{85},
  125128 (2012).

\bibitem{denlinger:14}
Denlinger, J.~D., Allen, J.~W., Kang, J.-S., Sun, K., Min, B.-I., Kim, D.-J.,
  and Fisk, Z.; \enquote{{SmB$_6$} photoemission: past and present}; \emph{JPS
  Conf. Proc.} \textbf{3}, 017038 (2014).

\bibitem{jiang:13}
Jiang, J., Li, S., Zhang, T., Sun, Z., Chen, F., Ye, Z.~R., Xu, M., Ge, Q.~Q.,
  Tan, S.~Y., Niu, X.~H., Xia, M., Xie, B.~P., Li, Y.~F., Chen, X.~H., Wen,
  H.~H., and Feng, D.~L.; \enquote{Observation of possible topological in-gap
  surface states in the Kondo insulator {SmB$_6$} by photoemission}; \emph{Nat.
  Commun.}  (2013).

\bibitem{xu:14}
Xu, N., Biswas, P.~K., Dil, J.~H., Dhaka, R.~S., Landolt, G., Muff, S., Matt,
  C.~E., Shi, X., Plumb, N.~C., Radovic, M., Pomjakushina, E., Conder, K.,
  Amato, A., Borisenko, S.~V., Yu, R., Weng, H.-M., Fang, Z., Dai, X., Mesot,
  J., Ding, H., and Shi, M.; \enquote{Direct observation of the spin texture in
  {SmB$_6$} as evidence of the topological {Kondo} insulator}; \emph{Nat.
  Commun.}  (2014).

\bibitem{roessler:14}
R\"ossler, S., Jang, T.-H., Kim, D.-J., Tjeng, L.~H., Fisk, Z., Steglich, F.,
  and Wirth, S.; \enquote{Hybridization gap and {Fano} resonance in {SmB$_6$}};
  \emph{PNAS} \textbf{111}, 4798 (2014).

\bibitem{matt:18}
Matt, C.~E., Pirie, H., Soumyanarayanan, A., Yee, M.~M., He, Y., Larson, D.~T.,
  Paz, W.~S., Palacios, J.~J., Hamidian, M.~H., and Hoffman, J.~E.;
  \enquote{Consistency between {ARPES} and {STM} measurements on {SmB$_6$}};
  \emph{arXiv:1810.13442}  (2018).

\bibitem{pirie:18}
Pirie, H., Liu, Y., Soumyanarayanan, A., Chen, P., He, Y., Yee, M.~M., Rosa, P.
  F.~S., Thompson, J.~D., Kim, D.-J., Fisk, Z., Wang, X., Paglione, J., Morr,
  D.~K., Hamidian, M.~H., and Hoffman, J.~E.; \enquote{Imaging emergent heavy
  {Dirac} fermions of a topological {Kondo} insulator}; \emph{arXiv:1810.13419}
   (2018).

\bibitem{watanuki:05}
Watanuki, R., Sato, G., Suzuki, K., Ishihara, M., Yanagisawa, T., Nemoto, Y.,
  and Goto, T.; \enquote{Geometrical quadrupolar frustration in {DyB$_4$}};
  \emph{J.~Phys. Soc. Jpn.} \textbf{74}, 2169 (2005).

\bibitem{inami:09}
Inami, T., Ohwada, K., Matsuda, Y.~H., Ouyang, Z.~W., Nojiri, H., Matsumura,
  T., Okuyama, D., and Murakami, Y.; \enquote{Resonant magnetic x-ray
  diffractio study of successive metamagnetic transitions in {TbB$_4$}};
  \emph{J.~Phys. Soc. Jpn.} \textbf{78}, 033707 (2009).

\bibitem{yoshii:08}
Yoshii, S., Yamamoto, T., Hagiwara, M., Michimura, S., Shigekawa, A., Iga, F.,
  Takabatake, T., and Kindo, K.; \enquote{Multistep magnetization plateaus in
  the {Shastry-Sutherland} system {TbB$_4$}}; \emph{Phys. Rev. Lett.}
  \textbf{101}, 087202 (2008).

\bibitem{matas:10}
Matas, S., Siemensmeyer, K., Wheeler, E., Wulf, E., Beyer, R.,
  Hermannsd\"orfer, T., Ignatchik, O., Uhlarz, M., Flachbart, K., Gabani, S.,
  Priputen, P., Efdokimova, A., and Shitsevalova, N.; \enquote{Magnetism of
  rare earth tetraborides}; \emph{J.~Phys.: Conf. Ser.} \textbf{200}, 032041
  (2009).

\bibitem{siemensmeyer:08}
Siemensmeyer, K., Wulf, E., Mikeska, H.-J., Flachbart, K., Gabani, S., Matas,
  S., Priputen, P., Efdokimova, A., and Shitsevalova, N.; \enquote{Fractional
  magnetization plateaus and magnetic order in the {Shastry-Sutherland} magnet
  {TmB$_4$}}; \emph{Phys. Rev. Lett.} \textbf{101}, 177201 (2008).

\bibitem{matsumura:11}
Matsumura, T., Okuyama, D., Mouri, T., and Murakami, Y.; \enquote{Successive
  magnetic phase transitions of component orderings in {DyB$_4$}};
  \emph{J.~Phys. Soc. Jpn.} \textbf{80}, 074701 (2011).

\bibitem{schmidt:17}
Schmidt, B. and Thalmeier, P.; \enquote{Frustrated two dimensional quantum
  magnets}; \emph{Phys. Rep.} \textbf{703}, 1 (2017).

\bibitem{huo:13}
Huo, L., Huang, W.~C., Yan, Z.~B., Jia, X.~T., Gao, X.~S., Qin, M.~H., and Liu,
  J.-M.; \enquote{The competing spin orders and fractional magnetization
  plateaus of the classical {Heisenberg} model on {Shastry-Sutherland} lattice:
  Consequence of long-range order}; \emph{J.~Appl. Phys.} \textbf{113}, 073908
  (2013).

\bibitem{yamauchi:99}
Yamauchi, H., Onodera, H., Ohoyama, K., Onimaru, T., Kosaka, M., Ohashi, M.,
  and Yamaguchi, Y.; \enquote{Antiferroquadrupolar ordering and magnetic
  properties of the tetragonal {DyB$_2$C$_2$} compound}; \emph{J.~Phys. Soc.
  Jpn.} \textbf{68}, 2057 (1999).

\bibitem{hirota:00}
Hirota, K., Oumi, N., Matsumura, T., Nakao, H., Wakabayashi, Y., Murakami, Y.,
  and Endoh, Y.; \enquote{Direct observation of antiferroquadrupolar ordering:
  Resonant x-ray scattering study of {DyB$_2$C$_2$}}; \emph{Phys. Rev. Lett.}
  \textbf{84}, 2706 (2000).

\bibitem{matsumura:02}
Matsumura, T., Oumi, N., Hirota, K., Nakao, H., Murakami, Y., Wakabayashi, Y.,
  Arima, T., Ishihara, S., and Endoh, Y.; \enquote{Observation of the
  antiferroquadrupolar order in {DyB$_2$C$_2$} by resonant x-ray scattering};
  \emph{Phys. Rev.~B} \textbf{65}, 094420 (2002).

\bibitem{kaneko:03}
Kaneko, K., Onodera, H., Yamauchi, H., Sakon, T., Motokawa, M., and Yamaguchi,
  Y.; \enquote{Magnetic phase diagrams with possible field-induced
  antiferroquadrupolar order in {TbB$_2$C$_2$}}; \emph{Phys. Rev.~B}
  \textbf{68}, 012401 (2003).

\bibitem{yanagisawa:05}
Yanagisawa, T., Goto, T., Nemoto, Y., Watanuki, R., Suzuki, K., Suzuki, O., and
  Kido, G.; \enquote{Magnetic phase diagram of antiferroquadrupole ordering in
  {HoB$_2$C$_2$}}; \emph{Phys. Rev.~B} \textbf{71}, 104416 (2005).

\end{thebibliography}

\printindex

\end{document}